\documentclass[aps,prb,onecolumn,superscriptaddress]{revtex4-1}
\usepackage{amsmath,amsthm,amssymb}
\usepackage[dvipdfmx]{graphicx}
\usepackage{amsmath,bm}
\usepackage{color}
\usepackage{mdwlist}

\newcommand{\1}{\mbox{1}\hspace{-0.25em}\mbox{l}}

\newlength{\figwidth}
\newlength{\figlarge}
\begin{document}
\title{
Exceptional band touching for strongly correlated systems in equilibrium
}
\author{Tsuneya Yoshida}
\affiliation{Department of Physics, University of Tsukuba, Ibaraki 305-8571, Japan}
\author{Robert Peters}
\affiliation{Department of Physics, Kyoto University, Kyoto 606-8502, Japan}
\author{Norio Kawakami}
\affiliation{Department of Physics, Kyoto University, Kyoto 606-8502, Japan}
\author{Yasuhiro Hatsugai}
\affiliation{Department of Physics, University of Tsukuba, Ibaraki 305-8571, Japan}
\date{\today}
\begin{abstract}
Quasi-particles described by Green's functions of equilibrium systems exhibit non-Hermitian topological phenomena because of their finite lifetime.
This non-Hermitian perspective on equilibrium systems provides new insights into correlated systems and attracts much interest because of its potential to solve open questions in correlated compounds.
In this paper, we provide a concise review of the non-Hermitian topological band structures for quantum many-body systems in equilibrium as well as their classification.
\end{abstract}
\pacs{
***
} 
\maketitle

\tableofcontents

\section{
Introduction
}
After the discovery of topological insulators/superconductors, the topological perspective of condensed matter is of growing importance~\cite{Hatsugai_PRL93,Kane_PRL05_1,Kane_PRL05_2,Bernevig_BHZ_Science06,Qi_TQFTZ2TI_PRB08,Hasan_RMP10,Qi_RMP10}.
While the notion of topology has been originally utilized to understand the band structure of a gapped quadratic Hamiltonian (i.e., free-fermion systems), it has been extended to gapless systems; it has elucidated that there exists topologically protected band touching for Weyl semi-metals~\cite{XWan_PRB11_Weyl,HWeng_PRX15_Weyl,SYXu_Science15_Weyl,BQLv_PRX15_Weyl} or nodal line superconductors~\cite{Ryu_Majoranaedge_PRL02}.
The notion of topological phases has been further extended to correlated systems where correlations and topology induce a variety of exotic phenomena~\cite{Raghu_CITI_PRL08,Mong_AFTI_PRB10,MHohenadler_PRL11_corr_topo,Yamaji_CorrTI_PRB11,SLYu_PRL11_corr_topo,Yoshida_TI_plusU_DMFT_PRB12,TBI_Mott_Tada,Gurarie_GFSymm_PRB11,Essin_GFSymm_PRB11,Hohenadler_Review_IOP_13,Rachel_Review_RPP18}, such as 
topologically ordered systems~\cite{Tsui_FQHEExp_PRL82,Laughlin_FQHE_PRL83,Jain_FQHE_PRL89,Wen_TopoOrder_SciAdv95,Kitaev_ToricCode_Elsevier03,Hamma_3DToricCode_PRB05,Kitaev_KitaevHoneyconmb_Elsevier06,Tang_FChern_PRL11,Sun_FChern_PRL11,Neupert_FChern_PRL11,Regnalt_FChen_PRX11,Sheng_FChern_NComm12,Bergholtz_FChern_IntJModPhysB13}, 
topological Mott insulators~\cite{Pesin_TM_NatPhys10,Manmana_TMI1D_PRB12,Yoshida_TMI1D_PRL14,Yoshida_TMI2D_PRB16,Kudo_HOTMI_PRL19}, 
and the reduction of topological classifications~\cite{Z_to_Zn_Fidkowski_10,Fidkowski_1Dclassificatin_11,Turner11,Ryu_Z_to_Z8_2013,YaoRyu_Z_to_Z8_2013,Qi_Z_to_Z8_2013,Lu_CS_2011,Levin_CS_2012,Hsieh_CS_CPT_2014,Wang_Potter_Senthil2014,Isobe_Fu2015,Yoshida_TCI_bosonization_PRB15,Morimoto_2015,Yoshida_ZtoZ16_superlattice_PRL17,Yoshida_ZtoZ4_coldatom_PRL18}.

Intriguingly, recent studies have elucidated that correlations induce even non-Hermitian topological phenomena~\cite{VKozii_nH_arXiv17,Zyuzin_nHEP_PRB18,Yoshida_EP_DMFT_PRB18,Yoshida_SPERs_PRB19,Papaji_nHEP_PRB19,Kimura_SPERs_PRB19,Michishta_EP_DMFT_arXiv19,Matsushita_ER_PRB19} 
which are extensively analyzed in various contexts~\cite{Hatano_nH_PRL96,Hatano_nH_PRB98,CMBender_nH_PRL98,Bender_nH_JMP99,Esaki_nHTI_PRB11,Sato_nHTI_PTEP12,Fukui_nHcorr_PRB98,TELeePRL16_Half_quantized,TKato_EP_book1966,HShen2017_non-Hermi,YXuPRL17_exceptional_ring,Carlstrom_nHKER_PRB19,SYao_nHSkin-1D_PRL18,SYao_nHSkin-1D_PRL18,KFlore_nHSkin_PRL18,EElizabet_PRBnHSkinHOTI_PRB19,Yokomizo_BBC_PRL19,Okuma_nHBBCpg_PRL19,Xiao_nHSkin_Exp_arXiv19,Alvarez_nHSkin_PRB18,SYao_nHSkin-1D_PRL18,Lee_Skin19,Zhang_BECskin19,Okuma_BECskin19,Gong_class_PRX18,Kawabata_Class_NatComm2019,Kawabata_gapped_PRX19,Zhou_gapped_class_PRB19,Budich_SPERs_PRB19,Okugawa_SPERs_PRB19,Yoshida_SPERs_PRB19,Zhou_SPERs_Optica19,Kawabata_gapless_PRL19,Ghatak_mechSkin_arXiv19,Scheibner_mechEP_arXiv20,McClarty_nHbEP_PRB19,Bergholtz_nHJunc_PRR19,Bergholtz_Review19}
(e.g., 
photonic systems~\cite{Guo_nHExp_PRL09,Ruter_nHExp_NatPhys10,Szameit_PRA11,Regensburger_nHExp_Nat12,BZhen_nH-PHC_Nat15,Hassan_EP_PRL17,Feng_nH-PHC_NatPhoto17,Takata_nH-PHC_PRL18,Zhou_BFarc_PHC_Science18,Takata_nH-PHC_OSA19,Ozawa_nHPHC_PMP19}, 
open quantum systems~\cite{Gong_ZenoHall_PRL17,Liu_nHSndTI_PRL19,Hatano_nHopen_MolPhys19,Yoshida_nHFQH19,Ashida_nHbHubb_PRA16,Ashida_PTcritical_NatComm17,Nakagawa_nHKondo_PRL18,Yamamoto_nHBCS_arXiv19,Shitaba_nHopen_PRB19,Scazza_2bdlossExp_NatPhys14,Pagano_2bdloss_PRL15,Hoefer_2bdlossExp_PRL15,Riegger_2bdlossExp_PRL18,Tomita_oneLoss_Science17}
 etc.).
In particular, Ref.~\onlinecite{VKozii_nH_arXiv17} has pointed out that the finite lifetime of quasi-particles induces an exceptional point (EP) in the Brillouin zone (BZ) which is a representative example of the non-Hermitian topological band structure.
Correspondingly, topologically protected band touching occurs both for the real and imaginary parts which we call exceptional band touching in this paper. 
The above EPs in many-body systems in equilibrium are connected by Fermi arcs, meaning that correlation induces the gapless excitations even for band insulators. 
The emergence of the EPs accompanied by the Fermi arcs is numerically demonstrated by applying the dynamical mean-field theory (DMFT) to heavy fermions~\cite{Yoshida_EP_DMFT_PRB18}.
The above non-Hermitian perspective of Green's functions for equilibrium systems has been further developed with symmetry of many-body Hamiltonians~\cite{Budich_SPERs_PRB19,Okugawa_SPERs_PRB19,Yoshida_SPERs_PRB19,Zhou_SPERs_Optica19,Kawabata_gapless_PRL19}; the interplay between symmetry and non-Hermiticity results in symmetry-protected exceptional rings (SPERs) in two dimensions~\cite{Yoshida_SPERs_PRB19} and symmetry-protected exceptional surfaces (SPESs) in three dimensions~\cite{Yoshida_SPERs_PRB19,Kimura_SPERs_PRB19}.
The above recently developed non-Hermitian perspective in equilibrium systems attracts much interest because it provides new insights into quasi-particle spectrums which potentially solve open questions in condensed matter physics~\cite{Horio_PGapCu_NatCom16,BTan_Science15_OscillationSmB6,ZXiang_OsciYbB12_Science18,HLiu_IOP18_OscillationYbB12}.

The aim of this article is to provide a concise review of these advances in the non-Hermitian perspective in correlated systems in equilibrium.
As a $2\times 2$ Hamiltonian describes the essential properties, we start with this simplest case and review numerical results demonstrating the emergence of exceptional band touching.

The rest of this paper is organized as follows. In Sec.~\ref{sec: EP}, we demonstrate the emergence of EPs for a heavy-fermion system by applying the DMFT to a heavy-fermion system. In Sec.~\ref{sec: SPER}, we show that SPERs and SPESs can emerge for correlated systems with chiral symmetry. In Sec.~\ref{sec: 10-fold way of SPEP}, we address the ten-fold way classification of the exceptional band touching for single-particle spectrum by taking into account $PT$- ($CP$-) and chiral symmetry, 
where $PT$- ($CP$-) symmetry denotes the symmetry under the product of time-reversal and inversion (charge conjugation and inversion), respectively.
A short summary and remaining open questions appear at the end of this paper.

\section{
Exceptional points for strongly correlated systems
}
\label{sec: EP}

In this section, we elucidate that the EPs emerge due to finite lifetimes of quasi-particles for strongly correlated systems~\cite{Yoshida_EP_DMFT_PRB18}.
Specifically, the origin of the above non-Hermitian topological phenomena is the imaginary part of the self-energy [see Eqs.~(\ref{eq: EP Dyson eq})~and~(\ref{eq: EP Heff in Awk})] which describes the lifetime of quasi-particles.
The emergence of EPs results in the significant difference of the single-particle spectrum.

In the following, after a brief explanation of EPs (Sec.~\ref{sec: EP gen}) and the single-particle Green's function (Sec.~\ref{sec: EP GF}), we demonstrate the emergence of EPs for heavy-fermion systems and see that EPs significantly change the single-particle spectrum.

\subsection{
Topological properties of EPs
}
\label{sec: EP gen}

\subsubsection{
Case of a $2\times 2$ Hamiltonian
}
\label{sec: EP 2x2}

Let us first analyze a non-Hermitian $2\times 2$ Hamiltonian, which elucidates the essential properties of EPs.

It is well-known that a generic $2\times 2$ matrix can be expanded by the Pauli matrices $\tau$'s and the identity matrix $\tau_0$
%
\begin{eqnarray}
H(\bm{k}) &=& \sum_\mu [b_\mu(\bm{k})+id_\mu(\bm{k})]\tau_\mu, \label{eq: 2x2 nonHermi} 
\end{eqnarray}
%
where $b_\mu$ and $d_\mu$ ($\mu=0,1,2,3$) are continuous functions taking real values. 

\begin{figure}[!h]
\begin{minipage}{0.3\hsize}
\begin{center}
\includegraphics[width=\hsize]{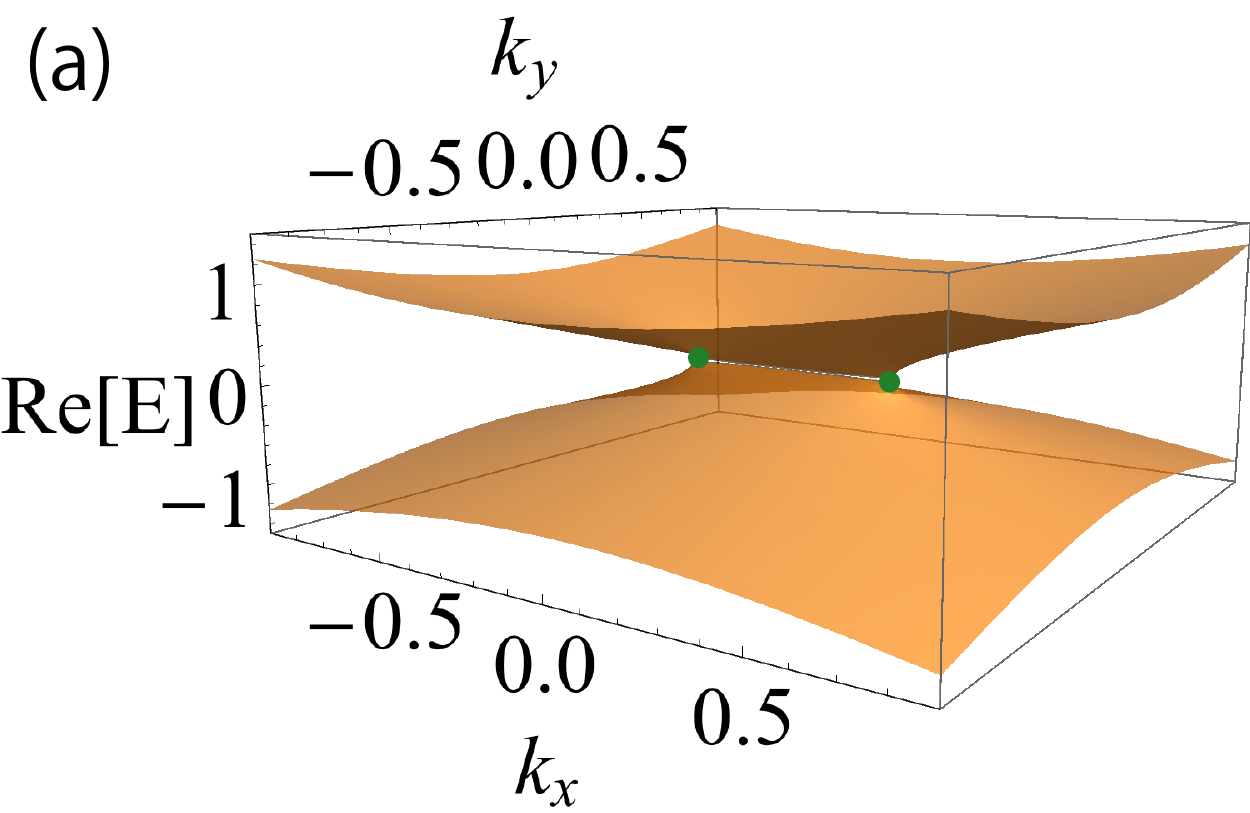}
\end{center}
\end{minipage}
\begin{minipage}{0.3\hsize}
\begin{center}
\includegraphics[width=\hsize]{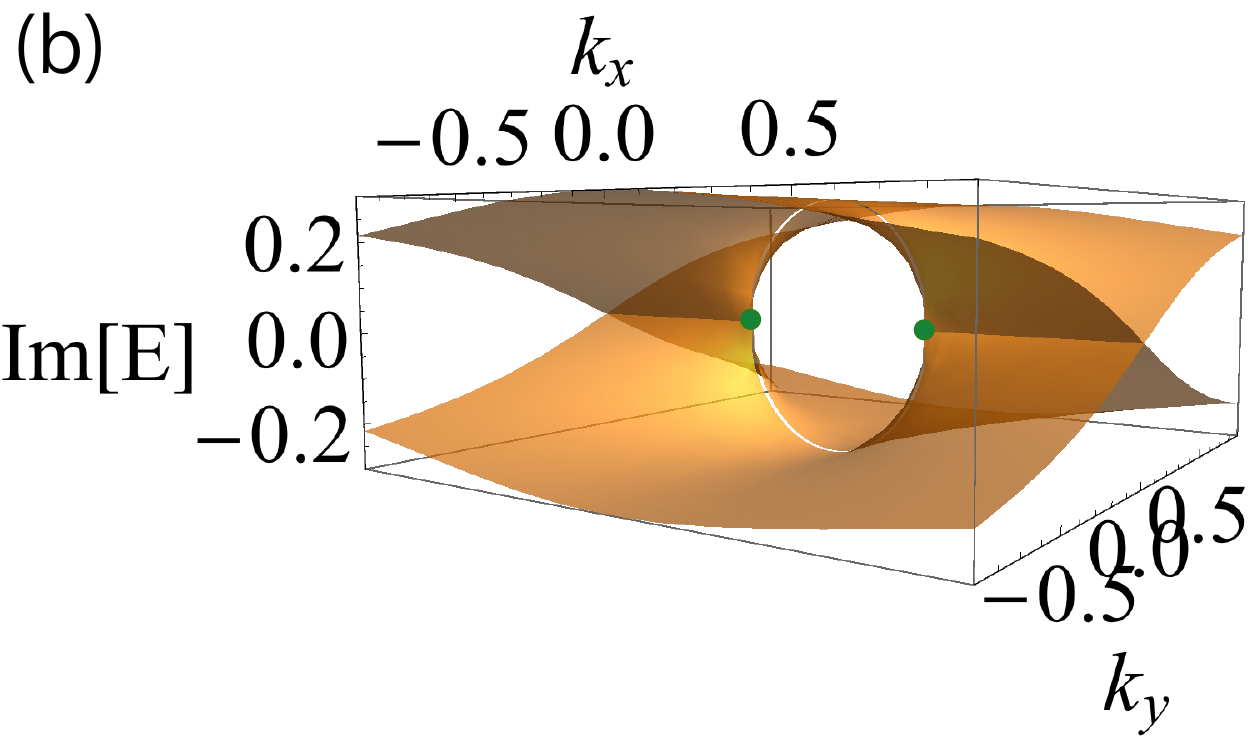}
\end{center}
\end{minipage}
\caption{
Band structure of a continuum model for
$
\left(
\begin{array}{cccc}
b_0, & b_1, & b_2, & b_3
\end{array}
\right)
=
\left(
\begin{array}{cccc}
0, & k_x, & 0, & k_y
\end{array}
\right)
$ 
and 
$
\left(
\begin{array}{cccc}
d_0, & d_1, & d_2, & d_3
\end{array}
\right)
=
\left(
\begin{array}{cccc}
0, & 0, & 0, & 0.3
\end{array}
\right)
$.
In panel (a) [(b)] the real (imaginary) part of the energy is plotted, respectively.
In these figures, the band touching points marked with green dots correspond to the EPs.
}
\label{fig: EP continu}
\end{figure}

One can numerically and analytically confirm that the above non-Hermitian matrix may show EPs. 
In Fig.~\ref{fig: EP continu}, energy eigenvalues taking complex numbers are plotted for a specific choice of $b$'s and $d$'s.
At the EPs, the Hamiltonian becomes non-diagonalizable. Correspondingly, as one can see in Fig.~\ref{fig: EP continu}, the band touching occurs both for the real and imaginary parts of the energy eigenvalues.

In order to see the details, we diagonalize the Hamiltonian~(\ref{eq: 2x2 nonHermi}), which yields
\begin{eqnarray}
E_{\pm}&=& b_0+id_0 \pm \sqrt{b^2-d^2+2i \bm{b}\cdot\bm{d}},
\label{eq: 2x2 Epm}
\end{eqnarray}
with $\bm{b}\cdot\bm{d}:=\sum_{j=1,2,3} b_j d_j$, $b^2=\bm{b}\cdot\bm{b}$ and $d^2=\bm{d}\cdot\bm{d}$.
The above equation indicates that band touching occurs both for the real and imaginary parts when the following conditions are satisfied
\begin{subequations}
\label{eq: cond EP 2x2}
\begin{eqnarray}
b^2-d^2 &=& 0, \\
\bm{b}\cdot \bm{d} &=& 0.
\end{eqnarray}
\end{subequations}
In other words, the above conditions are necessary conditions for the emergence of EPs. 
One can see that the above conditions are indeed sufficient conditions; supposing that Eq.~(\ref{eq: cond EP 2x2}) is satisfied, we can see  that the Hamiltonian can be rewritten as
\begin{eqnarray}
H(\bm{k}) &=& (b_0+id_0)\tau_0 +2d
\left(
\begin{array}{cc}
0 & 1 \\
0 & 0
\end{array}
\right),
\end{eqnarray}
with a proper choice of the basis.
In this basis, one can see that the Hamiltonian is generically non-diagonalizable for $d\neq 0$.

In the above, we have seen the following facts. At the EP, the $2\times2$ Hamiltonian~(\ref{eq: 2x2 nonHermi}) becomes non-diagonalizable, resulting in the exceptional band touching. For the $2\times2$ Hamiltonian, the EP emerges if and only if Eqs.~(\ref{eq: cond EP 2x2}a)~and~(\ref{eq: cond EP 2x2}b) are satisfied.
We note that the band touching is protected by non-trivial topology whose topological invariant is discussed in the next subsection.

\subsubsection{
Topological invariant characterizing EPs
}
\label{sec: EP vorticity}

As shown in Fig.~\ref{fig: EP continu}, the band touching occurs at the EPs.
Such band touching for two-dimensional systems can be topologically characterized by the vorticity, akin to the winding number;
\begin{eqnarray}
\label{eq: vorticity}
\nu &=& \frac{1}{4\pi i} \oint \! d^2\bm{k} \cdot \bm{\nabla}_{\bm{k}} \log \mathrm{det}[H(\bm{k})-E_0\1].
\end{eqnarray}
Here, we have considered that the band touching occurs at energy $E_{0}$. $H(\bm{k})$ ($\mathrm{dim} H \geq 2$) denotes a generic non-Hermitian matrix.
$\bm{\nabla}_{\bm{k}}:=(\partial_{k_x},\partial_{k_y})$.
The path of the integral is chosen so that it encloses the EP.
For $\mathrm{dim} H=2$, the vorticity can be written as~\cite{HShen2017_non-Hermi,shen_nu_ftnt}
\begin{eqnarray}
\label{eq: vorticity Shen app}
\nu &=& \frac{1}{2\pi} \oint \! d^2\bm{k}\cdot \bm{\nabla}_{\bm{k}} \mathrm{arg}[E_+(\bm{k})-E_-(\bm{k})],
\end{eqnarray}
where $E_\pm$ is the energy eigenvalue [see Eq.~(\ref{eq: 2x2 Epm})].

In the following, we see how the vorticity defined in Eq.~(\ref{eq: vorticity}) characterizes the EPs.
Consider a generic non-Hermitian matrix $H(\bm{k})$ with $\mathrm{dim} H \geq 2$ which shows the band touching at energy $E_{0}$.
The band touching point can be formulated as
\begin{eqnarray}
\mathrm{det}[H(\bm{k}_0)-E_0\1]&=& 0,
\end{eqnarray}
where $\bm{k}_0$ denotes the EP in the momentum space.
Mapping the non-Hermitian Hamiltonian to the Hermitian matrix $\tilde{H}$, we can rewrite the above condition as 
\begin{subequations}
\begin{eqnarray}
\label{eq: det H' pg}
\mathrm{det}[\tilde{H}(\bm{k}_0)]&=& 0,
\end{eqnarray}
with
\begin{eqnarray}
\label{eq: H' pg}
\tilde{H}&=& 
\left(
\begin{array}{cc}
0 &  H(\bm{k})-E_0\1 \\
H^\dagger(\bm{k})-E^*_0\1 & 0
\end{array}
\right)_\rho.
\end{eqnarray}
\end{subequations}
Here, we have extended the Hilbert space on which the Pauli matrices $\rho$'s act.
This can be easily confirmed by noticing that Eq.~(\ref{eq: det H' pg}) can be written as $|\mathrm{det}[H(\bm{k}_0)-E_0]|^2=0$. 
The above fact means that the exceptional band touching can be described by the zero modes of the Hermitian matrix $\tilde{H}$ which is chiral symmetric $\{\tilde{H}, \tilde{\Sigma} \}=0$ with $\tilde{\Sigma}:=\1\otimes \rho_3$.
Therefore, remembering that the zero modes of the chiral symmetric Hermitian Hamiltonian are characterized by the winding number,
\begin{eqnarray}
\label{eq: winding H'}
\nu_W &=& \frac{1}{4\pi i} \oint \! d^2\bm{k} \cdot \mathrm{tr}[\tilde{\Sigma} \tilde{H}^{-1}\bm{\nabla}_{\bm{k}} \tilde{H} ],
\end{eqnarray}
we can see that the EPs can be characterized by the vorticity~(\ref{eq: vorticity}); substituting $\tilde{\Sigma}=\1\otimes \rho_3$ to Eq.~(\ref{eq: winding H'}) yields Eq.~(\ref{eq: vorticity})~\cite{nu_w_to_nu_ftnt}.
We note that the vorticity is half-quantized due to the extra prefactor $1/2$, which is just a convention.

In this section, we have considered two-dimensional systems. We note, however, that the vorticity is well-defined along a one-dimensional path in the three-dimensional BZ. 
In this case, the vorticity characterizes exceptional loops in the BZ (see also Table~\ref{table: SPER SPES d=1,2,3}). 
For a $2\times 2$ Hamiltonian, there is complimentary understanding. 
The EPs appear when both of Eqs.~(\ref{eq: cond EP 2x2}a)~and~(\ref{eq: cond EP 2x2}b) are satisfied, meaning that one degree of freedom is left in the three dimensions. This remaining degree of freedom forms a loop which is nothing but the exceptional loop in three dimensions.

\subsection{
EPs appearing in the single-particle spectrum
}
\label{sec: EP GF}

In the above, we have seen that a non-Hermitian matrix may show EPs which are characterized by the vorticity~(\ref{eq: vorticity}).
In this section, we see that a non-Hermitian matrix governs the single-particle excitation spectrum of correlated systems in equilibrium (i.e., the energy is conserved).

Firstly, we define the retarded single-particle Green's function $G^{R}(t,\bm{k})$ whose imaginary part corresponds to the single-particle spectrum:
\begin{eqnarray}
\label{eq: defs of GR(t,k)}
G^{R}_{\alpha\beta}(t,\bm{k}) &=& -i\langle \hat{c}_{\bm{k}\alpha}(t)\hat{c}^\dagger_{\bm{k}\beta}(0)  + \hat{c}^\dagger_{\bm{k}\beta}(0)\hat{c}_{\bm{k}\alpha}(t) \rangle \theta(t),
\end{eqnarray}
where $\hat{c}^\dagger_{\bm{k}\alpha}$ creates a fermion with momentum $\bm{k}$ in the state $\alpha$ (spin, orbital etc.). $\langle{\, }\cdot{\, }\rangle$ denotes the expectation value for temperature $\beta^{-1}$ [$\langle{\, }\cdot{\, }\rangle:=\mathrm{tr}({\, }\cdot{\, }e^{-\beta \hat{H} })$].
$\hat{c}^\dagger_{\bm{k}\alpha}(t):=e^{i\hat{H}t} \hat{c}^\dagger_{\bm{k}\alpha} e^{-i\hat{H}t}$ with the second quantized Hamiltonian $\hat{H}$ describing the correlated system in equilibrium (i.e., $\hat{H}$ is a Hermitian operator). 
$\theta(t)$ takes $0$, $1/2$, and $1$ for $t<0$, $t=0$, and $t>0$, respectively.
Applying the Fourier transformation, we obtain the Dyson's equation~\cite{AGD}:
\begin{eqnarray}
\label{eq: EP Dyson eq}
G^{-1}(\omega+i\delta,\bm{k}) &=& g^{-1}(\omega+i\delta,\bm{k})-\Sigma(\omega+i\delta,\bm{k}),
\end{eqnarray}
which defines the self-energy $\Sigma(\omega+i\delta,\bm{k})$. Here, $g(\omega+i\delta,\bm{k})$ denotes the retarded Green's function for free fermions. $\delta$ is an infinitesimal constant ($\delta>0$).
With the Green's function, the single-particle spectral function is defined as $A(\omega,\bm{k})=-\mathrm{Im}\sum_\alpha G_{\alpha\alpha}(\omega+i\delta,\bm{k})/\pi$, which can be rewritten as
\begin{subequations}
\label{eq: EP Heff in Awk}
\begin{eqnarray}
A(\omega,\bm{k}) &=& -\frac{1}{\pi} \mathrm{Im} \, \mathrm{tr}[(\omega+i\delta) \1 -H_{\mathrm{eff}}(\omega,\bm{k})]^{-1},
\label{eq: Awk_Heff} \\
H_{\mathrm{eff}}(\omega,\bm{k}) &=& h(\bm{k})+\Sigma(\omega+i\delta,\bm{k}).
\label{eq: Heff}
\end{eqnarray}
\end{subequations}
Here, the matrix $h(\bm{k})$ is the Bloch Hamiltonian for free fermions with momentum $\bm{k}$. $\1$ denotes the identity matrix. 
We note that the self-energy $\Sigma(\omega+i\delta,\bm{k})$ is a non-Hermitian matrix, describing the lifetimes of quasi-particles.
Therefore, Eq.~(\ref{eq: Awk_Heff}) indicates that the single-particle excitations of energy $\omega$ are governed by the non-Hermitian matrix $H_{\mathrm{eff}}(\omega,\bm{k})$.

In addition to EPs, the non-Hermiticity of the effective Hamiltonian yields low energy excitations. 
The energy gap can be pure imaginary because of the non-Hermiticity of $H_{\mathrm{eff}}$. 
In this case, even when the Bloch Hamiltonian is gapped, the system may show Fermi arcs connecting EPs.

We finish this section by making a comment on an additional condition for the EPs in the single-particle spectrum. 
The effective Hamiltonian appears in the denominator of the spectral weight~(\ref{eq: Awk_Heff}), meaning that the EPs are seriously smeared when the denominator is large. 
Therefore, in order for EPs to emerge as a peak in the single-particle spectral function, the frequency $\omega$ should satisfy an additional condition [for instance see Eq.~(\ref{eq: EP KLM cond EP}a)].

\subsection{
EPs for two-dimensional heavy-fermion systems
}
\label{sec: EP KLM}

In this section, we demonstrate that EPs emerge in the single-particle spectrum of a heavy-fermion system, by employing the DMFT.
In particular, we analyze the Kondo lattice in two dimensions. The Hamiltonian reads,
\begin{eqnarray}
\label{eq: HKLM}
\hat{H} &=& \sum_{\langle i j\rangle \alpha,\beta} t_{i\alpha,j\beta} \hat{c}^\dagger_{i\alpha s} \hat{c}_{j\beta s} +J\sum_{i}\hat{\bm{s}}_{ib}\cdot \hat{\bm{S}}_{i},
\end{eqnarray}
where $\hat{c}^\dagger_{i\alpha s}$ creates an electron with spin $s=\uparrow,\downarrow$ in orbital $\alpha=a,b$ of site $i$.
$\hat{\bm{s}}_{ib}:=\frac{1}{2} \hat{c}^\dagger_{ib s}\bm{\sigma}_{ss'}\hat{c}_{ib s'}$ with the Pauli matrices $\sigma$'s acting on the spin space.
$\hat{\bm{S}}$ is the spin $1/2$ operator for the localized spins.
Here, the Kondo coupling of electrons in orbital $a$ is neglected for simplicity. 
The hopping $t_{i\alpha,j\beta}$ is defined so that the Bloch Hamiltonian is written as
\begin{eqnarray}
h(\bm{k})   &=& -2t'\sin k_y \tau_1 +[-\epsilon_0 -2t(\cos k_x + \cos k_y)] \tau_3,
\end{eqnarray}
%
where $\epsilon_0$, $t$, and $t'$ take real values, respectively.
The Pauli matrices $\tau$'s act on the orbital space.
In the non-interacting case, this model shows two Dirac cones for $t=1$ and $0<\epsilon_0<4$.

In order to analyze the above correlated electron system, we employ the DMFT~\cite{WMetznerPRL89_DMFT,MHartmannZP89_DMFT,Kotliar_DMFT_92,AGeorgesRMP96_DMFT} which treats local correlation exactly.
In the DMFT framework, the lattice model is mapped to an effective impurity model where the self-energy of spin $s$ 
$
[
\Sigma_s(\omega+i\delta):=
\mathrm{diag}
\left(
\begin{array}{cc}
0, & \Sigma_{bs}(\omega+i\delta)
\end{array}
\right)
]
$ 
is computed self-consistently~\cite{DMFT_ftnt}.
Here, $\Sigma_{bs}(\omega+i\delta)$ denotes the self-energy for orbital $b$ and spin $s$.
In order to compute the self-energy for the effective impurity model, we employ the numerical renormalization group method (NRG)~\cite{KWilsonRMP75_NRG,RPetersPRB06_NRG,RBullaRMP08_NRG}.
This method directly provides the single-particle spectral function, while other methods based on Monte Carlo calculations~\cite{Hirsh_QMC_PRL86,Werner_CTQMC_PRL2006,Werner_CTQMC_CTQMC2006} require the analytic continuation.

Once the self-energy is obtained,
the single-particle spectrum is obtained as
\begin{subequations}
\begin{eqnarray}
A(\omega,\bm{k})&=& -\frac{1}{\pi} \mathrm{Im} \, \mathrm{tr} [(\omega+i\delta)\1 -H_{\mathrm{eff}}(\omega,\bm{k}) ]^{-1}, \\
H_{\mathrm{eff}}(\omega,\bm{k}) &=& h(\bm{k}) +\Sigma(\omega+i\delta).
\end{eqnarray}
\end{subequations}
Here, we have omitted the subscript $s$
[$
\Sigma(\omega+i\delta):=
\mathrm{diag}
\left(
\begin{array}{cc}
0, & \Sigma_{b}(\omega+i\delta)
\end{array}
\right)
$] 
by assuming that the system is in the paramagnetic phase.
We note that the effective Hamiltonian is a $2\times2$ matrix.
Expanding it with the Pauli matrices as Eq.~(\ref{eq: 2x2 nonHermi}), we obtain the following coefficients
\begin{subequations}
\begin{eqnarray}
\left(
\begin{array}{cccc}
b_0, & b_1, & b_2, & b_3
\end{array}
\right)
&=&
\left(
\begin{array}{cccc}
\frac{\mathrm{Re}\Sigma_{b}(\omega+i\delta) }{2}, & 2t'\sin k_y, & 0, & -\epsilon_0 -2t(\cos k_x+\cos k_y) -\frac{\mathrm{Re} \Sigma_{b}(\omega+i\delta) }{2}
\end{array}
\right),\\
\left(
\begin{array}{cccc}
d_0, & d_1, & d_2, & d_3
\end{array}
\right)
&=&
\left(
\begin{array}{cccc}
\frac{\mathrm{Im}\Sigma_{b}(\omega+i\delta) }{2}, & 0, & 0, & -\frac{\mathrm{Im}\Sigma_{b}(\omega+i\delta) }{2}
\end{array}
\right).
\end{eqnarray}
\end{subequations}
Therefore, the conditions for EPs appearing as the peak of the single-particle spectral function are written as
\begin{subequations}
\label{eq: EP KLM cond EP}
\begin{eqnarray}
2\omega_0 -\mathrm{Re} \Sigma_{b}(\omega_0+i\delta) &=& 0, \\
-\epsilon_0 -2t(\cos k_{0x}+\cos k_{0y} ) -\frac{\mathrm{Re} \Sigma_{b}(\omega+i\delta) }{2}
&=& 0,\\
-\left[ \mathrm{Im} \Sigma_{b}(\omega_0+i\delta) \right]^2+
16t'^2\sin^2 k_{0y}
&=&
0.
\end{eqnarray}
\end{subequations}
Here, the second and the third equations are obtained from Eq.~(\ref{eq: cond EP 2x2}), specifying the position of the EP $\bm{k}_0$ in the BZ.  
The first equation specifies the energy $\omega_0$ where the EPs emerge as peaks of the spectral function.
We note that in the DMFT framework, the momentum dependence of the self-energy is neglected. 
However, the EPs should emerge even in calculations beyond the DMFT framework because they are topologically protected.

\begin{figure}[!h]
\begin{minipage}{0.75\hsize}
\begin{center}
\includegraphics[width=\hsize]{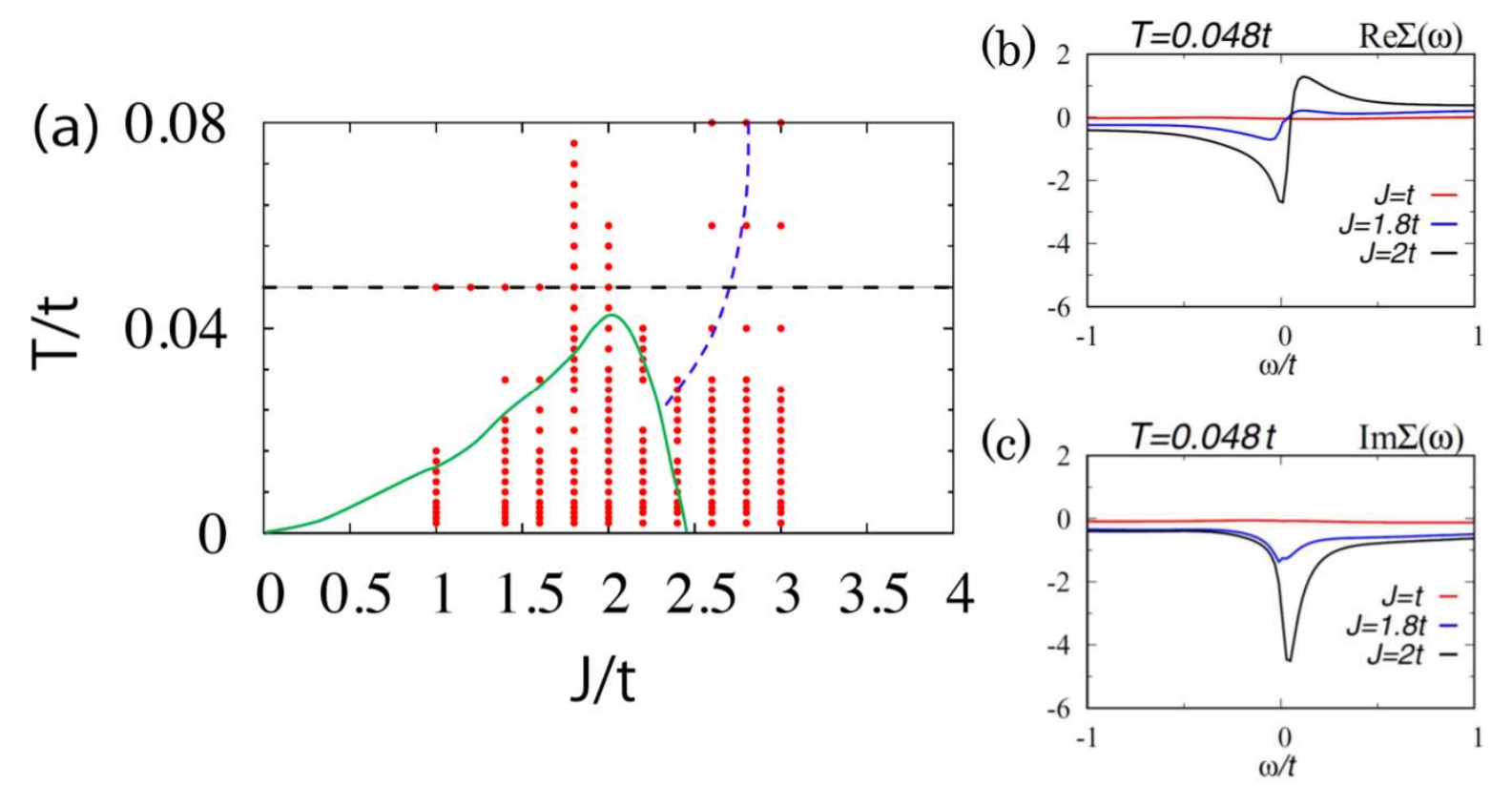}
\end{center}
\end{minipage}
\caption{
(a) Phase diagram of the Kondo coupling $J$ vs. the temperature $T$.
The Neel temperature is represented with the solid green line.
The Kondo temperature is represented with the dashed blue line.
The horizontal dashed line denotes the region of $T=0.048t$.
(b) [(c)] The real (imaginary) part of the self-energy for several values of the coupling $J$ at $T=0.048t$, respectively.
These figures are adapted with permission from Ref.~\onlinecite{Yoshida_EP_DMFT_PRB18}.
Copyright 2018 American Physical Society.
}
\label{fig: EP KLM phase}
\end{figure}

Let us now analyze the Kondo lattice model~(\ref{eq: HKLM}). 
In the rest of this section, we set the parameters to 
$
\left(
\begin{array}{ccc}
t, & t', & \epsilon_0
\end{array}
\right)
=
\left(
\begin{array}{ccc}
1, & 0.667, & 0.667
\end{array}
\right)
$. 
The obtained phase diagram is shown in Fig.~\ref{fig: EP KLM phase}. 
When the Kondo coupling is small, an anti-ferromagnetic phase emerges because the Ruderman-Kittel-Kasuya-Yosida interaction~\cite{Ruderman_RKKY_PR54,Kasuya_RKKY_PTP56,Yosida_PR57} becomes dominant. 
Increasing the interaction $J$, itinerant electrons and localized spins form singlets due to the Kondo effect. 
As a result, the anti-ferromagnetic phase is suppressed in the region of strong $J$.

\begin{figure}[!h]
\begin{minipage}{0.75\hsize}
\begin{center}
\includegraphics[width=\hsize]{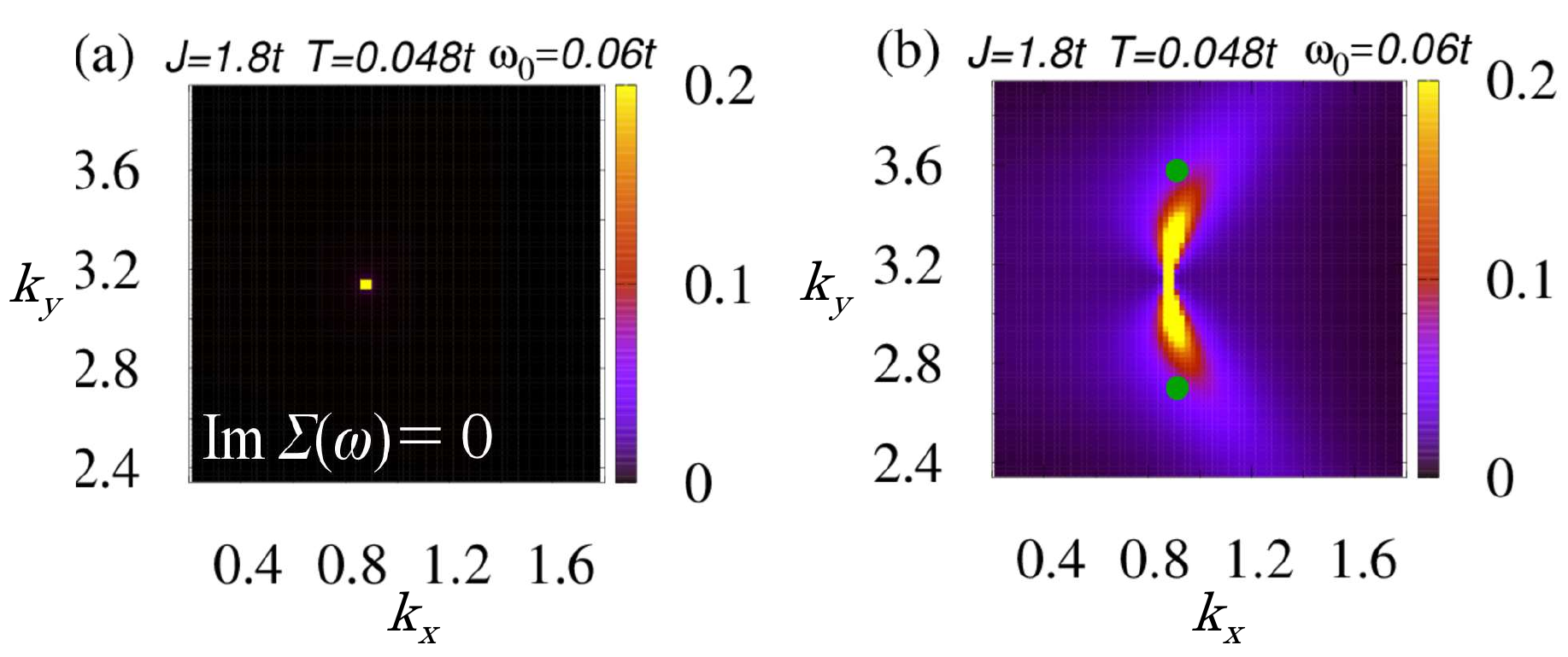}
\end{center}
\end{minipage}
\caption{
Single-particle spectral function $A(\omega_0,\bm{k})$ with $\omega_0=0.06t$ for $J=1.8t$ and $T=0.048t$.
The data are plotted around the boundary of the BZ ($k_y=\pi$). 
Panel (a) is plotted by setting the imaginary part of the self-energy to zero $\mathrm{Im}\Sigma_b(\omega_0)=0$.
In this figure, we can see that the peak for $k_y=\pi$, indicating the emergence of the Dirac cone with chiral symmetry.
Panel (b) shows that the Dirac cone splits into two EPs (green dots) because of the imaginary part of the self-energy.
These EPs are connected with Fermi arcs.
These figures are adapted with permission from Ref.~\onlinecite{Yoshida_EP_DMFT_PRB18}.
Copyright 2018 American Physical Society.
}
\label{fig: EP Akw J1.8}
\end{figure}

\begin{figure}[!h]
\begin{minipage}{0.75\hsize}
\begin{center}
\includegraphics[width=\hsize]{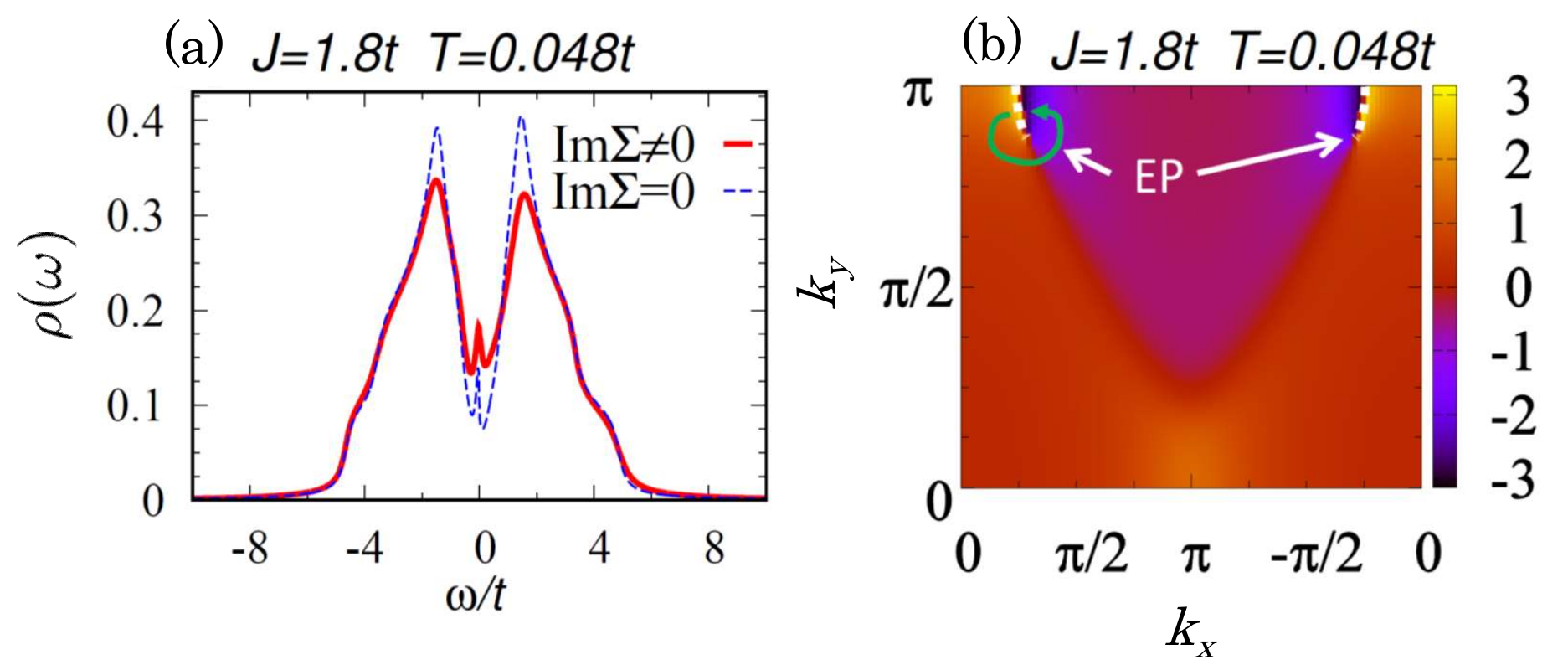}
\end{center}
\end{minipage}
\caption{
(a) The local density of states $\rho(\omega)={\displaystyle \sum_{\bm{k}} }A(\omega,\bm{k})/N$ for $J=2$ and $T=0.048t$. 
Here, $N$ denotes the number of unit cells.
The red line indicates the data computed with the obtained self-energy.
For comparison, we also plot the data obtained by setting $\mathrm{Im}\Sigma_b(\omega)=0$ (see blue line).
(b) Color map of $\mathrm{Arg}[\Delta^2_c(\omega_0,\bm{k})]$ with $\omega_0=0.06t$. 
On white dashed lines, the value $\mathrm{Arg}[\Delta^2_c(k_x,k_y)]$ jumps from $-\pi$ to $\pi$ which corresponds to the branch cut of $\Delta_c$.
These figures are adapted with permission from Ref.~\onlinecite{Yoshida_EP_DMFT_PRB18}.
Copyright 2018 American Physical Society.
}
\label{fig: EP LDOS arg J1.8}
\end{figure}

We numerically observe the EPs in the paramagnetic phase. 
The Kondo effect plays an important role for the emergence of EPs.
The self-energy is shown in Figs.~\ref{fig: EP KLM phase}(b)~and~\ref{fig: EP KLM phase}(c) for $T=0.048t$ which corresponds to the horizontal line in Fig.~\ref{fig: EP KLM phase}(a).
For small $J$ ($J=t$), the real and imaginary parts of the self-energy take small values because the electrons are almost decoupled from the localized spins.
Increasing the coupling $J$ enhances the Kondo effect, which results in a dip structure of $\mathrm{Im}\Sigma_b(\omega+i\delta)$ in the low-energy region (i.e., around $\omega\sim 0$).
This dip structure of the self-energy induces the EPs. The single-particle spectral function for $J=1.8t$ is plotted in Fig.~\ref{fig: EP Akw J1.8}.
Firstly, we show the data obtained by assuming that the imaginary part of the self-energy is zero [see Fig.~\ref{fig: EP Akw J1.8}(a)] in order to show that the imaginary part of the self-energy is essential for the EPs. 
In this figure, we can see a single peak due to the existence of a Dirac cone.
Fig.~\ref{fig: EP Akw J1.8}(b) shows the spectral function obtained by the DMFT. 
In this figure, we can see that the dip structure of the imaginary part splits the Dirac cone into two EPs as represented with green dots. 
Furthermore, we can see that the EPs are connected by the Fermi arc where the bulk gap $\Delta_c=E_+-E_-$ becomes pure imaginary. 
The emergence of the Fermi arc enhances the local density of states around $\omega \sim 0$ [see Fig.~\ref{fig: EP LDOS arg J1.8}(a)].
In the above, we have seen that the imaginary part of the self-energy splits each of two Dirac cones into a pair of EPs connected with the bulk Fermi arc.
As we see below, these bulk Fermi arcs are robust because the EPs are topologically protected.

Here, we address the characterization of the above EPs.
Because the vorticity is written as Eq.~(\ref{eq: vorticity Shen app}) for the $2\times 2$ Hamiltonian, we can compute its value by plotting the argument of $\Delta^2_c$ [see Fig.~\ref{fig: EP LDOS arg J1.8}(b)].
In this figure, the branch cut of $\Delta_c$ is represented with white dashed lines which end at EPs.
Therefore, taking the integral along the green line illustrated in Fig.~\ref{fig: EP LDOS arg J1.8}(b), we can see that the vorticity takes $\nu=-1/2$.
We note that the vorticity takes $\nu=1/2$ for the EP around $k_x=-\pi/2$.

\begin{figure}[!h]
\begin{minipage}{0.75\hsize}
\begin{center}
\includegraphics[width=\hsize]{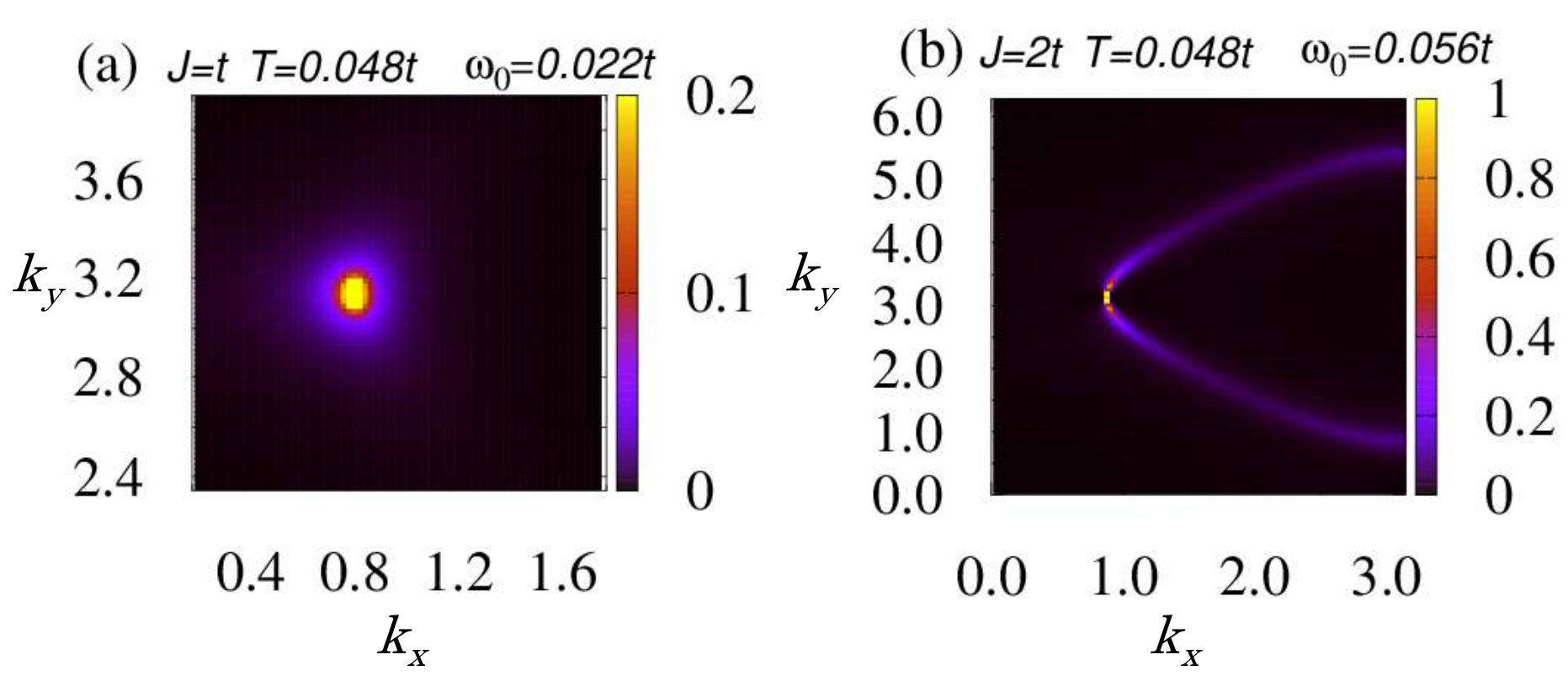}
\end{center}
\end{minipage}
\caption{
The single-particle spectral function $A(\omega_0,\bm{k})$ for $T=0.048t$.
Data for $J=t$ and $J=2t$ are plotted in panel (a) and (b), respectively.
Panel (a) indicates that the Fermi arc shrinks, corresponding to the fusion of two EPs.
Panel (b) indicates that the Fermi loop emerges because two EPs merge at the boundary of the BZ.
These figures are adapted with permission from Ref.~\onlinecite{Yoshida_EP_DMFT_PRB18}.
Copyright 2018 American Physical Society.
}
\label{fig: EP Akw J1.0 J2.0}
\end{figure}

\begin{figure}[!h]
\begin{minipage}{0.75\hsize}
\begin{center}
\includegraphics[width=\hsize]{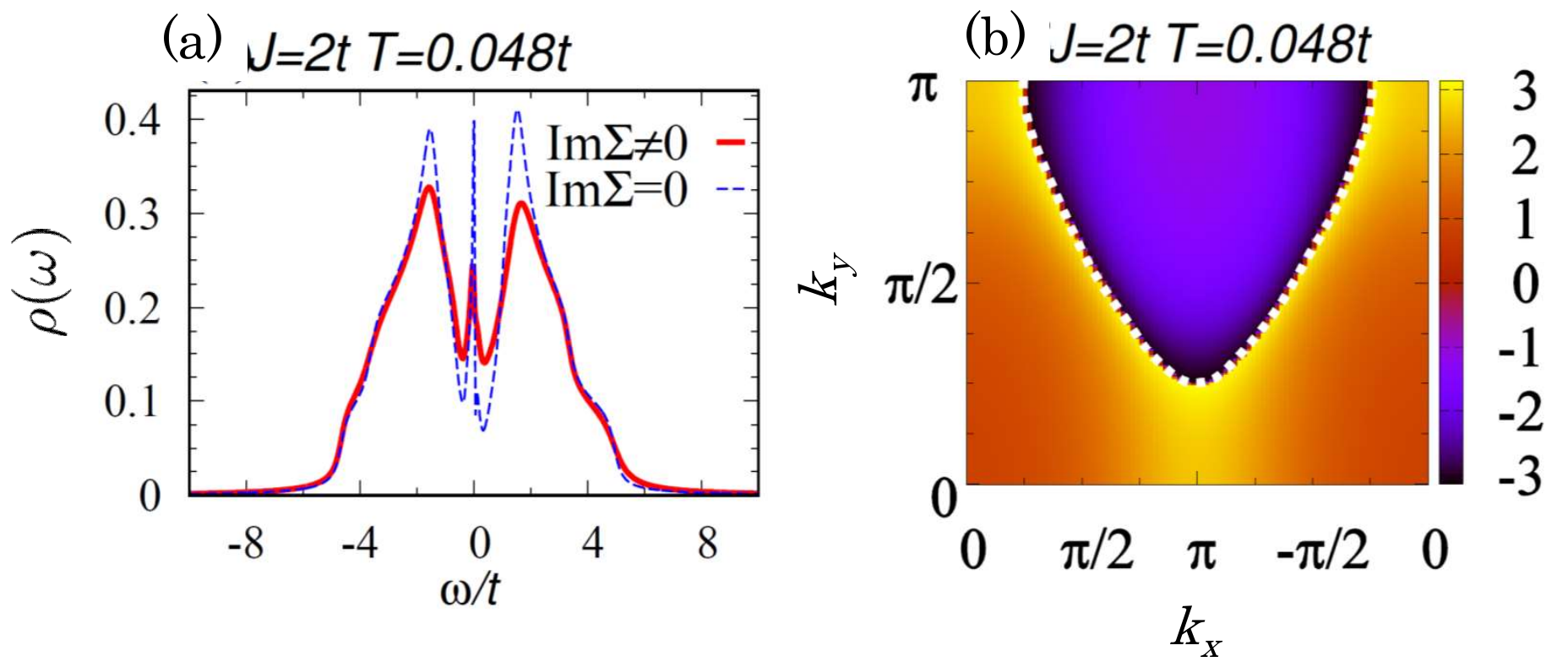}
\end{center}
\end{minipage}
\caption{
(a) The local density of states $\rho(\omega)={\displaystyle \sum_{\bm{k}} }A(\omega,\bm{k})/N$ for $J=2t$ and $T=0.048t$. 
Here, $N$ denotes the number of unit cells.
The red line indicates the data computed with the obtained self-energy.
For comparison, we plot the data obtained by setting $\mathrm{Im}\Sigma_b(\omega)=0$ (see blue line).
(b) Color map of $\mathrm{Arg}[\Delta^2_c(k_x,k_y)]$. On the white dashed lines, the value $\mathrm{Arg}[\Delta^2_c(k_x,k_y)]$ jumps from $-\pi$ to $\pi$.
For this parameter set, the white dashed line forms a closed loop.
These figures are adapted with permission from Ref.~\onlinecite{Yoshida_EP_DMFT_PRB18}.
Copyright 2018 American Physical Society.
}
\label{fig: LDOS arg J2.0}
\end{figure}
Changing the Kondo coupling results in pair annihilation of EPs. Here, we note that there are two scenarios: (i) a pair of EPs originating from a Dirac point are annihilated by themselves; (ii) two pairs EPs exchange the pairs and are annihilated.
The former scenario can be observed by decreasing the Kondo coupling $J$. In Fig.~\ref{fig: EP Akw J1.0 J2.0}(a), we can see that two EPs approach and are annihilated. 
Correspondingly, the Fermi arc vanishes. 
The latter scenario can be observed by increasing the interaction $J$. 
When the Kondo effect is enhanced, the EPs approach the boundary of the BZ specified by $k_x=\pi$. 
On this boundary, the pair of EPs arising from two distinct Dirac cones annihilate each other [see Fig.~\ref{fig: EP Akw J1.0 J2.0}(b)].
The qualitative difference from the previous case is that a Fermi loop emerges after the pair annihilation of EPs, enhancing the LDOS in the low-energy region [see Fig.~\ref{fig: LDOS arg J2.0}(a)].
The emergence of the Fermi loop is again due to the energy gap taking a value of pure imaginary [see Fig.~\ref{fig: LDOS arg J2.0}(b)].

\section{
Symmetry-protected exceptional rings and surfaces in correlated systems
}
\label{sec: SPER}

In the previous section, we have seen that electron correlations induce EPs in the absence of symmetry. 
In addition, it is well-known that the symmetry enriches the topological structures for Hermitian systems~\cite{Schnyder_classification_free_2008,Kitaev_classification_free_2009,Ryu_classification_free_2010,Chiu_class_RMP16}.
Therefore, it should be valuable to analyze the effects of symmetry on EPs, which is the main subject of this section.

Ref.~\onlinecite{Yoshida_SPERs_PRB19} has revealed that many-body chiral symmetry results in novel types of exceptional band touching, SPERs in two dimensions and SPESs in three dimensions.
In the following, after elucidating the topological properties of SPERs and SPESs, we demonstrate the emergence of them in correlated systems.

\subsection{
Symmetry-protection of exceptional band touching
}

\subsubsection{
Case of a $2\times 2$ Hamiltonian
}
\label{sec: 2x2 SPER}

Firstly, we analyze a case of the $2\times 2$ Hamiltonian [see Eq.~(\ref{eq: 2x2 nonHermi})] which captures the essential properties.
Here, let us suppose that the Hamiltonian for a two-dimensional system satisfies the following relation
%
\begin{eqnarray}
\label{eq: chiral 2x2}
\tau_3 H^\dagger(\bm{k}) \tau_3 &=& -H(\bm{k}),
\end{eqnarray}
%
which indicates that the system is chiral symmetric [see Eq.~(\ref{eq: G cond chiral})].
The above condition imposes the following symmetry condition on the coefficients, $b$'s and $d$'s:
\begin{eqnarray}
b_0=b_3=d_1=d_2=0.
\end{eqnarray}

Now, let us consider effects of the symmetry constraint on the EPs. 
As we have seen in Sec.~\ref{sec: EP 2x2}, EPs emerge when the two conditions Eqs.~(\ref{eq: cond EP 2x2}a)~and~(\ref{eq: cond EP 2x2}b) are satisfied.
We note, however, that one of the conditions, Eq.~(\ref{eq: cond EP 2x2}b), is always satisfied by the symmetry constraint, meaning that the number of the conditions for the EPs is reduced.
This fact indicates that for the two-dimensional BZ, fixing one degree of freedom is sufficient to obtain the EPs. 
Therefore, the remaining degree of freedom forms a ring of EPs which is denoted as a SPER~\cite{Yoshida_SPERs_PRB19}. 
On an arbitrary point of the SPERs, the band touching occurs both for the real and imaginary parts.

We can apply the same argument to a three-dimensional system where SPESs emerge~\cite{Yoshida_SPERs_PRB19}. In this case, the two degrees of freedom are left in the BZ.

\subsubsection{
Topological invariant characterizing SPERs and SPESs with chiral symmetry
}
\label{sec: SPERs 0th Ch}

In the above, we have seen that the symmetry constraint results in SPERs or SPESs where the exceptional band touching occurs.
In this section, we show that the band touching is topologically characterized by the zero-th Chern number, a zero-dimensional topological invariant.

Let us suppose that the $2n\times 2n$ Hamiltonian satisfies the following relation
\begin{eqnarray}
\label{eq: SPERs chiral symm generic}
U_\Gamma H^\dagger(\bm{k}) U^\dagger_\Gamma &=& -H(\bm{k}),
\end{eqnarray}
where $U_\Gamma$ is a unitary matrix satisfying $U^2_\Gamma=\1$.
The above equation is a generic form of the symmetry constraint~(\ref{eq: chiral 2x2}).
We now consider the following Hermitian Hamiltonian composed of $H(\bm{k})$;
\begin{eqnarray}
\tilde{H}(\bm{k}) &=& 
\left(
\begin{array}{cc}
0 & H(\bm{k})-E_0 \\
H^\dagger(\bm{k})-E^*_0 & 0
\end{array}
\right)_\rho,
\end{eqnarray}
where we have assumed that the exceptional band touching occurs at energy $E_0 \in i\mathbb{R}$.
In a similar way to the case of Sec.~\ref{sec: EP vorticity}, we can define the topological invariant characterizing the SPERs and SPESs by addressing topological characterization of zero energy excitations described by the Hermitian Hamiltonian~$\tilde{H}$.
The essential difference from the previous case (Sec.~\ref{sec: EP vorticity}) is that the Hermitian Hamiltonian preserves the two distinct constraints of chiral symmetry;
\begin{subequations}
\begin{eqnarray}
\tilde{\Sigma} \tilde{H}(\bm{k}) \tilde{\Sigma}^{-1} &=& -\tilde{H}(\bm{k}), \\
\tilde{U}_\Gamma \tilde{H}(\bm{k}) \tilde{U}^{-1}_\Gamma &=& -\tilde{H}(\bm{k}),
\end{eqnarray}
with 
\begin{eqnarray}
\tilde{\Sigma} &=& \1\otimes \rho_3, \\
\tilde{U}_\Gamma &=& U_\Gamma\otimes \rho_1.
\end{eqnarray}
\end{subequations}
The additional chiral symmetry allows us to define the zero-th Chern number.
Due to two distinct constraints of chiral symmetry, the Hamiltonian can be block-diagonalized with a unitary operator $\tilde{U}=i\tilde{\Sigma} \tilde{U}_\Gamma$ $(\tilde{U}^2=\1)$,
\begin{eqnarray}
\tilde{H}&=&
\left(
\begin{array}{cc}
H_+ & 0 \\
0 & H_-
\end{array}
\right).
\end{eqnarray}
Here, $H_+$ ($H_-$) denotes the Hamiltonian acting on the subspace where the operator $\tilde{U}$ is reduced to $\1$ ($-\1$), respectively. We denote these subspaces by plus and minus sectors.
We note that applying either $\tilde{\Sigma}$ or $\tilde{U}_\Gamma$ exchanges the plus and minus sectors because of the anti-commutation relation $\{\tilde{U}_\Gamma,\tilde{\Sigma}\}=\{\tilde{\Sigma},\tilde{U} \}=0$.
Namely, letting $|+\rangle$ be a state of the plus sector ($\tilde{U}|+\rangle=|+\rangle$), we obtain $\tilde{U} \tilde{\Sigma} |+\rangle=-\tilde{\Sigma}|+\rangle$, which means that $\tilde{\Sigma} |+\rangle$ belongs to the minus sector.
The above facts indicate that the block-diagonalized Hamiltonians $H_{+}$ and $H_{-}$ are related to each other and belong to symmetry class~A.
Therefore, the characterization of the zero energy excitations of $\tilde{H}(\bm{k})$ can be done with the zero-th Chern number for the plus sector which corresponds to the number of the eigenstates with negative eigenvalues of $H_+$. 
This fact suggests $\mathbb{Z}$ classification of zero-dimensional Hermitian systems belonging to class~A.

We note that the block-diagonalized Hamiltonian is rewritten as $H_{+}=iU_\Gamma H$, which can be seen as follows.
Noticing that the unitary matrix $V$ block-diagonalizes the unitary operator $\tilde{U}=U_\Gamma\otimes\rho_2$,
\begin{subequations}
\begin{eqnarray}
\tilde{V}^\dagger \tilde{U} \tilde{V} &=& 
\left(
\begin{array}{cc}
\1 & 0 \\
0 & -\1
\end{array}
\right),
\end{eqnarray}
with
\begin{eqnarray}
\tilde{V} &=& \frac{1}{\sqrt{2}} 
\left(
\begin{array}{cc}
\1 &  -i\1 \\
iU_\Gamma & -U_\Gamma 
\end{array}
\right)_\rho,
\end{eqnarray}
\end{subequations}
we can block-diagonalize the Hermitian Hamiltonian $\tilde{H}$:
\begin{eqnarray}
\tilde{V}^\dagger \tilde{H} \tilde{V} &=& 
\left(
\begin{array}{cc}
iHU_\Gamma &  0 \\
0 & -HU_\Gamma
\end{array}
\right).
\end{eqnarray}
Here, we have used the relation $U^2_\Gamma=\1$.

Therefore, the SPERs and the SPESs are characterized by the zero-th Chern number which is the number of negative eigenvalues of the Hermitian Hamiltonian $H_+(\bm{k})=iH(\bm{k})U_\Gamma$ at each point in the BZ.

The above result indicates that the dimension of the objects composed of the exceptional band touching becomes one-dimensional higher by chiral symmetry~[Eq.~(\ref{eq: SPERs chiral symm generic})] compared to system without the symmetry (see Table~\ref{table: SPER SPES d=1,2,3}).
\begin{table}[htb]
\begin{center}
  \begin{tabular}{cccc} \hline\hline
dimension & 1 & 2 & 3 \\ \hline
no symmetry & - & point & loop \\
with chiral symmetry & point & ring & surface \\
\hline\hline
  \end{tabular}
\end{center}
\caption{
Objects formed by EPs in the BZ for each case of spatial dimensions.
In the presence of chiral symmetry, exceptional band touching forms objects which are one-dimensional higher than the ones in the absence of symmetry.
}
\label{table: SPER SPES d=1,2,3}
\end{table}
Table~\ref{table: SPER SPES d=1,2,3} also indicates that the EPs emerging in one-dimensional systems are either unstable or symmetry-protected.

We finish this section with the complementary understanding for the $2\times 2$ Hamiltonian.
Namely, exceptional band touching appears at points where both of Eqs.~(\ref{eq: cond EP 2x2}a)~and~(\ref{eq: cond EP 2x2}b) are satisfied.
Thus, in the absence of symmetry, exceptional band touching forms ($d-2$)-dimensional objects in the $d$-dimensional BZ.
On the other hand, in the presence of chiral symmetry, exceptional band touching forms ($d-1$)-dimensional objects because Eq.~(\ref{eq: cond EP 2x2}b) is always satisfied by symmetry.

\subsection{
SPERs for a correlated honeycomb lattice
}

\begin{figure}[!h]
\begin{minipage}{0.75\hsize}
\begin{center}
\includegraphics[width=\hsize,clip]{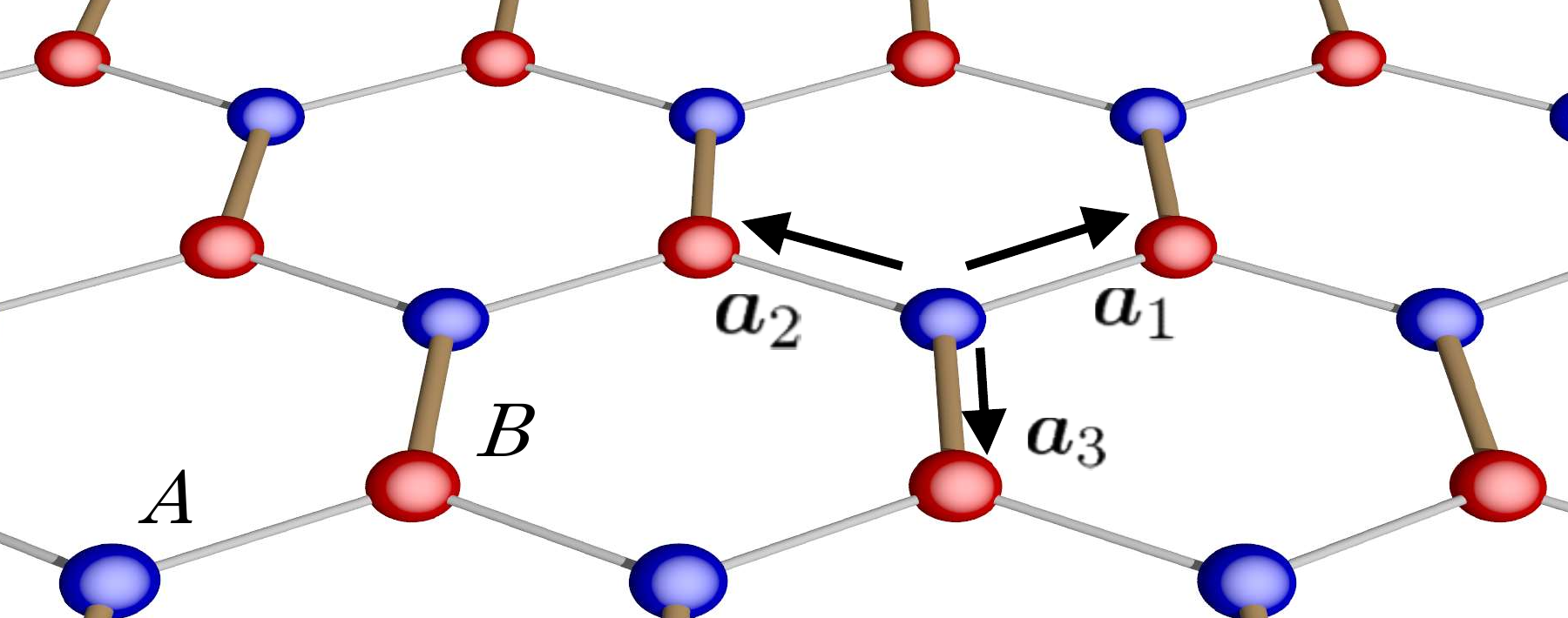}
\end{center}
\end{minipage}
\caption{
Sketch of the honeycomb Hubbard model. The $A$- ($B$-) sublattice is illustrated with blue (red) spheres, respectively.
Vectors $\bm{a}_1$, $\bm{a}_2$, and $\bm{a}_3$ specify the neighboring sites; $\bm{a}_1:=(\sqrt{3},1)/2$, $\bm{a}_2:=(-\sqrt{3},1)/2$, and $\bm{a}_3:=(0,-1)$.
Nearest-neighbor hopping with $t$ ($rt$) is represented with gray (brown) bonds, respectively.
This figure is adapted with permission from Ref.~\onlinecite{Yoshida_SPERs_PRB19}.
Copyright 2019 American Physical Society.
}
\label{fig: SPERs honey_Hubb}
\end{figure}

The SPERs can emerge for strongly correlated systems in equilibrium.
In order to demonstrate the emergence of the SPERs, we apply the DMFT+NRG to a Hubbard model of a honeycomb lattice. The Hamiltonian reads
%
\begin{eqnarray}
\label{eq: honey Hubb}
\hat{H}&=& \sum_{\langle ij \rangle \alpha\beta} t_{i\alpha,j\beta} \hat{c}^\dagger_{i\alpha s} \hat{c}_{j\beta s'} +\sum_{i\alpha} U_{\alpha} \left( \hat{n}_{i\alpha \uparrow} -\frac{1}{2} \right) \left( \hat{n}_{i\alpha \downarrow} -\frac{1}{2} \right),
\end{eqnarray}
%
where $\hat{c}^\dagger_{i\alpha s}$ creates a fermion with spin $s=\uparrow,\downarrow$ at site $i$ and sublattice $\alpha=A,B$. $\hat{n}_{i\alpha \uparrow}:=\hat{c}^\dagger_{i\alpha s}\hat{c}_{i\alpha s}$. The first term describes the nearest-neighbor hopping with $t_{i\alpha,j\beta}\in \mathbb{R}$. 
The second term describes the on-site repulsion ($U_\alpha \geq 0$).
Applying the Fourier transformation for $U_\alpha=0$, we obtain the Bloch Hamiltonian which is written as
\begin{subequations}
\begin{eqnarray}
h(\bm{k}) &=& b_1(\bm{k})\tau_1 + b_2(\bm{k})\tau_2,
\end{eqnarray}
where $b_1(\bm{k})$ and $b_2(\bm{k})$ [$b_1(\bm{k}),b_2(\bm{k}) \in \mathbb{R}$] are defined as
\begin{eqnarray}
b_1(\bm{k}) -i b_2(\bm{k})&=& 
te^{i\bm{k}\cdot \bm{a}_1}+te^{i\bm{k}\cdot \bm{a}_2} +t'e^{i\bm{k}\cdot \bm{a}_3}.
\end{eqnarray}
\end{subequations}
Here, we have assumed the hopping $t$ ($t':=rt$) between sites connected with gray (brown) lines, respectively.
Vectors $\bm{a}$'s are illustrated in Fig.~\ref{fig: SPERs honey_Hubb}.
We consider that this model can be fabricated for cold atoms because the inhomogeneous Hubbard interaction is implemented with  the optical Feshbach resonance ~\cite{RYamazaki_PRL10,LWClark_PRL15}.

The above model preserves the chiral symmetry for an arbitrary value of the interaction $U_\alpha$:
\begin{subequations}
\label{eq: chiral many body}
\begin{eqnarray}
 \hat{\Gamma} \hat{H} \hat{\Gamma}^{-1} &=& \hat{H},
\end{eqnarray}
\begin{eqnarray}
 \hat{\Gamma} &=& \prod_{j\alpha} \left( \hat{c}^\dagger_{j\alpha\uparrow} +\mathrm{sgn}(\alpha) \hat{c}_{j\alpha\uparrow} \right) \left( \hat{c}^\dagger_{j\alpha\downarrow} +\mathrm{sgn}(\alpha) \hat{c}_{j\alpha\downarrow} \right),
\end{eqnarray}
\end{subequations}
with $\mathrm{sgn}(\alpha)$ taking $1$ ($-1$) for $\alpha=A$ ($\alpha=B$), respectively.
This symmetry imposes the following constraint on the Green's function
\begin{subequations}
\begin{eqnarray}
\label{eq: G cond chiral}
\tau_3 G(-\omega+i\delta,\bm{k})\tau_3&=& -G(\omega+i\delta,\bm{k}).
\end{eqnarray}
In particular, for $\omega=0$, the above condition can be rewritten as 
\begin{eqnarray}
\label{eq: Heff cond chiral at w=0}
\tau_3 H^\dagger_{\mathrm{eff}}(0,\bm{k}) \tau_3 &=& -H_{\mathrm{eff}}(0,\bm{k}),
\end{eqnarray}
\end{subequations}
in terms of the effective Hamiltonian $H_{\mathrm{eff}}(\omega,\bm{k})$. This constraint is nothing but the symmetry discussed in the previous section [see Eq.~(\ref{eq: chiral 2x2})].
Therefore, the chiral symmetry of the correlated systems~(\ref{eq: chiral many body}) protectes the SPERs emerging in the single-particle spectrum.
%
\begin{figure}[!h]
\begin{minipage}{0.75\hsize}
\begin{center}
\includegraphics[width=\hsize,clip]{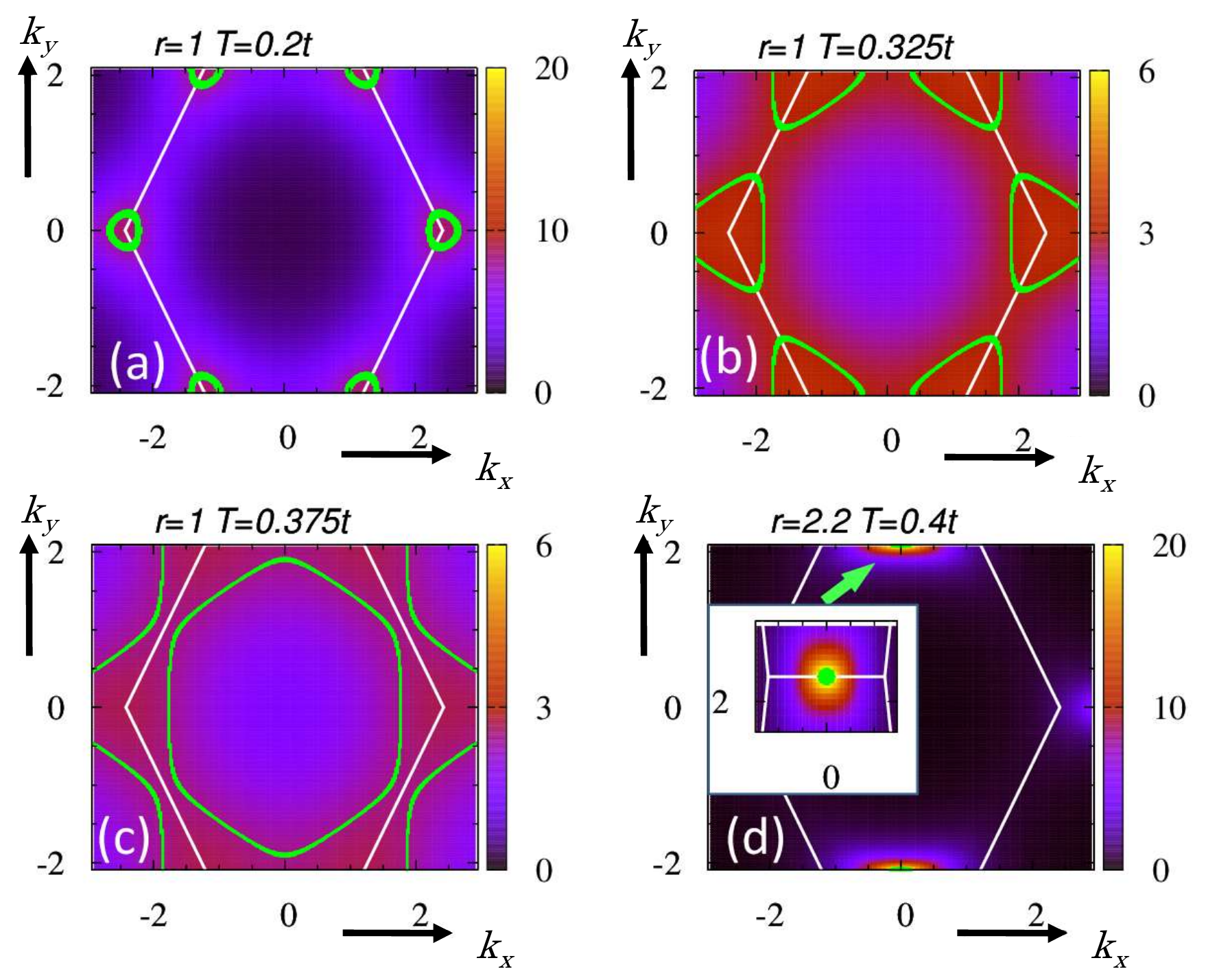}
\end{center}
\end{minipage}
\caption{
The single-particle spectral function [$A(\omega=0,\bm{k})$] for several values of temperature. Panels (a)-(c) show data for $r=1$ while panel (d) shows data for $r=2.2$ where the system does not host Dirac cones.
These data are obtained for $U_A=10t$ and $U_B=5t$.
In these figures, the SPERs are shown with green lines. In the region enclosed with the rings, the energy gap becomes pure imaginary.
The BZ is illustrated with the white hexagon.
These figures are adapted with permission from Ref.~\onlinecite{Yoshida_SPERs_PRB19}.
Copyright 2019 American Physical Society.
}
\label{fig: SPERs honey Ak}
\end{figure}

The DMFT results elucidate the emergence of the SPERs. In Fig.~\ref{fig: SPERs honey Ak}, the spectrum at $\omega=0$ is plotted for several values of the temperature. 
In the non-interacting case, it is well-known that the Dirac cones appear at the corners of the BZ illustrated with the white hexagon. In the presence of the correlations, the Dirac cones split into rings [see green rings in Fig.~\ref{fig: SPERs honey Ak}(a)]. 
Increasing the temperature suppresses the lifetimes of quasi-particles. 
Correspondingly, the SPERs become large [Fig.~\ref{fig: SPERs honey Ak}(b)].
In this figure, we can also see the effect of symmetry on the Fermi arcs shown in Fig.~\ref{fig: EP Akw J1.8}(b). Because of the chiral symmetry, the Fermi arcs change into the Fermi planes. 
This is because the energy eigenvalues $E_n$ appear in a pair $(E_n,-E^*_n)$ or become pure imaginary in the presence of the chiral symmetry~(\ref{eq: Heff cond chiral at w=0})~\cite{gamma_En_ftnt}.

\begin{figure}[!h]
\begin{minipage}{0.75\hsize}
\begin{center}
\includegraphics[width=1\hsize]{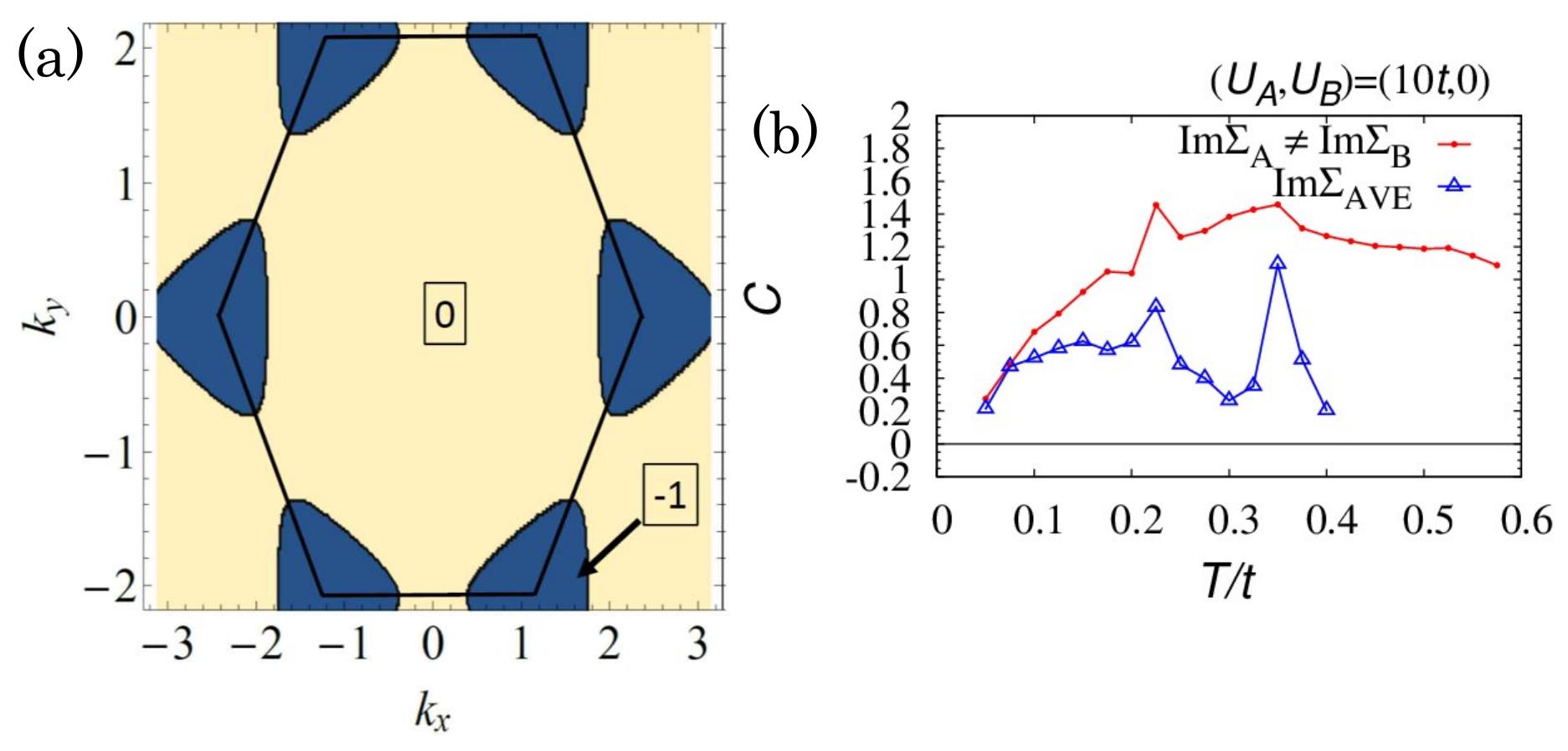}
\end{center}
\end{minipage}
\caption{
(a) 
Colormap of the zero-th Chern number for $r=1$, $T=0.0325t$, $U_A=10t$, and $U_B=5t$.
Black lines represent the SPERs separating domains where the zero-th Chern number takes distinct values.
The black hexagon illustrates the BZ.
The numbers enclosed with black squares denote the zero-th Chern number which is defined so that it takes zero for the Hermitian case $H_{\mathrm{eff}}=b_1\tau_1+ b_2\tau_2$.
(b) Temperature dependence of the specific heat for $r=1$, $U_A=10t$, and $U_B=0$.
Data plotted with the blue line is for comparison; the data are obtained by assuming that the imaginary part of the self-energy takes the same value $\mathrm{Im} [\Sigma_A(\omega)+ \Sigma_B(\omega) ]/2$. 
Namely, the data of blue line do not show the SPERs.
Fermi planes emerge for $T\gtrsim 0.1$.
These figures are adapted with permission from Ref.~\onlinecite{Yoshida_SPERs_PRB19}.
Copyright 2019 American Physical Society.
}
\label{fig: SPERs Ch0 and Cv}
\end{figure}

For higher temperatures, the SPERs, arising from distinct Dirac cones, merge into the single loop [see Fig.~\ref{fig: EP Akw J1.8}(c)]. 
We also note that the presence of the Dirac cones is not a necessary condition for the SPERs. 
Introducing the anisotropy of the hopping, the Dirac cones disappear because of the pair annihilation. 
Even in the absence of the Dirac cones the SPERs emerge [see Fig.~\ref{fig: SPERs honey Ak}(d)].
We note that the SPERs are topologically stable; Fig.~\ref{fig: SPERs Ch0 and Cv}(a) shows that the numerical characterization of the SPERs with the zero-th Chern number can be done.

Finally we show that the emergence of Fermi plane accompanying the SPER enhances the specific heat $C=d\langle H \rangle/d T$ because the Fermi plane induces additional low energy excitations.
In Fig.~\ref{fig: SPERs Ch0 and Cv}(b), the specific heat is shown with the red line. 
For comparison, we also plot data with the blue line by assuming that 
the imaginary part of self-energy for $A$- and $B$-sublattices takes the average value $\mathrm{Im}[\Sigma_A(\omega+i\delta) +\Sigma_B(\omega+i\delta)]/2$. 
We note that the system does not show SPERs when the imaginary part for $A$-sublattice is identical to that for $B$-sublattice.
In this figure, we can see that the specific heat is enhanced because of the Fermi planes accompanying SPERs.

\subsection{
SPESs for a correlated diamond lattice
}

\begin{figure}[!h]
\begin{minipage}{0.75\hsize}
\begin{center}
\includegraphics[width=\hsize]{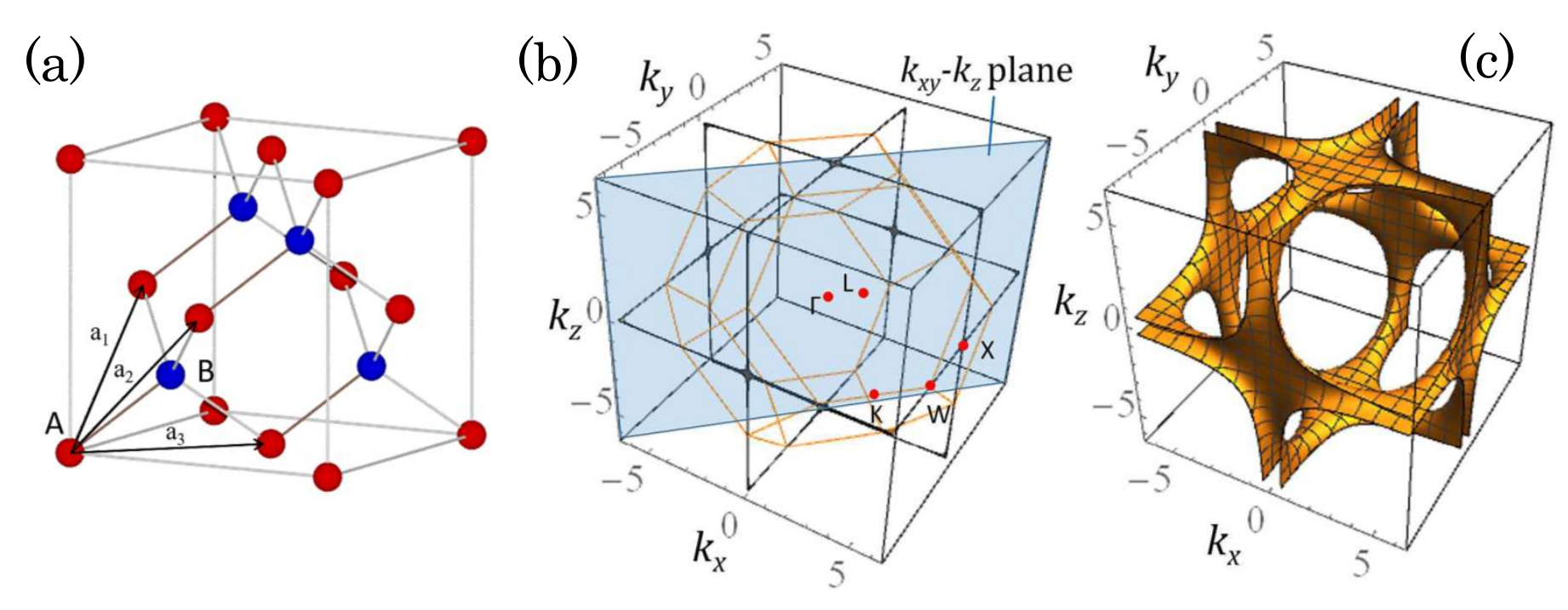}
\end{center}
\end{minipage}
\caption{(Color Online).
(a) Sketch of the diamond lattice. This lattice is composed of two sublattices, $A$ and $B$ which are illustrated with red and blue spheres, respectively. 
We assume that interaction for $A$-sublattice is stronger than that of $B$-sublattice.
(b) The BZ and the high symmetric points for the diamond lattice. The BZ is illustrated with orange lines.
(c) Exceptional surface for $U_A=8t$, $U_B=0$, and $T=0.8t$.
These figures are adapted with permission from Ref.~\onlinecite{Kimura_SPERs_PRB19}.
Copyright 2019 American Physical Society.
}
\label{fig: SPES diamond summ}
\end{figure}

The emergence of the SPESs can also be demonstrated by applying the DMFT to a Hubbard model of a diamond lattice, which is a three-dimensional extension of the honeycomb Hubbard model~(\ref{eq: honey Hubb}).
The lattice structure and the BZ is shown in Fig.~\ref{fig: SPES diamond summ}(a)~and~\ref{fig: SPES diamond summ}(b), respectively. 
In a similar way as the previous section, we introduce an inhomogeneity of the interaction.

\begin{figure}[!h]
\begin{minipage}{0.75\hsize}
\begin{center}
\includegraphics[width=\hsize]{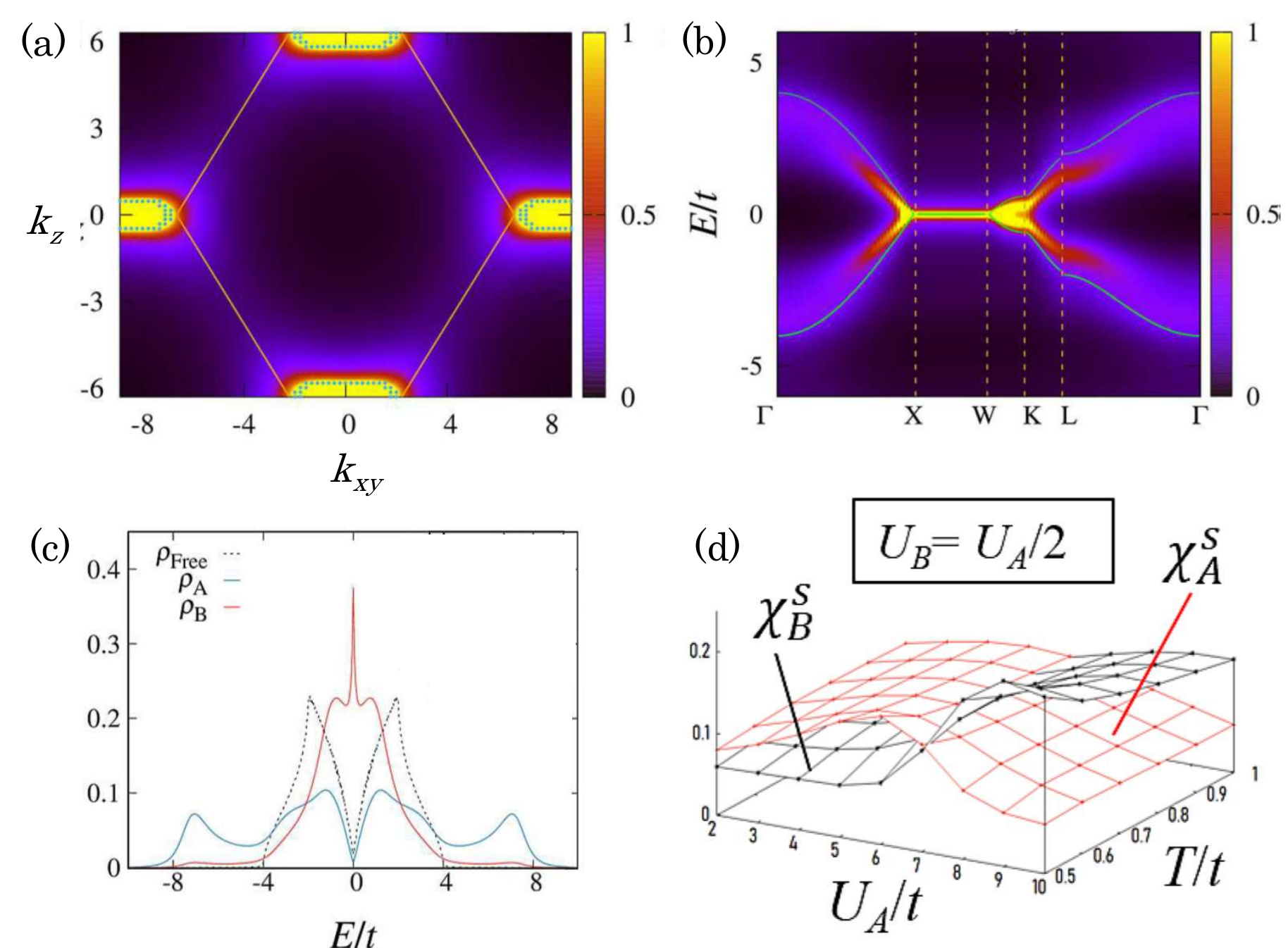}
\end{center}
\end{minipage}
\caption{(Color Online).
Spectral properties and the magnetic response. 
(a) The single-particle spectral function $A(\omega=0,\bm{k})$ for the $k_{xy}$-$k_z$ plane shown in Fig.~\ref{fig: SPES diamond summ}(b). 
(b) The single-particle spectral function $A(\omega,\bm{k})$ along the lines connecting high symmetry points in the BZ.  The blue lines in this panel illustrate the dispersion relation for the non-interacting case.
(c) The local density of states $\rho_\alpha(\omega)=-\sum_{\bm{k}}\mathrm{Im} G_{\alpha\alpha}(\omega,\bm{k})/(\pi N)$ with $N$ denoting the number of unit cells. 
Data shown in panels (a), (b), and (c) are obtained for $U_A=8t$, $U_B=0$, and $T=0.8t$.
(d)
Magnetic susceptibility $\chi^s_\alpha$ against interaction $U_A$ and temperature $T$. Here, $U_B$ is set to $U_B=U_A/2$.
For the detailed calculation of the susceptibility, see footnote~\onlinecite{suscep_RPA_ftnt}.
These figures are adapted with permission from Ref.~\onlinecite{Kimura_SPERs_PRB19}.
Copyright 2019 American Physical Society.
}
\label{fig: SPES LDOS Xi}
\end{figure}

In the following, we see the details. 
For $U_A=8t$, $U_B=0$, and $T=0.8t$, the SPESs emerge as shown in Fig.~\ref{fig: SPES diamond summ}(c). Here, we have employed the iterative perturbation method~\cite{Zhang_IPT_93,Kajueter_IPT_96} as the impurity solver of the DMFT. In the following, we see the results in detail.
In Fig.~\ref{fig: SPES LDOS Xi}(a), the single-particle spectral function at zero energy $A(\omega=0,\bm{k})$ is plotted for the $k_{xy}$-$k_z$ plane [i.e., the blue plane in Fig.~\ref{fig: SPES diamond summ}(b)]. The green dots plotted in Fig.~\ref{fig: SPES LDOS Xi}(a) correspond to the sections of SPESs. We note that in the region enclosed with the SPESs, the energy gap becomes pure imaginary, meaning that the zero energy excitations appear in this region. Thus, the Fermi volume appears instead of the Fermi arc discussed in Sec.~\ref{sec: EP KLM}.
Fig.~\ref{fig: SPES LDOS Xi}(b) shows the single-particle spectral function $A(\omega,\bm{k})$ along the lines connecting the high symmetry points in the BZ.
In this figure, we can confirm the emergence of the Fermi volume by the presence of the zero energy excitations between X and K points.
Outside of the SPESs, the zero energy excitations disappear.

We finish this section with a comment concerning the effect of SPESs on the magnetic response.
As shown in Fig.~\ref{fig: SPES LDOS Xi}(c), the LDOS of the $B$-sublattice is enhanced by the Fermi volume accompanying the SPESs.
We note that the LDOS of the $A$-sublattice is just renormalized. 
This imbalance of the LDOS can induce a counterintuitive behavior of the local magnetic susceptibility. 
In Fig.~\ref{fig: SPES LDOS Xi}(b), the local magnetic susceptibility computed with the random-phase approximation (RPA)~\cite{suscep_RPA_ftnt} is plotted.
As shown in Fig.~\ref{fig: SPES LDOS Xi}(d), due to the imbalance of the LDOS, the magnetic susceptibility of the $B$-sublattice becomes larger than that of the $A$-sublattice, although the interaction of the $B$-sublattice is weaker than that of the $A$-sublattice.

\section{
Ten-fold way classification of the exceptional band touching in equilibrium systems
}
\label{sec: 10-fold way of SPEP}

In Sec.~\ref{sec: SPER}, we have seen that the correlated systems with chiral symmetry may show SPERs and SPESs in two and three dimensions, respectively. 
These SPERs and SPESs are characterized by the zero-th Chern number, a zero-dimensional topological invariant taking an arbitrary integer (see Sec.~\ref{sec: SPERs 0th Ch}). In other words, the topological classification of the exceptional band touching is $\mathbb{Z}$ for the system with chiral symmetry. 

In this section, by generalizing the argument in Sec.~\ref{sec: SPER}, we address the topological classification of the exceptional band touching. Specifically, we carry out the ten-fold way classification~\cite{Schnyder_classification_free_2008,Kitaev_classification_free_2009,Ryu_classification_free_2010,Chiu_class_RMP16} of exceptional band touching in the presence/absence of $PT$-, $CP$-, and chiral symmetry for correlated systems. 
This is because $PT$- ($CP$-) symmetry is closed at each point in the BZ as well as the chiral symmetry (i.e., the corresponding symmetry transformation does no flip the momentum).
We note that the 38-fold way classification for exceptional band touching is carried out in Ref.~\onlinecite{Kawabata_gapless_PRL19} for a generic Bloch Hamiltonian. 
However, our analysis clarifies which symmetry classes are relevant for correlated systems.
Our ten-fold way classification is consistent with the corresponding classification results for 38 symmetry classes.

In what follows, we address the classification of exceptional band touching after a brief description of the relevant symmetry.
\subsection{
Symmetry constraints
}

\subsubsection{
$PT$-symmetry
}
\label{sec: class TP symm}

For the correlated systems preserving $PT$-symmetry (i.e., symmetry under the product of time-reversal and spatial inversion), the second quantized Hamiltonian $\hat{H}$ satisfies
\begin{eqnarray}
\label{eq: THT=H 2nd q}
\widehat{PT} \hat{H} \widehat{PT}^{-1} &=& \hat{H}.
\end{eqnarray}
Here, the anti-unitary operator $\widehat{PT}$ is written as
%
\begin{subequations}
\label{eq: defs of PT and UTP}
\begin{eqnarray}
\widehat{PT}&=&\hat{U}_{PT} \mathcal{K}, \\
\hat{U}_{PT} \hat{c}^\dagger_{i\alpha} \hat{U}^\dagger_{PT} &=& \sum_{\beta} \hat{c}^\dagger_{-i\beta} U_{PT,\beta\alpha},
\end{eqnarray}
\end{subequations}
%
where $\hat{c}^\dagger_{i\alpha}$ creates a fermion with state $\alpha$ at site $i$.
$\hat{U}_{PT}$ is a unitary operator. $\mathcal{K}$ is an operator taking complex conjugation.
$U_{PT}$ is a matrix satisfying $U_{PT}U^*_{PT}=\pm \1$.
Here we have supposed that under the inversion, site $j$ is mapped to $-j$.

For $PT$-symmetric systems, the Green's function satisfies~\cite{Gurarie_GFSymm_PRB11}
\begin{subequations}
\begin{eqnarray}
\label{eq: Class PT symm GF}
G(\omega+i\delta,\bm{k})
&=& 
U_{PT} G^T(\omega+i\delta,\bm{k}) U^\dagger_{PT},
\end{eqnarray}
which can be rewritten as
\begin{eqnarray}
\label{eq: Class PT symm Heff}
H_{\mathrm{eff}}(\omega+i\delta,\bm{k})
&=& 
U_{PT} H^T_{\mathrm{eff}}(\omega+i\delta,\bm{k}) U^\dagger_{PT}.
\end{eqnarray}
\end{subequations}
Eq.~(\ref{eq: Class PT symm GF}) can be seen by a straightforward calculation~\cite{TP_GF_ftnt}.

\subsubsection{
$CP$-symmetry
}
\label{sec: class CP symm}
For correlated systems preserving $CP$-symmetry (i.e., symmetry under the product of charge conjugation and inversion), the second quantized Hamiltonian $\hat{H}$ satisfies
\begin{eqnarray}
\widehat{CP}\hat{H}\widehat{CP}^{-1}&=& \hat{H},
\end{eqnarray}
with $\widehat{CP}$ corresponding to the unitary operator $\widehat{CP}=\hat{U}_{CP}$ which transforms $\hat{c}_{i\alpha}$ as
\begin{eqnarray}
\hat{U}_{CP} \hat{c}_{i\alpha}  \hat{U}^\dagger_{CP}&=& \sum_{\beta} \hat{c}^\dagger_{-i\beta}U_{CP,\beta\alpha}.\nonumber 
\end{eqnarray}
%
Here, $U_{CP}$ is a unitary matrix satisfying $U_{CP}U^*_{CP}=\pm \1$.

For $CP$-symmetric systems, the Green's function satisfies
\begin{subequations}
\begin{eqnarray}
\label{eq: Class CP symm GF}
G(\omega+i\delta,\bm{k})
&=& 
-U_{CP} G^*(-\omega+i\delta,\bm{k}) U^\dagger_{CP},
\end{eqnarray}
which can be rewritten as
\begin{eqnarray}
\label{eq: Class CP symm Heff}
H_{\mathrm{eff}}(\omega+i\delta,\bm{k})
&=& 
-U_{CP} H^*_{\mathrm{eff}}(-\omega+i\delta,\bm{k}) U^\dagger_{CP}.
\end{eqnarray}
\end{subequations}
Eq.~(\ref{eq: Class CP symm GF}) can be obtained by using the following relations:
\begin{eqnarray}
\label{eq: CP GR to GA}
G(\omega+i\delta)&=& - U_{CP} G^T(-\omega-i\delta) U^\dagger_{CP},
\end{eqnarray}
and
\begin{eqnarray}
\label{eq: GAdag=GR}
G^*_{\alpha\beta}(\omega-i\delta,\bm{k})
&=& 
G_{\beta\alpha}(\omega+i\delta,\bm{k}).
\end{eqnarray}
We note that applying the Fourier transformation, $G(-\omega-i\delta,\bm{k})$ is rewritten as $G^A(-t,\bm{k})$ which is defined as
\begin{eqnarray}
\label{eq: def of GA}
G^A_{\alpha\beta}(t,\bm{k})&:=& i \langle \hat{c}_{\bm{k}\alpha}(t)\hat{c}^\dagger_{\bm{k}\beta}+\hat{c}^\dagger_{\bm{k}\beta}\hat{c}_{\bm{k}\alpha}(t) \rangle \theta(-t).
\end{eqnarray}
Eqs.~(\ref{eq: CP GR to GA})~and~(\ref{eq: GAdag=GR}) are obtained by straightforward calculations~\cite{CP_GF_ftnt,GAGR_ftnt}.

\subsubsection{
Chiral symmetry
}
\label{sec: class chiral symm}

For the correlated systems preserving chiral symmetry, the second quantized Hamiltonian $\hat{H}$ satisfies
\begin{subequations}
\begin{eqnarray}
\hat{\Gamma} \hat{H} \hat{\Gamma}^{-1} &=& \hat{H},
\end{eqnarray}
with
\begin{eqnarray}
\hat{\Gamma}&=&\hat{U}_{\Gamma} \mathcal{K}. 
\end{eqnarray}
Here, $\hat{U}_\Gamma$ is a unitary operator transforming the annihilation operator as 
\begin{eqnarray}
\hat{U}_{\Gamma} \hat{c}_{i\alpha} \hat{U}^\dagger_{\Gamma} &=& \sum_{\beta}\hat{c}^\dagger_{i\beta} U_{\Gamma,\beta\alpha},
\end{eqnarray}
\end{subequations}
where $U_{\Gamma}$ is a matrix satisfying $U^2_{\Gamma}=\1$.

For chiral symmetric systems, the Green's function satisfies
\begin{subequations}
\begin{eqnarray}
\label{eq: Class Gamma symm GF}
G(\omega+i\delta,\bm{k})
&=& 
-U_{\Gamma} G^\dagger(-\omega+i\delta,\bm{k}) U^\dagger_{\Gamma},
\end{eqnarray}
which can be rewritten as
\begin{eqnarray}
\label{eq: Class chiral symm Heff}
H_{\mathrm{eff}}(\omega+i\delta,\bm{k})
&=& 
-U_{\Gamma} H^\dagger_{\mathrm{eff}}(-\omega+i\delta,\bm{k}) U^\dagger_{\Gamma}.
\end{eqnarray}
\end{subequations}
Eq.~(\ref{eq: Class PT symm GF}) can be obtained by a straightforward calculation~\cite{Gamma_GF_ftnt}.
This equation can also be obtained from Eqs.~(\ref{eq: Class PT symm GF})~and~(\ref{eq: Class CP symm GF}) by noticing that applying the operator $\hat{\Gamma}$ is equivalent to applying the product of the operators $\widehat{PT}$ and $\widehat{CP}$.

\subsection{
Ten-fold way classification
}

Prior to the topological classification of exceptional band touching, we note the following two facts.
(i) Exceptional band touching of the non-Hermitian Hamiltonian $H_{\mathrm{eff}}(\omega=0,\bm{k})$ can be described by a Hermitian Hamiltonian satisfying $\{\tilde{H},\tilde{\Sigma} \}=0$ with $\tilde{\Sigma}=\1\otimes \rho_3$ [see e.g., Eq.~(\ref{eq: H' pg})]~\cite{Gong_class_PRX18,Kawabata_gapped_PRX19,Zhou_gapped_class_PRB19,Kawabata_gapless_PRL19,Yoshida_SPERs_mech19,Liu_MirroPointClass_PRB19}. 
(ii) For Hermitian systems, the classification of $d_{\mathrm{EP}}$-dimensional gapless excitations in $d$ spatial dimensions is accomplished by classifying the $\delta-1$ dimensional gapped Hermitian Hamiltonian with $\delta=d-d_{\mathrm{EP}}$~\cite{Teo_disloc_class_PRB10,Chiu_gapless_class_PRB14}.

Thus, the problem is reduced to classifying gapless excitations of the Hermitian Hamiltonian $\tilde{H}$ in the presence/absence of the following symmetry constraints:
\begin{subequations}
\begin{eqnarray}
\tilde{U}_{PT} \tilde{H}^*(\bm{k}) \tilde{U}^\dagger_{PT}&=& \tilde{H}(\bm{k}), \\
\tilde{U}_{CP} \tilde{H}^*(\bm{k}) \tilde{U}^\dagger_{CP}&=& -\tilde{H}(\bm{k}), \\
\tilde{U}_{\Gamma} \tilde{H}^*(\bm{k}) \tilde{U}^\dagger_{\Gamma}&=& -\tilde{H}(\bm{k}),
\end{eqnarray}
with
\begin{eqnarray}
\tilde{H} &=& 
\left(
\begin{array}{cc}
0 & H_{\mathrm{eff}}(0,\bm{k}) \\
H^\dagger_{\mathrm{eff}}(0,\bm{k}) & 0
\end{array}
\right)_\rho,
\\
\tilde{U}_{PT} &=& 
\left(
\begin{array}{cc}
0 & U_{PT} \\
U^\dagger_{PT} & 0
\end{array}
\right)_\rho,
\\
\tilde{U}_{CP} &=& U_{CP}\rho_0,
\\
\tilde{U}_{\Gamma} &=& 
\left(
\begin{array}{cc}
0 & U_{\Gamma} \\
U^\dagger_{\Gamma} & 0
\end{array}
\right)_\rho,
\end{eqnarray}
\end{subequations}
and 
$\tilde{U}_{PT}\tilde{U}^{*}_{PT}=\pm \1$, $\tilde{U}_{CP}\tilde{U}^{*}_{CP}=\pm \1$, and $\tilde{U}_{\Gamma}\tilde{U}_{\Gamma}=\1$. 

The above relation can also be written with the two anti-unitary operators ($\widetilde{PT}=\tilde{U}_{PT}\mathcal{K}$, and $\widetilde{CP}=\tilde{U}_{CP}\mathcal{K}$) and a unitary operator ($\tilde{\Gamma}:=\tilde{U}_\Gamma$).
We note that the above unitary matrices ($\tilde{U}_{PT}$, $\tilde{U}_{CP}$, and $\tilde{U}_{\Gamma}$) satisfy the following commutation/anti-commutation relations:
\begin{subequations}
\label{eq: class comm anticomm of Sigma}
\begin{eqnarray}
\{ \tilde{U}_{PT}, \tilde{\Sigma} \}&=& 0, \\
{}[ \tilde{U}_{CP}, \tilde{\Sigma} ]&=& 0, \\
\{ \tilde{U}_{\Gamma}, \tilde{\Sigma} \} &=& 0.
\end{eqnarray}
\end{subequations}
%

Therefore, exceptional band touching can be classified by addressing the classification of gapless excitations in Hermitian systems with additional chiral symmetry whose operator $\tilde{\Sigma}$ satisfies Eq.~(\ref{eq: class comm anticomm of Sigma}). 
We address the classification based on the method of the Clifford algebra~\cite{Kitaev_classification_free_2009,TMorimoto_AZclass_PRB13}.
The specific procedure of the classification is summarized in Sec.~\ref{sec: class Hermi H}.
In the next section, we discuss the classification results.

\subsubsection{
Classification results
}

Classification results of $d_{\mathrm{EP}}$-dimensional exceptional band touching for $H_{\mathrm{eff}}(\omega=0,\bm{k})$ are summarized in Table~\ref{table: Hermi_class_AZ+I}. Here, we consider the $d$-dimensional BZ.

For each case of $\delta=d-d_{\mathrm{EP}}$ and symmetry class, this table elucidates the presence/absence of the $\delta-1$ dimensional topological invariant in the BZ;``$\mathbb{Z}$" (``$\mathbb{Z}_2$") indicates the presence of a topological invariant taking an arbitrary integer ($0$ or $1$), respectively; ``0" appearing as the classification result (i.e., from sixth to 13-th column) indicates the absence of such topological invariants.

These classification results explain the exceptional band touching reported so far.
For instance, this table indicates the $\mathbb{Z}$ classification for class~A with $\delta=2$, meaning that there exists exceptional band touching characterized by a one-dimensional topological invariant. This classification result explains the presence of EPs observed in Fig.~\ref{fig: EP Akw J1.8}(b) ($d=2$ and $d_{\mathrm{EP}}=0$).
We note that the emergence of EPs for class~A is also reported for systems with disorder~\cite{Zyuzin_nHEP_PRB18,Papaji_nHEP_PRB19} or electron-phonon coupling~\cite{VKozii_nH_arXiv17}.
With $d=3$ and $d_{EP}=1$, we obtain the same $\delta$, resulting in the $\mathbb{Z}$ classification for class~A. 
This fact also explains the emergence of exceptional loops in three-dimensional systems~\cite{Matsushita_ER_PRB19}.
The classification results for symmetry classes AI, AII, D, and C elucidate the stability of these band touching points in the presence/absence of $PT$- or $CP$-symmetry.

The emergence of SPERs observed in Fig.~\ref{fig: SPERs honey Ak} is also consistent with Table~\ref{table: Hermi_class_AZ+I} ($d=2$ and $d_{\mathrm{EP}}=1$). For class~{AIII} with $\delta=1$, we obtain the $\mathbb{Z}$ classification, implying the presence of the zero-th Chern number.
The $\mathbb{Z}$ classification for class~{AIII} with $\delta=1$ is also consistent with the emergence of SPESs observed in Fig.~\ref{fig: SPES diamond summ}(c) ($d=3$ and $d_{\mathrm{EP}}=2$).
We note that the classification results for symmetry classes BDI, DIII, CII, and CI elucidate the stability of the exceptional band touching in the presence of $PT$- or $CP$-symmetry.

While we have mainly analyzed exceptional band touching for symmetry class~A~or~{AIII} in the previous sections, the classification results summarized in Table~\ref{table: Hermi_class_AZ+I} imply the existence of novel exceptional band touching. 
The verification of exceptional band touching for other cases of symmetry is still missing as well as the material realization.

It is also worth noting that the above table may explain the exceptional band touching away from $\omega=0$ for class~A by recognizing the frequency as an additional momentum, although we have restricted ourselves to $\omega=0$ so far.
Indeed, the emergence of exceptional rings in the $\omega$-$\bm{k}$ space has been demonstrated for two-dimensional heavy fermions~\cite{Michishta_EP_DMFT_arXiv19} ($d=3$ and $d_{\mathrm{EP}}=1$), which is consistent with the $\mathbb{Z}$ classification for symmetry class~A with $\delta=2$.
The above fact allows us to interpret the $\mathbb{Z}$ classification for class~A with $\delta=4$; it implies the presence of novel EPs in the $\omega$-$\bm{k}$ space for three spatial dimensions. 
Further analysis in this direction should be addressed.

\begin{table}[htb]
\begin{center}
\begin{tabular}{c c c c c r c c c c c c c c c}  \hline\hline
symmetry class     & $PT$ & $CP$ & $\Gamma$         & homotopy                           & $\delta=1$      & 2              &    3           & 4              & 5              & 6              & 7              &    8            & $\quad$ & Clifford generators  \\ \hline
A                  & $0$  & $0$  & $0$              &     $\pi_0(C_{\delta})$            &    0            & $\mathbb{Z}$   &    0           & $\mathbb{Z}$   & 0              & $\mathbb{Z}$   & 0              & $\mathbb{Z}$    & $\quad$ & $\{\gamma_0,\cdots,\gamma_{\delta-1},\tilde{\Sigma}\}$ \\ 
AIII               & $0$  & $0$  & $1$              &     $\pi_0(C_{\delta+1})$          & $\mathbb{Z}$    &    0           & $\mathbb{Z}$   & 0              & $\mathbb{Z}$   & 0              & $\mathbb{Z}$   & 0               & $\quad$ & $\{\gamma_0,\cdots,\gamma_{\delta-1},\tilde{\Gamma},\tilde{\Sigma}\}$  \\\hline
AI                 & $1$  & $0$  & $0$              &     $\pi_0(R_{\delta+6})$          &    0            & $\mathbb{Z}$   & $\mathbb{Z}_2$ & $\mathbb{Z}_2$ & 0              & $\mathbb{Z}$   &    0           &    0            & $\quad$ & $\{J\gamma_0,J\gamma_1,\cdots,J\gamma_{\delta-1};\widetilde{PT},J\widetilde{PT},\tilde{\Sigma} \}$ \\
BDI                & $1$  & $1$  & $1$              &     $\pi_0(R_{\delta+7})$          &  $\mathbb{Z}$   & $\mathbb{Z}_2$ & $\mathbb{Z}_2$ &    0           & $\mathbb{Z}$   &    0           &    0           &    0            & $\quad$ & $\{J\gamma_0,J\gamma_1,\cdots,J\gamma_{\delta-1},J\tilde{\Gamma}; \widetilde{PT},J\widetilde{PT},\tilde{\Sigma} \}$  \\
D                  & $0$  & $1$  & $0$              &     $\pi_0(R_{\delta})$            &  $\mathbb{Z}_2$ & $\mathbb{Z}_2$ &    0           & $\mathbb{Z}$   &    0           &    0           &    0           &  $\mathbb{Z}$   & $\quad$ & $\{J\tilde{\Sigma};\gamma_0,\gamma_1,\cdots,\gamma_{\delta-1},\widetilde{CP},J\widetilde{CP} \}$ \\
DIII               & $-1$ & $1$  & $1$              &     $\pi_0(R_{\delta+1})$          &  $\mathbb{Z}_2$ &    0           & $\mathbb{Z}$   &    0           &    0           &    0           &  $\mathbb{Z}$  &  $\mathbb{Z}_2$ & $\quad$ & $\{J\gamma_0,J\gamma_1.\cdots,J\gamma_{\delta-1},\widetilde{PT},J\widetilde{PT}; J\tilde{\Gamma},\tilde{\Sigma} \}$  \\
AII                & $-1$ & $0$  & $0$              &     $\pi_0(R_{\delta+2})$          &    0            & $\mathbb{Z}$   &    0           &    0           &    0           &  $\mathbb{Z}$  &  $\mathbb{Z}_2$&  $\mathbb{Z}_2$ & $\quad$ & $\{J\gamma_0,J\gamma_1,\cdots,J\gamma_{\delta-1},\widetilde{PT},J\widetilde{PT};\tilde{\Sigma} \}$  \\
CII                & $-1$ & $-1$ & $1$              &     $\pi_0(R_{\delta+3})$          &  $\mathbb{Z}$   &    0           &    0           &    0           &  $\mathbb{Z}$  &  $\mathbb{Z}_2$&  $\mathbb{Z}_2$&    0            & $\quad$ & $\{J\gamma_0,J\gamma_1,\cdots,J\gamma_{\delta-1},\widetilde{PT},J\widetilde{PT},J\tilde{\Gamma};\tilde{\Sigma} \}$  \\
C                  & $0$  & $-1$ & $0$              &     $\pi_0(R_{\delta+4})$          &    0            &    0           &    0           &  $\mathbb{Z}$  &  $\mathbb{Z}_2$&  $\mathbb{Z}_2$&    0           &  $\mathbb{Z}$   & $\quad$ & $\{\widetilde{CP},J\widetilde{CP},J\tilde{\Sigma};\gamma_0,\gamma_1,\cdots,\gamma_{\delta-1} \}$  \\
CI                 & $1$  & $-1$ & $1$              &     $\pi_0(R_{\delta+5})$          &    0            &    0           & $\mathbb{Z}$   & $\mathbb{Z}_2$ &  $\mathbb{Z}_2$&   0            &  $\mathbb{Z}$  &    0            & $\quad$ & $\{J\gamma_0,J\gamma_1,\cdots,J\gamma_{\delta-1}; \widetilde{PT}, J\widetilde{PT}, J\tilde{\Gamma},\tilde{\Sigma} \}$  \\
\hline \hline
\end{tabular}
\end{center}
\caption{
Classification results of the $d_{\mathrm{EP}}$-dimensional exceptional band touching for the $d$-dimensional non-Hermitian Hamiltonian $H_{\mathrm{eff}}(\omega=0,\bm{k})$.
Here $\delta:=d-d_{\mathrm{EP}}$ denotes the codimension.
The second, third, and fourth columns specify the symmetry class where $\pm 1$ in the second (third) column indicates the sign of $\widetilde{PT}^2=\pm 1$ ($\widetilde{CP}^2=\pm 1$), respectively.
Here, ``$0$" in these columns indicates the absence of the corresponding symmetry. 
For a given $\delta$, $\mathbb{Z}$ ($\mathbb{Z}_2$) classification indicates the presence of $\delta-1$-dimensional topological invariant taking an arbitrary integer ($0$ or $1$), respectively.
``$0$" in these columns indicates the absence of such topological invariants.
The homotopy of the classifying space $C_q$ or $R_q$ is shown in the fifth column.
The classifying space can be obtained by considering the extension problem where the relevant generators of the Clifford algebra are shown in the last column (see Sec.~\ref{sec: class Hermi H}).
Here, we assume that $[\widehat{PT},\widehat{CP}]=0$ which is satisfied with a proper choice of the gauge.
$\tilde{\Gamma}$ for class BDI, DIII, CII, and CI is defined as the product of $\widetilde{PT}$ and $\widetilde{CP}$ with the prefactor satisfying $\tilde{\Gamma}^2=\1$.
}
\label{table: Hermi_class_AZ+I}
\end{table}

\subsubsection{
Details of the classification for the Hermitian Hamiltonian
}
\label{sec: class Hermi H}

As discussed in the beginning of this section, classification of the $d_{\mathrm{EP}}$-dimensional exceptional band touching in $d$ spatial dimensions is accomplished by classifying the $\delta-1$-dimensional gapped Hermitian Hamiltonian with additional chiral symmetry satisfying Eq.~(\ref{eq: class comm anticomm of Sigma}).
Here, $\delta$ denotes codimension ($\delta=d-d_{\mathrm{EP}}$).
In this section, we address the classification of the gapped Hermitian Hamiltonian based on the method of the Clifford algebra~\cite{Kitaev_classification_free_2009,TMorimoto_AZclass_PRB13}.

In what follows are technical details of the derivation of Table~\ref{table: Hermi_class_AZ+I}. 
Thus, readers, who are interested in physical interpretation of the classification results rather than the technical details, can skip this section.

Specifically, the topological classification based on the Clifford algebra can be carried out by the following steps~\cite{Kitaev_classification_free_2009,TMorimoto_AZclass_PRB13}.

(i) Deform the Hermitian Hamiltonian $\tilde{H}$ to the Hermitian Dirac Hamiltonian $H_0$
\begin{eqnarray}
H_0(\bm{k}) &=& \sum_{j=1,\cdots,\delta-1} k_j \gamma_j +m \gamma_0,
\end{eqnarray}
where $\gamma$'s satisfy $\{ \gamma_i, \gamma_j \}=2\delta_{i,j}$ for $i,j=0.\cdots,\delta-1$. 
Because such deformation is possible for an arbitrary gapped Hamiltonian, the problem is reduced to classifying the possible mass term $\gamma_0$.

(ii) Consider a Clifford algebra $Cl_q$ or $Cl_{p,q}$ with the matrices $\gamma$'s and the symmetry operators. 
$Cl_q$ denotes the Clifford algebra composed of $q$ generators,
\begin{eqnarray}
\{ e_1, e_2,\cdots, e_q \}, 
\end{eqnarray}
where the generator $e_i$ satisfies $e^2_i=1$ for $i=1,\cdots,q$.
$Cl_{p,q}$ represents the Clifford algebra composed of $p+q$ generators,
\begin{eqnarray}
\{ e_1, e_2,\cdots, e_p; e_{p+1},\cdots, e_{p+q}\}, 
\end{eqnarray}
where the generator $e_i$ satisfies $e^2_i=-1$ ($e^2_i=1$) for $i=1,\cdots,p$ ($i=p+1,\cdots,p+q$), respectively.
We note that an operator $J$, satisfying $\{\widetilde{PT},J\}=\{\widetilde{CP},J\}=[H_0(\bm{k}),J]=0$, needs to be introduced in the presence of $PT$- or $CP$-symmetry. This is because $\widetilde{PT}$ and $\widetilde{CP}$ are anti-unitary operators.

(iii) By adding the mass term $\gamma_0$, consider the extension problem to obtain the corresponding classifying space which turns out to be $C_q$ ($R_{q-p}$) when the extension problem is $Cl_{q}\to Cl_{q+1}$ ($Cl_{p,q}\to Cl_{p,q+1}$), respectively.
Here, we note that the corresponding classifying space of the extension problem $Cl_{p,q}\to Cl_{p+1,q}$ is $R_{2+p-q}$~\cite{TMorimoto_AZclass_PRB13}.

(iv) By making use of the relation summarized in Table~\ref{table: Class homotopy}, obtain the classification result $\pi_0(C_q)$ [$\pi_0(R_q)$]. We note that the relations $\pi_0(C_{q+2})=\pi_0(C_{q})$ and $\pi_0(R_{q+8})=\pi_0(R_{q})$ hold, which are known as the Bott periodicity.

With the above procedure, (i)-(iv), we can obtain the classification results shown in Table~\ref{table: Hermi_class_AZ+I}. In the last column, the Clifford algebra, which is generated by the mass term, the kinetic terms and symmetry operators, is shown for each symmetry class. Although one can reproduce the classification results from the last column, we explicitly apply the above procedure for class~A and AII as examples.

\textit{class~A--.} Remembering that the Hamiltonian $\tilde{H}$ in $\delta-1$ dimensions is chiral symmetric $\{ \tilde{H}, \tilde{\Sigma} \}=0$, we obtain the Clifford algebra $C_{\delta}$ generated by
\begin{eqnarray}
\{ \gamma_1,\cdots, \gamma_{\delta-1}, \tilde{\Sigma}\}.
\end{eqnarray}
Introducing the mass term $\gamma_0$ results in the extension problem which is written as $Cl_{\delta} \to Cl_{\delta+1}$.
Here, the Clifford algebra $Cl_{\delta+1}$ is generated by 
\begin{eqnarray}
\{ \gamma_0, \gamma_1,\cdots, \gamma_{\delta-1}, \tilde{\Sigma}\}, 
\end{eqnarray}
which is shown in the last column of Table~\ref{table: Hermi_class_AZ+I}.
Therefore, the corresponding classifying space is $C_\delta$, which indicates that the classification result is computed with $\pi_0(C_\delta)$. 
By making use of the Bott periodicity and the relation summarized in Table~\ref{table: Class homotopy}, we obtain the classification results for $\delta=1,\cdots,8$.

\textit{class~{AII}--.} Firstly, we note that $\widetilde{PT}^2=- 1$ holds. Remembering that the Hamiltonian $\tilde{H}$ in $\delta-1$ dimensions is chiral symmetric $\{ \tilde{H}, \tilde{\Sigma} \}=0$, we obtain the Clifford algebra $C_{\delta}$ generated by
\begin{eqnarray}
\{ J\gamma_1, \cdots, J\gamma_{\delta-1}, \widetilde{PT}, J\widetilde{PT}; \tilde{\Sigma}\}.
\end{eqnarray}
Introducing the mass term $\gamma_0$ results in the extension problem which is written as $Cl_{\delta+1,1} \to Cl_{\delta+2,1}$.
Here, the Clifford algebra $Cl_{\delta+2,1}$ is generated by
\begin{eqnarray}
\{  J\gamma_0, J\gamma_1, \cdots, J\gamma_{\delta-1}, \widetilde{PT}, J\widetilde{PT}; \tilde{\Sigma}\},
\end{eqnarray}
which is shown in the last column of Table~\ref{table: Hermi_class_AZ+I}.
Therefore, the corresponding classifying space is $R_{2+\delta}$, which indicates that the classification result is computed with $\pi_0(R_{2+\delta})$. 
By making use of the Bott periodicity and the relation summarized in Table~\ref{table: Class homotopy}, we obtain the classification results for $\delta=1,\cdots,8$.

\begin{table}[htb]
\begin{center}
\begin{tabular}{c c c c c c c c c c c c c} \hline\hline
classifying space            & $\quad$ & $C_0$        & $C_1$ & $\quad$ & $R_0$        & $R_1$          & $R_2$          & $R_3$ & $R_4$            & $R_5$ & $R_6$ & $R_7$\\ \hline
$\pi_0(C_q)$ or $\pi_0(R_q)$ & $\quad$ & $\mathbb{Z}$ & $0$   & $\quad$ & $\mathbb{Z}$ & $\mathbb{Z}_2$ & $\mathbb{Z}_2$ & $0$   &  $\mathbb{Z}_2$  & $0$   & $0$   & $0$  \\ 
\hline \hline
\end{tabular}
\end{center}
\caption{
Classifying space ($C_q$ and $R_q$) and the corresponding homotopy group [$\pi_0(C_q)$ and $\pi_0(R_q)$].
}
\label{table: Class homotopy}
\end{table}

We note that for $\delta=d+1$, Table~\ref{table: Hermi_class_AZ+I} indicates the classification results for the $d$-dimensional gapped Hamiltonian with additional chiral symmetry satisfying Eq.~(\ref{eq: class comm anticomm of Sigma}).
In this case, the classification results are given by the homotopy group $\pi_0(C_{q-1+d})$ or $\pi_0(C_{q-1+d})$ with an integer $q$ while the original ten-fold way classification for topological insulators/superconductors is given by $\pi_0(C_{q-d})$ or $\pi_0(C_{q-d})$. 
This is due to the fact that applying $\widetilde{PT}$ or $\widetilde{CP}$ does not flip the momentum $\bm{k}$~\cite{KShiozaki_AZclass_PRB14,Bzdusek_AZ+I_PRB17,Yoshida_SPERs_mech19} while applying time-reversal or particle-hole operator does ($\bm{k}\to-\bm{k}$).

\section{
Summary and outlook
}

In this paper, we have briefly reviewed the recently developed non-Hermitian perspective of the band structure in equilibrium systems.
We have seen that the finite lifetime of quasi-particles induces EPs. In addition, we have seen that the symmetry of the many-body Hamiltonian results in SPERs (SPESs) in two (three) dimensions, respectively.
While the above non-Hermitian perspective has been developed recently, there are several open questions to be addressed.

For instance, effects of EPs on transport properties should be further analyzed. 
As seen in this paper, the exceptional band touching induces low energy excitations such as Fermi arcs. The emergence of these low energy excitations may change the conductivity or other electromagnetic responses.

The experimental observation of EPs in electronic systems is also a crucial issue along this direction.
Topological Kondo insulators such as $\mathrm{SmB}_6$~\cite{Takimoto_SmB6_JPSJ11,Neupane_SmB6_NatComm13,Jiang_SmB6_NatComm13,Xu_SmB6_PRB13,Peters_SmB6_PRB16,Peters_SmB6mag_PRB18,Thunstrom_SmB6_arXiv19} and $\mathrm{YbB}_{12}$~\cite{Weng_YbB12_PRL14,Hagiwara_YbB12_NatComm16} might serve as a platform of the EPs because they are strongly correlated materials and show Dirac cones at surfaces. 
Prior to the experimental observation, the quantitative analysis such as LDA+DMFT calculations should be carried out as well as the theoretical proposal of how to experimentally observe the EPs.

While this paper focuses on exceptional band touching, non-Hermiticity induces richer topological physics. Non-Hermitian skin effect is the another representative unique phenomenon~\cite{Alvarez_nHSkin_PRB18,SYao_nHSkin-1D_PRL18,Helbig_elecirSkin_19,Hofmann_ExpRecipSkin_19,Xiao_nHSkin_Exp_arXiv19,Lee_Skin19,Zhang_BECskin19,Okuma_BECskin19,Yoshida_MSkin_arXiv19,Jiang_SkinDisord_PRB19}; the energy spectrum of a non-Hermitian matrix significantly depends on the boundary condition when the skin effect occurs.
Elucidating whether the non-Hermiticity by the finite lifetimes induces the skin effect is an intriguing theoretical open question to be addressed.

Finally, we comment on another significant issue of non-Hermiticity and correlations.
Recently, a fractional quantum Hall phase, a topologically ordered phase, has been extended to non-Hermitian systems~\cite{Yoshida_nHFQH19}.
The extension of topologically ordered phases to non-Hermitian systems is further addressed for a non-Hermitian toric code~\cite{Matsumoto_nHtoric_arXiv19,Guo_nHToric_arXiv20}.
Developing the effective field theory to describe these non-Hermitian topologically ordered phases should be addressed as well as extending them to systems with symmetry (e.g., time-reversal symmetry).

\section{
Acknowledgements
}

This work is partly supported by JSPS KAKENHI Grants No.~JP15H05855, No.~JP16K13845, No.~JP17H06138, No.~JP18H01140, No.~JP18H04316, No.~JP18K03511, No.~JP18H05842, No.~JP19K21032, and No~JP19H01838 and by JST CREST Grant No.~JPMJCR19T1.
A part of numerical data plotted in this paper were obtained on the supercomputer at the ISSP in the University of Tokyo.


\begin{thebibliography}{181}%
\makeatletter
\providecommand \@ifxundefined [1]{%
 \@ifx{#1\undefined}
}%
\providecommand \@ifnum [1]{%
 \ifnum #1\expandafter \@firstoftwo
 \else \expandafter \@secondoftwo
 \fi
}%
\providecommand \@ifx [1]{%
 \ifx #1\expandafter \@firstoftwo
 \else \expandafter \@secondoftwo
 \fi
}%
\providecommand \natexlab [1]{#1}%
\providecommand \enquote  [1]{``#1''}%
\providecommand \bibnamefont  [1]{#1}%
\providecommand \bibfnamefont [1]{#1}%
\providecommand \citenamefont [1]{#1}%
\providecommand \href@noop [0]{\@secondoftwo}%
\providecommand \href [0]{\begingroup \@sanitize@url \@href}%
\providecommand \@href[1]{\@@startlink{#1}\@@href}%
\providecommand \@@href[1]{\endgroup#1\@@endlink}%
\providecommand \@sanitize@url [0]{\catcode `\\12\catcode `\$12\catcode
  `\&12\catcode `\#12\catcode `\^12\catcode `\_12\catcode `\%12\relax}%
\providecommand \@@startlink[1]{}%
\providecommand \@@endlink[0]{}%
\providecommand \url  [0]{\begingroup\@sanitize@url \@url }%
\providecommand \@url [1]{\endgroup\@href {#1}{\urlprefix }}%
\providecommand \urlprefix  [0]{URL }%
\providecommand \Eprint [0]{\href }%
\providecommand \doibase [0]{http://dx.doi.org/}%
\providecommand \selectlanguage [0]{\@gobble}%
\providecommand \bibinfo  [0]{\@secondoftwo}%
\providecommand \bibfield  [0]{\@secondoftwo}%
\providecommand \translation [1]{[#1]}%
\providecommand \BibitemOpen [0]{}%
\providecommand \bibitemStop [0]{}%
\providecommand \bibitemNoStop [0]{.\EOS\space}%
\providecommand \EOS [0]{\spacefactor3000\relax}%
\providecommand \BibitemShut  [1]{\csname bibitem#1\endcsname}%
\let\auto@bib@innerbib\@empty
\bibitem [{\citenamefont {Hatsugai}(1993)}]{Hatsugai_PRL93}%
  \BibitemOpen
  \bibfield  {author} {\bibinfo {author} {\bibfnamefont {Y.}~\bibnamefont
  {Hatsugai}},\ }\href {\doibase 10.1103/PhysRevLett.71.3697} {\bibfield
  {journal} {\bibinfo  {journal} {Phys. Rev. Lett.}\ }\textbf {\bibinfo
  {volume} {71}},\ \bibinfo {pages} {3697} (\bibinfo {year}
  {1993})}\BibitemShut {NoStop}%
\bibitem [{\citenamefont {Kane}\ and\ \citenamefont
  {Mele}(2005{\natexlab{a}})}]{Kane_PRL05_1}%
  \BibitemOpen
  \bibfield  {author} {\bibinfo {author} {\bibfnamefont {C.~L.}\ \bibnamefont
  {Kane}}\ and\ \bibinfo {author} {\bibfnamefont {E.~J.}\ \bibnamefont
  {Mele}},\ }\href {\doibase 10.1103/PhysRevLett.95.146802} {\bibfield
  {journal} {\bibinfo  {journal} {Phys. Rev. Lett.}\ }\textbf {\bibinfo
  {volume} {95}},\ \bibinfo {pages} {146802} (\bibinfo {year}
  {2005}{\natexlab{a}})}\BibitemShut {NoStop}%
\bibitem [{\citenamefont {Kane}\ and\ \citenamefont
  {Mele}(2005{\natexlab{b}})}]{Kane_PRL05_2}%
  \BibitemOpen
  \bibfield  {author} {\bibinfo {author} {\bibfnamefont {C.~L.}\ \bibnamefont
  {Kane}}\ and\ \bibinfo {author} {\bibfnamefont {E.~J.}\ \bibnamefont
  {Mele}},\ }\href {\doibase 10.1103/PhysRevLett.95.226801} {\bibfield
  {journal} {\bibinfo  {journal} {Phys. Rev. Lett.}\ }\textbf {\bibinfo
  {volume} {95}},\ \bibinfo {pages} {226801} (\bibinfo {year}
  {2005}{\natexlab{b}})}\BibitemShut {NoStop}%
\bibitem [{\citenamefont {Bernevig}\ \emph {et~al.}(2006)\citenamefont
  {Bernevig}, \citenamefont {Hughes},\ and\ \citenamefont
  {Zhang}}]{Bernevig_BHZ_Science06}%
  \BibitemOpen
  \bibfield  {author} {\bibinfo {author} {\bibfnamefont {B.~A.}\ \bibnamefont
  {Bernevig}}, \bibinfo {author} {\bibfnamefont {T.~L.}\ \bibnamefont
  {Hughes}}, \ and\ \bibinfo {author} {\bibfnamefont {S.-C.}\ \bibnamefont
  {Zhang}},\ }\href {\doibase 10.1126/science.1133734} {\ \textbf {\bibinfo
  {volume} {314}},\ \bibinfo {pages} {1757} (\bibinfo {year}
  {2006})}\BibitemShut {NoStop}%
\bibitem [{\citenamefont {Qi}\ \emph {et~al.}(2008)\citenamefont {Qi},
  \citenamefont {Hughes},\ and\ \citenamefont {Zhang}}]{Qi_TQFTZ2TI_PRB08}%
  \BibitemOpen
  \bibfield  {author} {\bibinfo {author} {\bibfnamefont {X.-L.}\ \bibnamefont
  {Qi}}, \bibinfo {author} {\bibfnamefont {T.~L.}\ \bibnamefont {Hughes}}, \
  and\ \bibinfo {author} {\bibfnamefont {S.-C.}\ \bibnamefont {Zhang}},\ }\href
  {\doibase 10.1103/PhysRevB.78.195424} {\bibfield  {journal} {\bibinfo
  {journal} {Phys. Rev. B}\ }\textbf {\bibinfo {volume} {78}},\ \bibinfo
  {pages} {195424} (\bibinfo {year} {2008})}\BibitemShut {NoStop}%
\bibitem [{\citenamefont {Hasan}\ and\ \citenamefont
  {Kane}(2010)}]{Hasan_RMP10}%
  \BibitemOpen
  \bibfield  {author} {\bibinfo {author} {\bibfnamefont {M.~Z.}\ \bibnamefont
  {Hasan}}\ and\ \bibinfo {author} {\bibfnamefont {C.~L.}\ \bibnamefont
  {Kane}},\ }\href {\doibase 10.1103/RevModPhys.82.3045} {\bibfield  {journal}
  {\bibinfo  {journal} {Rev. Mod. Phys.}\ }\textbf {\bibinfo {volume} {82}},\
  \bibinfo {pages} {3045} (\bibinfo {year} {2010})}\BibitemShut {NoStop}%
\bibitem [{\citenamefont {Qi}\ and\ \citenamefont {Zhang}(2011)}]{Qi_RMP10}%
  \BibitemOpen
  \bibfield  {author} {\bibinfo {author} {\bibfnamefont {X.-L.}\ \bibnamefont
  {Qi}}\ and\ \bibinfo {author} {\bibfnamefont {S.-C.}\ \bibnamefont {Zhang}},\
  }\href {\doibase 10.1103/RevModPhys.83.1057} {\bibfield  {journal} {\bibinfo
  {journal} {Rev. Mod. Phys.}\ }\textbf {\bibinfo {volume} {83}},\ \bibinfo
  {pages} {1057} (\bibinfo {year} {2011})}\BibitemShut {NoStop}%
\bibitem [{\citenamefont {Wan}\ \emph {et~al.}(2011)\citenamefont {Wan},
  \citenamefont {Turner}, \citenamefont {Vishwanath},\ and\ \citenamefont
  {Savrasov}}]{XWan_PRB11_Weyl}%
  \BibitemOpen
  \bibfield  {author} {\bibinfo {author} {\bibfnamefont {X.}~\bibnamefont
  {Wan}}, \bibinfo {author} {\bibfnamefont {A.~M.}\ \bibnamefont {Turner}},
  \bibinfo {author} {\bibfnamefont {A.}~\bibnamefont {Vishwanath}}, \ and\
  \bibinfo {author} {\bibfnamefont {S.~Y.}\ \bibnamefont {Savrasov}},\ }\href
  {\doibase 10.1103/PhysRevB.83.205101} {\bibfield  {journal} {\bibinfo
  {journal} {Phys. Rev. B}\ }\textbf {\bibinfo {volume} {83}},\ \bibinfo
  {pages} {205101} (\bibinfo {year} {2011})}\BibitemShut {NoStop}%
\bibitem [{\citenamefont {Weng}\ \emph {et~al.}(2015)\citenamefont {Weng},
  \citenamefont {Fang}, \citenamefont {Fang}, \citenamefont {Bernevig},\ and\
  \citenamefont {Dai}}]{HWeng_PRX15_Weyl}%
  \BibitemOpen
  \bibfield  {author} {\bibinfo {author} {\bibfnamefont {H.}~\bibnamefont
  {Weng}}, \bibinfo {author} {\bibfnamefont {C.}~\bibnamefont {Fang}}, \bibinfo
  {author} {\bibfnamefont {Z.}~\bibnamefont {Fang}}, \bibinfo {author}
  {\bibfnamefont {B.~A.}\ \bibnamefont {Bernevig}}, \ and\ \bibinfo {author}
  {\bibfnamefont {X.}~\bibnamefont {Dai}},\ }\href {\doibase
  10.1103/PhysRevX.5.011029} {\bibfield  {journal} {\bibinfo  {journal} {Phys.
  Rev. X}\ }\textbf {\bibinfo {volume} {5}},\ \bibinfo {pages} {011029}
  (\bibinfo {year} {2015})}\BibitemShut {NoStop}%
\bibitem [{\citenamefont {Xu}\ \emph {et~al.}(2015)\citenamefont {Xu},
  \citenamefont {Belopolski}, \citenamefont {Alidoust}, \citenamefont
  {Neupane}, \citenamefont {Bian}, \citenamefont {Zhang}, \citenamefont
  {Sankar}, \citenamefont {Chang}, \citenamefont {Yuan}, \citenamefont {Lee},
  \citenamefont {Huang}, \citenamefont {Zheng}, \citenamefont {Ma},
  \citenamefont {Sanchez}, \citenamefont {Wang}, \citenamefont {Bansil},
  \citenamefont {Chou}, \citenamefont {Shibayev}, \citenamefont {Lin},
  \citenamefont {Jia},\ and\ \citenamefont {Hasan}}]{SYXu_Science15_Weyl}%
  \BibitemOpen
  \bibfield  {author} {\bibinfo {author} {\bibfnamefont {S.-Y.}\ \bibnamefont
  {Xu}}, \bibinfo {author} {\bibfnamefont {I.}~\bibnamefont {Belopolski}},
  \bibinfo {author} {\bibfnamefont {N.}~\bibnamefont {Alidoust}}, \bibinfo
  {author} {\bibfnamefont {M.}~\bibnamefont {Neupane}}, \bibinfo {author}
  {\bibfnamefont {G.}~\bibnamefont {Bian}}, \bibinfo {author} {\bibfnamefont
  {C.}~\bibnamefont {Zhang}}, \bibinfo {author} {\bibfnamefont
  {R.}~\bibnamefont {Sankar}}, \bibinfo {author} {\bibfnamefont
  {G.}~\bibnamefont {Chang}}, \bibinfo {author} {\bibfnamefont
  {Z.}~\bibnamefont {Yuan}}, \bibinfo {author} {\bibfnamefont {C.-C.}\
  \bibnamefont {Lee}}, \bibinfo {author} {\bibfnamefont {S.-M.}\ \bibnamefont
  {Huang}}, \bibinfo {author} {\bibfnamefont {H.}~\bibnamefont {Zheng}},
  \bibinfo {author} {\bibfnamefont {J.}~\bibnamefont {Ma}}, \bibinfo {author}
  {\bibfnamefont {D.~S.}\ \bibnamefont {Sanchez}}, \bibinfo {author}
  {\bibfnamefont {B.}~\bibnamefont {Wang}}, \bibinfo {author} {\bibfnamefont
  {A.}~\bibnamefont {Bansil}}, \bibinfo {author} {\bibfnamefont
  {F.}~\bibnamefont {Chou}}, \bibinfo {author} {\bibfnamefont {P.~P.}\
  \bibnamefont {Shibayev}}, \bibinfo {author} {\bibfnamefont {H.}~\bibnamefont
  {Lin}}, \bibinfo {author} {\bibfnamefont {S.}~\bibnamefont {Jia}}, \ and\
  \bibinfo {author} {\bibfnamefont {M.~Z.}\ \bibnamefont {Hasan}},\ }\href
  {\doibase 10.1126/science.aaa9297} {\ \textbf {\bibinfo {volume} {349}},\
  \bibinfo {pages} {613} (\bibinfo {year} {2015})}\BibitemShut {NoStop}%
\bibitem [{\citenamefont {Lv}\ \emph {et~al.}(2015)\citenamefont {Lv},
  \citenamefont {Weng}, \citenamefont {Fu}, \citenamefont {Wang}, \citenamefont
  {Miao}, \citenamefont {Ma}, \citenamefont {Richard}, \citenamefont {Huang},
  \citenamefont {Zhao}, \citenamefont {Chen}, \citenamefont {Fang},
  \citenamefont {Dai}, \citenamefont {Qian},\ and\ \citenamefont
  {Ding}}]{BQLv_PRX15_Weyl}%
  \BibitemOpen
  \bibfield  {author} {\bibinfo {author} {\bibfnamefont {B.~Q.}\ \bibnamefont
  {Lv}}, \bibinfo {author} {\bibfnamefont {H.~M.}\ \bibnamefont {Weng}},
  \bibinfo {author} {\bibfnamefont {B.~B.}\ \bibnamefont {Fu}}, \bibinfo
  {author} {\bibfnamefont {X.~P.}\ \bibnamefont {Wang}}, \bibinfo {author}
  {\bibfnamefont {H.}~\bibnamefont {Miao}}, \bibinfo {author} {\bibfnamefont
  {J.}~\bibnamefont {Ma}}, \bibinfo {author} {\bibfnamefont {P.}~\bibnamefont
  {Richard}}, \bibinfo {author} {\bibfnamefont {X.~C.}\ \bibnamefont {Huang}},
  \bibinfo {author} {\bibfnamefont {L.~X.}\ \bibnamefont {Zhao}}, \bibinfo
  {author} {\bibfnamefont {G.~F.}\ \bibnamefont {Chen}}, \bibinfo {author}
  {\bibfnamefont {Z.}~\bibnamefont {Fang}}, \bibinfo {author} {\bibfnamefont
  {X.}~\bibnamefont {Dai}}, \bibinfo {author} {\bibfnamefont {T.}~\bibnamefont
  {Qian}}, \ and\ \bibinfo {author} {\bibfnamefont {H.}~\bibnamefont {Ding}},\
  }\href {\doibase 10.1103/PhysRevX.5.031013} {\bibfield  {journal} {\bibinfo
  {journal} {Phys. Rev. X}\ }\textbf {\bibinfo {volume} {5}},\ \bibinfo {pages}
  {031013} (\bibinfo {year} {2015})}\BibitemShut {NoStop}%
\bibitem [{\citenamefont {Ryu}\ and\ \citenamefont
  {Hatsugai}(2002)}]{Ryu_Majoranaedge_PRL02}%
  \BibitemOpen
  \bibfield  {author} {\bibinfo {author} {\bibfnamefont {S.}~\bibnamefont
  {Ryu}}\ and\ \bibinfo {author} {\bibfnamefont {Y.}~\bibnamefont {Hatsugai}},\
  }\href {\doibase 10.1103/PhysRevLett.89.077002} {\bibfield  {journal}
  {\bibinfo  {journal} {Phys. Rev. Lett.}\ }\textbf {\bibinfo {volume} {89}},\
  \bibinfo {pages} {077002} (\bibinfo {year} {2002})}\BibitemShut {NoStop}%
\bibitem [{\citenamefont {Raghu}\ \emph {et~al.}(2008)\citenamefont {Raghu},
  \citenamefont {Qi}, \citenamefont {Honerkamp},\ and\ \citenamefont
  {Zhang}}]{Raghu_CITI_PRL08}%
  \BibitemOpen
  \bibfield  {author} {\bibinfo {author} {\bibfnamefont {S.}~\bibnamefont
  {Raghu}}, \bibinfo {author} {\bibfnamefont {X.-L.}\ \bibnamefont {Qi}},
  \bibinfo {author} {\bibfnamefont {C.}~\bibnamefont {Honerkamp}}, \ and\
  \bibinfo {author} {\bibfnamefont {S.-C.}\ \bibnamefont {Zhang}},\ }\href
  {\doibase 10.1103/PhysRevLett.100.156401} {\bibfield  {journal} {\bibinfo
  {journal} {Phys. Rev. Lett.}\ }\textbf {\bibinfo {volume} {100}},\ \bibinfo
  {pages} {156401} (\bibinfo {year} {2008})}\BibitemShut {NoStop}%
\bibitem [{\citenamefont {Mong}\ \emph {et~al.}(2010)\citenamefont {Mong},
  \citenamefont {Essin},\ and\ \citenamefont {Moore}}]{Mong_AFTI_PRB10}%
  \BibitemOpen
  \bibfield  {author} {\bibinfo {author} {\bibfnamefont {R.~S.~K.}\
  \bibnamefont {Mong}}, \bibinfo {author} {\bibfnamefont {A.~M.}\ \bibnamefont
  {Essin}}, \ and\ \bibinfo {author} {\bibfnamefont {J.~E.}\ \bibnamefont
  {Moore}},\ }\href {\doibase 10.1103/PhysRevB.81.245209} {\bibfield  {journal}
  {\bibinfo  {journal} {Phys. Rev. B}\ }\textbf {\bibinfo {volume} {81}},\
  \bibinfo {pages} {245209} (\bibinfo {year} {2010})}\BibitemShut {NoStop}%
\bibitem [{\citenamefont {Hohenadler}\ \emph {et~al.}(2011)\citenamefont
  {Hohenadler}, \citenamefont {Lang},\ and\ \citenamefont
  {Assaad}}]{MHohenadler_PRL11_corr_topo}%
  \BibitemOpen
  \bibfield  {author} {\bibinfo {author} {\bibfnamefont {M.}~\bibnamefont
  {Hohenadler}}, \bibinfo {author} {\bibfnamefont {T.~C.}\ \bibnamefont
  {Lang}}, \ and\ \bibinfo {author} {\bibfnamefont {F.~F.}\ \bibnamefont
  {Assaad}},\ }\href {\doibase 10.1103/PhysRevLett.106.100403} {\bibfield
  {journal} {\bibinfo  {journal} {Phys. Rev. Lett.}\ }\textbf {\bibinfo
  {volume} {106}},\ \bibinfo {pages} {100403} (\bibinfo {year}
  {2011})}\BibitemShut {NoStop}%
\bibitem [{\citenamefont {Yamaji}\ and\ \citenamefont
  {Imada}(2011)}]{Yamaji_CorrTI_PRB11}%
  \BibitemOpen
  \bibfield  {author} {\bibinfo {author} {\bibfnamefont {Y.}~\bibnamefont
  {Yamaji}}\ and\ \bibinfo {author} {\bibfnamefont {M.}~\bibnamefont {Imada}},\
  }\href {\doibase 10.1103/PhysRevB.83.205122} {\bibfield  {journal} {\bibinfo
  {journal} {Phys. Rev. B}\ }\textbf {\bibinfo {volume} {83}},\ \bibinfo
  {pages} {205122} (\bibinfo {year} {2011})}\BibitemShut {NoStop}%
\bibitem [{\citenamefont {Yu}\ \emph {et~al.}(2011)\citenamefont {Yu},
  \citenamefont {Xie},\ and\ \citenamefont {Li}}]{SLYu_PRL11_corr_topo}%
  \BibitemOpen
  \bibfield  {author} {\bibinfo {author} {\bibfnamefont {S.-L.}\ \bibnamefont
  {Yu}}, \bibinfo {author} {\bibfnamefont {X.~C.}\ \bibnamefont {Xie}}, \ and\
  \bibinfo {author} {\bibfnamefont {J.-X.}\ \bibnamefont {Li}},\ }\href
  {\doibase 10.1103/PhysRevLett.107.010401} {\bibfield  {journal} {\bibinfo
  {journal} {Phys. Rev. Lett.}\ }\textbf {\bibinfo {volume} {107}},\ \bibinfo
  {pages} {010401} (\bibinfo {year} {2011})}\BibitemShut {NoStop}%
\bibitem [{\citenamefont {Yoshida}\ \emph {et~al.}(2012)\citenamefont
  {Yoshida}, \citenamefont {Fujimoto},\ and\ \citenamefont
  {Kawakami}}]{Yoshida_TI_plusU_DMFT_PRB12}%
  \BibitemOpen
  \bibfield  {author} {\bibinfo {author} {\bibfnamefont {T.}~\bibnamefont
  {Yoshida}}, \bibinfo {author} {\bibfnamefont {S.}~\bibnamefont {Fujimoto}}, \
  and\ \bibinfo {author} {\bibfnamefont {N.}~\bibnamefont {Kawakami}},\ }\href
  {\doibase 10.1103/PhysRevB.85.125113} {\bibfield  {journal} {\bibinfo
  {journal} {Phys. Rev. B}\ }\textbf {\bibinfo {volume} {85}},\ \bibinfo
  {pages} {125113} (\bibinfo {year} {2012})}\BibitemShut {NoStop}%
\bibitem [{\citenamefont {Tada}\ \emph {et~al.}(2012)\citenamefont {Tada},
  \citenamefont {Peters}, \citenamefont {Oshikawa}, \citenamefont {Koga},
  \citenamefont {Kawakami},\ and\ \citenamefont {Fujimoto}}]{TBI_Mott_Tada}%
  \BibitemOpen
  \bibfield  {author} {\bibinfo {author} {\bibfnamefont {Y.}~\bibnamefont
  {Tada}}, \bibinfo {author} {\bibfnamefont {R.}~\bibnamefont {Peters}},
  \bibinfo {author} {\bibfnamefont {M.}~\bibnamefont {Oshikawa}}, \bibinfo
  {author} {\bibfnamefont {A.}~\bibnamefont {Koga}}, \bibinfo {author}
  {\bibfnamefont {N.}~\bibnamefont {Kawakami}}, \ and\ \bibinfo {author}
  {\bibfnamefont {S.}~\bibnamefont {Fujimoto}},\ }\href {\doibase
  10.1103/PhysRevB.85.165138} {\bibfield  {journal} {\bibinfo  {journal} {Phys.
  Rev. B}\ }\textbf {\bibinfo {volume} {85}},\ \bibinfo {pages} {165138}
  (\bibinfo {year} {2012})}\BibitemShut {NoStop}%
\bibitem [{\citenamefont {Gurarie}(2011)}]{Gurarie_GFSymm_PRB11}%
  \BibitemOpen
  \bibfield  {author} {\bibinfo {author} {\bibfnamefont {V.}~\bibnamefont
  {Gurarie}},\ }\href {\doibase 10.1103/PhysRevB.83.085426} {\bibfield
  {journal} {\bibinfo  {journal} {Phys. Rev. B}\ }\textbf {\bibinfo {volume}
  {83}},\ \bibinfo {pages} {085426} (\bibinfo {year} {2011})}\BibitemShut
  {NoStop}%
\bibitem [{\citenamefont {Essin}\ and\ \citenamefont
  {Gurarie}(2011)}]{Essin_GFSymm_PRB11}%
  \BibitemOpen
  \bibfield  {author} {\bibinfo {author} {\bibfnamefont {A.~M.}\ \bibnamefont
  {Essin}}\ and\ \bibinfo {author} {\bibfnamefont {V.}~\bibnamefont
  {Gurarie}},\ }\href {\doibase 10.1103/PhysRevB.84.125132} {\bibfield
  {journal} {\bibinfo  {journal} {Phys. Rev. B}\ }\textbf {\bibinfo {volume}
  {84}},\ \bibinfo {pages} {125132} (\bibinfo {year} {2011})}\BibitemShut
  {NoStop}%
\bibitem [{\citenamefont {Hohenadler}\ and\ \citenamefont
  {Assaad}(2013)}]{Hohenadler_Review_IOP_13}%
  \BibitemOpen
  \bibfield  {author} {\bibinfo {author} {\bibfnamefont {M.}~\bibnamefont
  {Hohenadler}}\ and\ \bibinfo {author} {\bibfnamefont {F.~F.}\ \bibnamefont
  {Assaad}},\ }\href {\doibase 10.1088/0953-8984/25/14/143201} {\bibfield
  {journal} {\bibinfo  {journal} {Journal of Physics: Condensed Matter}\
  }\textbf {\bibinfo {volume} {25}},\ \bibinfo {pages} {143201} (\bibinfo
  {year} {2013})}\BibitemShut {NoStop}%
\bibitem [{\citenamefont {Rachel}(2018)}]{Rachel_Review_RPP18}%
  \BibitemOpen
  \bibfield  {author} {\bibinfo {author} {\bibfnamefont {S.}~\bibnamefont
  {Rachel}},\ }\href {\doibase 10.1088/1361-6633/aad6a6} {\bibfield  {journal}
  {\bibinfo  {journal} {Reports on Progress in Physics}\ }\textbf {\bibinfo
  {volume} {81}},\ \bibinfo {pages} {116501} (\bibinfo {year}
  {2018})}\BibitemShut {NoStop}%
\bibitem [{\citenamefont {Tsui}\ \emph {et~al.}(1982)\citenamefont {Tsui},
  \citenamefont {Stormer},\ and\ \citenamefont {Gossard}}]{Tsui_FQHEExp_PRL82}%
  \BibitemOpen
  \bibfield  {author} {\bibinfo {author} {\bibfnamefont {D.~C.}\ \bibnamefont
  {Tsui}}, \bibinfo {author} {\bibfnamefont {H.~L.}\ \bibnamefont {Stormer}}, \
  and\ \bibinfo {author} {\bibfnamefont {A.~C.}\ \bibnamefont {Gossard}},\
  }\href {\doibase 10.1103/PhysRevLett.48.1559} {\bibfield  {journal} {\bibinfo
   {journal} {Phys. Rev. Lett.}\ }\textbf {\bibinfo {volume} {48}},\ \bibinfo
  {pages} {1559} (\bibinfo {year} {1982})}\BibitemShut {NoStop}%
\bibitem [{\citenamefont {Laughlin}(1983)}]{Laughlin_FQHE_PRL83}%
  \BibitemOpen
  \bibfield  {author} {\bibinfo {author} {\bibfnamefont {R.~B.}\ \bibnamefont
  {Laughlin}},\ }\href {\doibase 10.1103/PhysRevLett.50.1395} {\bibfield
  {journal} {\bibinfo  {journal} {Phys. Rev. Lett.}\ }\textbf {\bibinfo
  {volume} {50}},\ \bibinfo {pages} {1395} (\bibinfo {year}
  {1983})}\BibitemShut {NoStop}%
\bibitem [{\citenamefont {Jain}(1989)}]{Jain_FQHE_PRL89}%
  \BibitemOpen
  \bibfield  {author} {\bibinfo {author} {\bibfnamefont {J.~K.}\ \bibnamefont
  {Jain}},\ }\href {\doibase 10.1103/PhysRevLett.63.199} {\bibfield  {journal}
  {\bibinfo  {journal} {Phys. Rev. Lett.}\ }\textbf {\bibinfo {volume} {63}},\
  \bibinfo {pages} {199} (\bibinfo {year} {1989})}\BibitemShut {NoStop}%
\bibitem [{\citenamefont {Wen}(1995)}]{Wen_TopoOrder_SciAdv95}%
  \BibitemOpen
  \bibfield  {author} {\bibinfo {author} {\bibfnamefont {X.-G.}\ \bibnamefont
  {Wen}},\ }\href {\doibase 10.1080/00018739500101566} {\bibfield  {journal}
  {\bibinfo  {journal} {Advances in Physics}\ }\textbf {\bibinfo {volume}
  {44}},\ \bibinfo {pages} {405} (\bibinfo {year} {1995})}\BibitemShut
  {NoStop}%
\bibitem [{\citenamefont {Kitaev}(2003)}]{Kitaev_ToricCode_Elsevier03}%
  \BibitemOpen
  \bibfield  {author} {\bibinfo {author} {\bibfnamefont {A.}~\bibnamefont
  {Kitaev}},\ }\href {\doibase https://doi.org/10.1016/S0003-4916(02)00018-0}
  {\bibfield  {journal} {\bibinfo  {journal} {Annals of Physics}\ }\textbf
  {\bibinfo {volume} {303}},\ \bibinfo {pages} {2 } (\bibinfo {year}
  {2003})}\BibitemShut {NoStop}%
\bibitem [{\citenamefont {Hamma}\ \emph {et~al.}(2005)\citenamefont {Hamma},
  \citenamefont {Zanardi},\ and\ \citenamefont
  {Wen}}]{Hamma_3DToricCode_PRB05}%
  \BibitemOpen
  \bibfield  {author} {\bibinfo {author} {\bibfnamefont {A.}~\bibnamefont
  {Hamma}}, \bibinfo {author} {\bibfnamefont {P.}~\bibnamefont {Zanardi}}, \
  and\ \bibinfo {author} {\bibfnamefont {X.-G.}\ \bibnamefont {Wen}},\ }\href
  {\doibase 10.1103/PhysRevB.72.035307} {\bibfield  {journal} {\bibinfo
  {journal} {Phys. Rev. B}\ }\textbf {\bibinfo {volume} {72}},\ \bibinfo
  {pages} {035307} (\bibinfo {year} {2005})}\BibitemShut {NoStop}%
\bibitem [{\citenamefont {Kitaev}(2006)}]{Kitaev_KitaevHoneyconmb_Elsevier06}%
  \BibitemOpen
  \bibfield  {author} {\bibinfo {author} {\bibfnamefont {A.}~\bibnamefont
  {Kitaev}},\ }\href {\doibase https://doi.org/10.1016/j.aop.2005.10.005}
  {\bibfield  {journal} {\bibinfo  {journal} {Annals of Physics}\ }\textbf
  {\bibinfo {volume} {321}},\ \bibinfo {pages} {2 } (\bibinfo {year} {2006})},\
  \bibinfo {note} {january Special Issue}\BibitemShut {NoStop}%
\bibitem [{\citenamefont {Tang}\ \emph {et~al.}(2011)\citenamefont {Tang},
  \citenamefont {Mei},\ and\ \citenamefont {Wen}}]{Tang_FChern_PRL11}%
  \BibitemOpen
  \bibfield  {author} {\bibinfo {author} {\bibfnamefont {E.}~\bibnamefont
  {Tang}}, \bibinfo {author} {\bibfnamefont {J.-W.}\ \bibnamefont {Mei}}, \
  and\ \bibinfo {author} {\bibfnamefont {X.-G.}\ \bibnamefont {Wen}},\ }\href
  {\doibase 10.1103/PhysRevLett.106.236802} {\bibfield  {journal} {\bibinfo
  {journal} {Phys. Rev. Lett.}\ }\textbf {\bibinfo {volume} {106}},\ \bibinfo
  {pages} {236802} (\bibinfo {year} {2011})}\BibitemShut {NoStop}%
\bibitem [{\citenamefont {Sun}\ \emph {et~al.}(2011)\citenamefont {Sun},
  \citenamefont {Gu}, \citenamefont {Katsura},\ and\ \citenamefont
  {Das~Sarma}}]{Sun_FChern_PRL11}%
  \BibitemOpen
  \bibfield  {author} {\bibinfo {author} {\bibfnamefont {K.}~\bibnamefont
  {Sun}}, \bibinfo {author} {\bibfnamefont {Z.}~\bibnamefont {Gu}}, \bibinfo
  {author} {\bibfnamefont {H.}~\bibnamefont {Katsura}}, \ and\ \bibinfo
  {author} {\bibfnamefont {S.}~\bibnamefont {Das~Sarma}},\ }\href {\doibase
  10.1103/PhysRevLett.106.236803} {\bibfield  {journal} {\bibinfo  {journal}
  {Phys. Rev. Lett.}\ }\textbf {\bibinfo {volume} {106}},\ \bibinfo {pages}
  {236803} (\bibinfo {year} {2011})}\BibitemShut {NoStop}%
\bibitem [{\citenamefont {Neupert}\ \emph {et~al.}(2011)\citenamefont
  {Neupert}, \citenamefont {Santos}, \citenamefont {Chamon},\ and\
  \citenamefont {Mudry}}]{Neupert_FChern_PRL11}%
  \BibitemOpen
  \bibfield  {author} {\bibinfo {author} {\bibfnamefont {T.}~\bibnamefont
  {Neupert}}, \bibinfo {author} {\bibfnamefont {L.}~\bibnamefont {Santos}},
  \bibinfo {author} {\bibfnamefont {C.}~\bibnamefont {Chamon}}, \ and\ \bibinfo
  {author} {\bibfnamefont {C.}~\bibnamefont {Mudry}},\ }\href {\doibase
  10.1103/PhysRevLett.106.236804} {\bibfield  {journal} {\bibinfo  {journal}
  {Phys. Rev. Lett.}\ }\textbf {\bibinfo {volume} {106}},\ \bibinfo {pages}
  {236804} (\bibinfo {year} {2011})}\BibitemShut {NoStop}%
\bibitem [{\citenamefont {Regnault}\ and\ \citenamefont
  {Bernevig}(2011)}]{Regnalt_FChen_PRX11}%
  \BibitemOpen
  \bibfield  {author} {\bibinfo {author} {\bibfnamefont {N.}~\bibnamefont
  {Regnault}}\ and\ \bibinfo {author} {\bibfnamefont {B.~A.}\ \bibnamefont
  {Bernevig}},\ }\href {\doibase 10.1103/PhysRevX.1.021014} {\bibfield
  {journal} {\bibinfo  {journal} {Phys. Rev. X}\ }\textbf {\bibinfo {volume}
  {1}},\ \bibinfo {pages} {021014} (\bibinfo {year} {2011})}\BibitemShut
  {NoStop}%
\bibitem [{\citenamefont {Sheng}\ \emph {et~al.}(2011)\citenamefont {Sheng},
  \citenamefont {Gu}, \citenamefont {Sun},\ and\ \citenamefont
  {Sheng}}]{Sheng_FChern_NComm12}%
  \BibitemOpen
  \bibfield  {author} {\bibinfo {author} {\bibfnamefont {D.~N.}\ \bibnamefont
  {Sheng}}, \bibinfo {author} {\bibfnamefont {Z.-C.}\ \bibnamefont {Gu}},
  \bibinfo {author} {\bibfnamefont {K.}~\bibnamefont {Sun}}, \ and\ \bibinfo
  {author} {\bibfnamefont {L.}~\bibnamefont {Sheng}},\ }\href
  {https://doi.org/10.1038/ncomms1380} {\bibfield  {journal} {\bibinfo
  {journal} {Nature Communications}\ }\textbf {\bibinfo {volume} {2}},\
  \bibinfo {pages} {389 EP } (\bibinfo {year} {2011})},\ \bibinfo {note}
  {article}\BibitemShut {NoStop}%
\bibitem [{\citenamefont {Bergholtz}\ and\ \citenamefont
  {Liu}(2013)}]{Bergholtz_FChern_IntJModPhysB13}%
  \BibitemOpen
  \bibfield  {author} {\bibinfo {author} {\bibfnamefont {E.~J.}\ \bibnamefont
  {Bergholtz}}\ and\ \bibinfo {author} {\bibfnamefont {Z.}~\bibnamefont
  {Liu}},\ }\href {\doibase 10.1142/S021797921330017X} {\bibfield  {journal}
  {\bibinfo  {journal} {International Journal of Modern Physics B}\ }\textbf
  {\bibinfo {volume} {27}},\ \bibinfo {pages} {1330017} (\bibinfo {year}
  {2013})}\BibitemShut {NoStop}%
\bibitem [{\citenamefont {Pesin}\ and\ \citenamefont
  {Balents}(2010)}]{Pesin_TM_NatPhys10}%
  \BibitemOpen
  \bibfield  {author} {\bibinfo {author} {\bibfnamefont {D.}~\bibnamefont
  {Pesin}}\ and\ \bibinfo {author} {\bibfnamefont {L.}~\bibnamefont
  {Balents}},\ }\href {https://doi.org/10.1038/nphys1606} {\bibfield  {journal}
  {\bibinfo  {journal} {Nature Physics}\ }\textbf {\bibinfo {volume} {6}},\
  \bibinfo {pages} {376 EP } (\bibinfo {year} {2010})},\ \bibinfo {note}
  {article}\BibitemShut {NoStop}%
\bibitem [{\citenamefont {Manmana}\ \emph {et~al.}(2012)\citenamefont
  {Manmana}, \citenamefont {Essin}, \citenamefont {Noack},\ and\ \citenamefont
  {Gurarie}}]{Manmana_TMI1D_PRB12}%
  \BibitemOpen
  \bibfield  {author} {\bibinfo {author} {\bibfnamefont {S.~R.}\ \bibnamefont
  {Manmana}}, \bibinfo {author} {\bibfnamefont {A.~M.}\ \bibnamefont {Essin}},
  \bibinfo {author} {\bibfnamefont {R.~M.}\ \bibnamefont {Noack}}, \ and\
  \bibinfo {author} {\bibfnamefont {V.}~\bibnamefont {Gurarie}},\ }\href
  {\doibase 10.1103/PhysRevB.86.205119} {\bibfield  {journal} {\bibinfo
  {journal} {Phys. Rev. B}\ }\textbf {\bibinfo {volume} {86}},\ \bibinfo
  {pages} {205119} (\bibinfo {year} {2012})}\BibitemShut {NoStop}%
\bibitem [{\citenamefont {Yoshida}\ \emph {et~al.}(2014)\citenamefont
  {Yoshida}, \citenamefont {Peters}, \citenamefont {Fujimoto},\ and\
  \citenamefont {Kawakami}}]{Yoshida_TMI1D_PRL14}%
  \BibitemOpen
  \bibfield  {author} {\bibinfo {author} {\bibfnamefont {T.}~\bibnamefont
  {Yoshida}}, \bibinfo {author} {\bibfnamefont {R.}~\bibnamefont {Peters}},
  \bibinfo {author} {\bibfnamefont {S.}~\bibnamefont {Fujimoto}}, \ and\
  \bibinfo {author} {\bibfnamefont {N.}~\bibnamefont {Kawakami}},\ }\href
  {\doibase 10.1103/PhysRevLett.112.196404} {\bibfield  {journal} {\bibinfo
  {journal} {Phys. Rev. Lett.}\ }\textbf {\bibinfo {volume} {112}},\ \bibinfo
  {pages} {196404} (\bibinfo {year} {2014})}\BibitemShut {NoStop}%
\bibitem [{\citenamefont {Yoshida}\ and\ \citenamefont
  {Kawakami}(2016)}]{Yoshida_TMI2D_PRB16}%
  \BibitemOpen
  \bibfield  {author} {\bibinfo {author} {\bibfnamefont {T.}~\bibnamefont
  {Yoshida}}\ and\ \bibinfo {author} {\bibfnamefont {N.}~\bibnamefont
  {Kawakami}},\ }\href {\doibase 10.1103/PhysRevB.94.085149} {\bibfield
  {journal} {\bibinfo  {journal} {Phys. Rev. B}\ }\textbf {\bibinfo {volume}
  {94}},\ \bibinfo {pages} {085149} (\bibinfo {year} {2016})}\BibitemShut
  {NoStop}%
\bibitem [{\citenamefont {Kudo}\ \emph {et~al.}(2019)\citenamefont {Kudo},
  \citenamefont {Yoshida},\ and\ \citenamefont {Hatsugai}}]{Kudo_HOTMI_PRL19}%
  \BibitemOpen
  \bibfield  {author} {\bibinfo {author} {\bibfnamefont {K.}~\bibnamefont
  {Kudo}}, \bibinfo {author} {\bibfnamefont {T.}~\bibnamefont {Yoshida}}, \
  and\ \bibinfo {author} {\bibfnamefont {Y.}~\bibnamefont {Hatsugai}},\ }\href
  {\doibase 10.1103/PhysRevLett.123.196402} {\bibfield  {journal} {\bibinfo
  {journal} {Phys. Rev. Lett.}\ }\textbf {\bibinfo {volume} {123}},\ \bibinfo
  {pages} {196402} (\bibinfo {year} {2019})}\BibitemShut {NoStop}%
\bibitem [{\citenamefont {Fidkowski}\ and\ \citenamefont
  {Kitaev}(2010)}]{Z_to_Zn_Fidkowski_10}%
  \BibitemOpen
  \bibfield  {author} {\bibinfo {author} {\bibfnamefont {L.}~\bibnamefont
  {Fidkowski}}\ and\ \bibinfo {author} {\bibfnamefont {A.}~\bibnamefont
  {Kitaev}},\ }\href {\doibase 10.1103/PhysRevB.81.134509} {\bibfield
  {journal} {\bibinfo  {journal} {Phys. Rev. B}\ }\textbf {\bibinfo {volume}
  {81}},\ \bibinfo {pages} {134509} (\bibinfo {year} {2010})}\BibitemShut
  {NoStop}%
\bibitem [{\citenamefont {Fidkowski}\ and\ \citenamefont
  {Kitaev}(2011)}]{Fidkowski_1Dclassificatin_11}%
  \BibitemOpen
  \bibfield  {author} {\bibinfo {author} {\bibfnamefont {L.}~\bibnamefont
  {Fidkowski}}\ and\ \bibinfo {author} {\bibfnamefont {A.}~\bibnamefont
  {Kitaev}},\ }\href {\doibase 10.1103/PhysRevB.83.075103} {\bibfield
  {journal} {\bibinfo  {journal} {Phys. Rev. B}\ }\textbf {\bibinfo {volume}
  {83}},\ \bibinfo {pages} {075103} (\bibinfo {year} {2011})}\BibitemShut
  {NoStop}%
\bibitem [{\citenamefont {Turner}\ \emph {et~al.}(2011)\citenamefont {Turner},
  \citenamefont {Pollmann},\ and\ \citenamefont {Berg}}]{Turner11}%
  \BibitemOpen
  \bibfield  {author} {\bibinfo {author} {\bibfnamefont {A.~M.}\ \bibnamefont
  {Turner}}, \bibinfo {author} {\bibfnamefont {F.}~\bibnamefont {Pollmann}}, \
  and\ \bibinfo {author} {\bibfnamefont {E.}~\bibnamefont {Berg}},\ }\href
  {\doibase 10.1103/PhysRevB.83.075102} {\bibfield  {journal} {\bibinfo
  {journal} {Phys. Rev. B}\ }\textbf {\bibinfo {volume} {83}},\ \bibinfo
  {pages} {075102} (\bibinfo {year} {2011})}\BibitemShut {NoStop}%
\bibitem [{\citenamefont {Ryu}\ and\ \citenamefont
  {Zhang}(2012)}]{Ryu_Z_to_Z8_2013}%
  \BibitemOpen
  \bibfield  {author} {\bibinfo {author} {\bibfnamefont {S.}~\bibnamefont
  {Ryu}}\ and\ \bibinfo {author} {\bibfnamefont {S.-C.}\ \bibnamefont
  {Zhang}},\ }\href {\doibase 10.1103/PhysRevB.85.245132} {\bibfield  {journal}
  {\bibinfo  {journal} {Phys. Rev. B}\ }\textbf {\bibinfo {volume} {85}},\
  \bibinfo {pages} {245132} (\bibinfo {year} {2012})}\BibitemShut {NoStop}%
\bibitem [{\citenamefont {Yao}\ and\ \citenamefont
  {Ryu}(2013)}]{YaoRyu_Z_to_Z8_2013}%
  \BibitemOpen
  \bibfield  {author} {\bibinfo {author} {\bibfnamefont {H.}~\bibnamefont
  {Yao}}\ and\ \bibinfo {author} {\bibfnamefont {S.}~\bibnamefont {Ryu}},\
  }\href {\doibase 10.1103/PhysRevB.88.064507} {\bibfield  {journal} {\bibinfo
  {journal} {Phys. Rev. B}\ }\textbf {\bibinfo {volume} {88}},\ \bibinfo
  {pages} {064507} (\bibinfo {year} {2013})}\BibitemShut {NoStop}%
\bibitem [{\citenamefont {Qi}(2013)}]{Qi_Z_to_Z8_2013}%
  \BibitemOpen
  \bibfield  {author} {\bibinfo {author} {\bibfnamefont {X.-L.}\ \bibnamefont
  {Qi}},\ }\href {\doibase 10.1088/1367-2630/15/6/065002} {\bibfield  {journal}
  {\bibinfo  {journal} {New J. Phys.}\ }\textbf {\bibinfo {volume} {15}},\
  \bibinfo {pages} {065002} (\bibinfo {year} {2013})}\BibitemShut {NoStop}%
\bibitem [{\citenamefont {Lu}\ and\ \citenamefont
  {Vishwanath}(2012)}]{Lu_CS_2011}%
  \BibitemOpen
  \bibfield  {author} {\bibinfo {author} {\bibfnamefont {Y.-M.}\ \bibnamefont
  {Lu}}\ and\ \bibinfo {author} {\bibfnamefont {A.}~\bibnamefont
  {Vishwanath}},\ }\href {\doibase 10.1103/PhysRevB.86.125119} {\bibfield
  {journal} {\bibinfo  {journal} {Phys. Rev. B}\ }\textbf {\bibinfo {volume}
  {86}},\ \bibinfo {pages} {125119} (\bibinfo {year} {2012})}\BibitemShut
  {NoStop}%
\bibitem [{\citenamefont {Levin}\ and\ \citenamefont
  {Stern}(2012)}]{Levin_CS_2012}%
  \BibitemOpen
  \bibfield  {author} {\bibinfo {author} {\bibfnamefont {M.}~\bibnamefont
  {Levin}}\ and\ \bibinfo {author} {\bibfnamefont {A.}~\bibnamefont {Stern}},\
  }\href {\doibase 10.1103/PhysRevB.86.115131} {\bibfield  {journal} {\bibinfo
  {journal} {Phys. Rev. B}\ }\textbf {\bibinfo {volume} {86}},\ \bibinfo
  {pages} {115131} (\bibinfo {year} {2012})}\BibitemShut {NoStop}%
\bibitem [{\citenamefont {Hsieh}\ \emph {et~al.}(2014)\citenamefont {Hsieh},
  \citenamefont {Morimoto},\ and\ \citenamefont {Ryu}}]{Hsieh_CS_CPT_2014}%
  \BibitemOpen
  \bibfield  {author} {\bibinfo {author} {\bibfnamefont {C.-T.}\ \bibnamefont
  {Hsieh}}, \bibinfo {author} {\bibfnamefont {T.}~\bibnamefont {Morimoto}}, \
  and\ \bibinfo {author} {\bibfnamefont {S.}~\bibnamefont {Ryu}},\ }\href
  {\doibase 10.1103/PhysRevB.90.245111} {\bibfield  {journal} {\bibinfo
  {journal} {Phys. Rev. B}\ }\textbf {\bibinfo {volume} {90}},\ \bibinfo
  {pages} {245111} (\bibinfo {year} {2014})}\BibitemShut {NoStop}%
\bibitem [{\citenamefont {Wang}\ \emph {et~al.}(2014)\citenamefont {Wang},
  \citenamefont {Potter},\ and\ \citenamefont
  {Senthil}}]{Wang_Potter_Senthil2014}%
  \BibitemOpen
  \bibfield  {author} {\bibinfo {author} {\bibfnamefont {C.}~\bibnamefont
  {Wang}}, \bibinfo {author} {\bibfnamefont {A.~C.}\ \bibnamefont {Potter}}, \
  and\ \bibinfo {author} {\bibfnamefont {T.}~\bibnamefont {Senthil}},\ }\href
  {\doibase 10.1126/science.1243326} {\bibfield  {journal} {\bibinfo  {journal}
  {Science}\ }\textbf {\bibinfo {volume} {343}},\ \bibinfo {pages} {629}
  (\bibinfo {year} {2014})}\BibitemShut {NoStop}%
\bibitem [{\citenamefont {Isobe}\ and\ \citenamefont
  {Fu}(2015)}]{Isobe_Fu2015}%
  \BibitemOpen
  \bibfield  {author} {\bibinfo {author} {\bibfnamefont {H.}~\bibnamefont
  {Isobe}}\ and\ \bibinfo {author} {\bibfnamefont {L.}~\bibnamefont {Fu}},\
  }\href {\doibase 10.1103/PhysRevB.92.081304} {\bibfield  {journal} {\bibinfo
  {journal} {Phys. Rev. B}\ }\textbf {\bibinfo {volume} {92}},\ \bibinfo
  {pages} {081304} (\bibinfo {year} {2015})}\BibitemShut {NoStop}%
\bibitem [{\citenamefont {Yoshida}\ and\ \citenamefont
  {Furusaki}(2015)}]{Yoshida_TCI_bosonization_PRB15}%
  \BibitemOpen
  \bibfield  {author} {\bibinfo {author} {\bibfnamefont {T.}~\bibnamefont
  {Yoshida}}\ and\ \bibinfo {author} {\bibfnamefont {A.}~\bibnamefont
  {Furusaki}},\ }\href {\doibase 10.1103/PhysRevB.92.085114} {\bibfield
  {journal} {\bibinfo  {journal} {Phys. Rev. B}\ }\textbf {\bibinfo {volume}
  {92}},\ \bibinfo {pages} {085114} (\bibinfo {year} {2015})}\BibitemShut
  {NoStop}%
\bibitem [{\citenamefont {Morimoto}\ \emph {et~al.}(2015)\citenamefont
  {Morimoto}, \citenamefont {Furusaki},\ and\ \citenamefont
  {Mudry}}]{Morimoto_2015}%
  \BibitemOpen
  \bibfield  {author} {\bibinfo {author} {\bibfnamefont {T.}~\bibnamefont
  {Morimoto}}, \bibinfo {author} {\bibfnamefont {A.}~\bibnamefont {Furusaki}},
  \ and\ \bibinfo {author} {\bibfnamefont {C.}~\bibnamefont {Mudry}},\ }\href
  {\doibase 10.1103/PhysRevB.92.125104} {\bibfield  {journal} {\bibinfo
  {journal} {Phys. Rev. B}\ }\textbf {\bibinfo {volume} {92}},\ \bibinfo
  {pages} {125104} (\bibinfo {year} {2015})}\BibitemShut {NoStop}%
\bibitem [{\citenamefont {Yoshida}\ \emph {et~al.}(2017)\citenamefont
  {Yoshida}, \citenamefont {Daido}, \citenamefont {Yanase},\ and\ \citenamefont
  {Kawakami}}]{Yoshida_ZtoZ16_superlattice_PRL17}%
  \BibitemOpen
  \bibfield  {author} {\bibinfo {author} {\bibfnamefont {T.}~\bibnamefont
  {Yoshida}}, \bibinfo {author} {\bibfnamefont {A.}~\bibnamefont {Daido}},
  \bibinfo {author} {\bibfnamefont {Y.}~\bibnamefont {Yanase}}, \ and\ \bibinfo
  {author} {\bibfnamefont {N.}~\bibnamefont {Kawakami}},\ }\href {\doibase
  10.1103/PhysRevLett.118.147001} {\bibfield  {journal} {\bibinfo  {journal}
  {Phys. Rev. Lett.}\ }\textbf {\bibinfo {volume} {118}},\ \bibinfo {pages}
  {147001} (\bibinfo {year} {2017})}\BibitemShut {NoStop}%
\bibitem [{\citenamefont {Yoshida}\ \emph
  {et~al.}(2018{\natexlab{a}})\citenamefont {Yoshida}, \citenamefont
  {Danshita}, \citenamefont {Peters},\ and\ \citenamefont
  {Kawakami}}]{Yoshida_ZtoZ4_coldatom_PRL18}%
  \BibitemOpen
  \bibfield  {author} {\bibinfo {author} {\bibfnamefont {T.}~\bibnamefont
  {Yoshida}}, \bibinfo {author} {\bibfnamefont {I.}~\bibnamefont {Danshita}},
  \bibinfo {author} {\bibfnamefont {R.}~\bibnamefont {Peters}}, \ and\ \bibinfo
  {author} {\bibfnamefont {N.}~\bibnamefont {Kawakami}},\ }\href {\doibase
  10.1103/PhysRevLett.121.025301} {\bibfield  {journal} {\bibinfo  {journal}
  {Phys. Rev. Lett.}\ }\textbf {\bibinfo {volume} {121}},\ \bibinfo {pages}
  {025301} (\bibinfo {year} {2018}{\natexlab{a}})}\BibitemShut {NoStop}%
\bibitem [{\citenamefont {Kozii}\ and\ \citenamefont
  {Fu}(2017)}]{VKozii_nH_arXiv17}%
  \BibitemOpen
  \bibfield  {author} {\bibinfo {author} {\bibfnamefont {V.}~\bibnamefont
  {Kozii}}\ and\ \bibinfo {author} {\bibfnamefont {L.}~\bibnamefont {Fu}},\
  }\href@noop {} {\bibfield  {journal} {\bibinfo  {journal} {arXiv preprint
  arXiv:1708.05841}\ } (\bibinfo {year} {2017})}\BibitemShut {NoStop}%
\bibitem [{\citenamefont {Zyuzin}\ and\ \citenamefont
  {Zyuzin}(2018)}]{Zyuzin_nHEP_PRB18}%
  \BibitemOpen
  \bibfield  {author} {\bibinfo {author} {\bibfnamefont {A.~A.}\ \bibnamefont
  {Zyuzin}}\ and\ \bibinfo {author} {\bibfnamefont {A.~Y.}\ \bibnamefont
  {Zyuzin}},\ }\href {\doibase 10.1103/PhysRevB.97.041203} {\bibfield
  {journal} {\bibinfo  {journal} {Phys. Rev. B}\ }\textbf {\bibinfo {volume}
  {97}},\ \bibinfo {pages} {041203} (\bibinfo {year} {2018})}\BibitemShut
  {NoStop}%
\bibitem [{\citenamefont {Yoshida}\ \emph
  {et~al.}(2018{\natexlab{b}})\citenamefont {Yoshida}, \citenamefont {Peters},\
  and\ \citenamefont {Kawakami}}]{Yoshida_EP_DMFT_PRB18}%
  \BibitemOpen
  \bibfield  {author} {\bibinfo {author} {\bibfnamefont {T.}~\bibnamefont
  {Yoshida}}, \bibinfo {author} {\bibfnamefont {R.}~\bibnamefont {Peters}}, \
  and\ \bibinfo {author} {\bibfnamefont {N.}~\bibnamefont {Kawakami}},\ }\href
  {\doibase 10.1103/PhysRevB.98.035141} {\bibfield  {journal} {\bibinfo
  {journal} {Phys. Rev. B}\ }\textbf {\bibinfo {volume} {98}},\ \bibinfo
  {pages} {035141} (\bibinfo {year} {2018}{\natexlab{b}})}\BibitemShut
  {NoStop}%
\bibitem [{\citenamefont {Yoshida}\ \emph
  {et~al.}(2019{\natexlab{a}})\citenamefont {Yoshida}, \citenamefont {Peters},
  \citenamefont {Kawakami},\ and\ \citenamefont
  {Hatsugai}}]{Yoshida_SPERs_PRB19}%
  \BibitemOpen
  \bibfield  {author} {\bibinfo {author} {\bibfnamefont {T.}~\bibnamefont
  {Yoshida}}, \bibinfo {author} {\bibfnamefont {R.}~\bibnamefont {Peters}},
  \bibinfo {author} {\bibfnamefont {N.}~\bibnamefont {Kawakami}}, \ and\
  \bibinfo {author} {\bibfnamefont {Y.}~\bibnamefont {Hatsugai}},\ }\href
  {\doibase 10.1103/PhysRevB.99.121101} {\bibfield  {journal} {\bibinfo
  {journal} {Phys. Rev. B}\ }\textbf {\bibinfo {volume} {99}},\ \bibinfo
  {pages} {121101} (\bibinfo {year} {2019}{\natexlab{a}})}\BibitemShut
  {NoStop}%
\bibitem [{\citenamefont {Papaj}\ \emph {et~al.}(2019)\citenamefont {Papaj},
  \citenamefont {Isobe},\ and\ \citenamefont {Fu}}]{Papaji_nHEP_PRB19}%
  \BibitemOpen
  \bibfield  {author} {\bibinfo {author} {\bibfnamefont {M.}~\bibnamefont
  {Papaj}}, \bibinfo {author} {\bibfnamefont {H.}~\bibnamefont {Isobe}}, \ and\
  \bibinfo {author} {\bibfnamefont {L.}~\bibnamefont {Fu}},\ }\href {\doibase
  10.1103/PhysRevB.99.201107} {\bibfield  {journal} {\bibinfo  {journal} {Phys.
  Rev. B}\ }\textbf {\bibinfo {volume} {99}},\ \bibinfo {pages} {201107}
  (\bibinfo {year} {2019})}\BibitemShut {NoStop}%
\bibitem [{\citenamefont {Kimura}\ \emph {et~al.}(2019)\citenamefont {Kimura},
  \citenamefont {Yoshida},\ and\ \citenamefont
  {Kawakami}}]{Kimura_SPERs_PRB19}%
  \BibitemOpen
  \bibfield  {author} {\bibinfo {author} {\bibfnamefont {K.}~\bibnamefont
  {Kimura}}, \bibinfo {author} {\bibfnamefont {T.}~\bibnamefont {Yoshida}}, \
  and\ \bibinfo {author} {\bibfnamefont {N.}~\bibnamefont {Kawakami}},\ }\href
  {\doibase 10.1103/PhysRevB.100.115124} {\bibfield  {journal} {\bibinfo
  {journal} {Phys. Rev. B}\ }\textbf {\bibinfo {volume} {100}},\ \bibinfo
  {pages} {115124} (\bibinfo {year} {2019})}\BibitemShut {NoStop}%
\bibitem [{\citenamefont {Michishita}\ \emph {et~al.}(2019)\citenamefont
  {Michishita}, \citenamefont {Yoshida},\ and\ \citenamefont
  {Peters}}]{Michishta_EP_DMFT_arXiv19}%
  \BibitemOpen
  \bibfield  {author} {\bibinfo {author} {\bibfnamefont {Y.}~\bibnamefont
  {Michishita}}, \bibinfo {author} {\bibfnamefont {T.}~\bibnamefont {Yoshida}},
  \ and\ \bibinfo {author} {\bibfnamefont {R.}~\bibnamefont {Peters}},\
  }\href@noop {} {\bibfield  {journal} {\bibinfo  {journal} {arXiv preprint
  arXiv:1905.12287}\ } (\bibinfo {year} {2019})}\BibitemShut {NoStop}%
\bibitem [{\citenamefont {Matsushita}\ \emph {et~al.}(2019)\citenamefont
  {Matsushita}, \citenamefont {Nagai},\ and\ \citenamefont
  {Fujimoto}}]{Matsushita_ER_PRB19}%
  \BibitemOpen
  \bibfield  {author} {\bibinfo {author} {\bibfnamefont {T.}~\bibnamefont
  {Matsushita}}, \bibinfo {author} {\bibfnamefont {Y.}~\bibnamefont {Nagai}}, \
  and\ \bibinfo {author} {\bibfnamefont {S.}~\bibnamefont {Fujimoto}},\ }\href
  {\doibase 10.1103/PhysRevB.100.245205} {\bibfield  {journal} {\bibinfo
  {journal} {Phys. Rev. B}\ }\textbf {\bibinfo {volume} {100}},\ \bibinfo
  {pages} {245205} (\bibinfo {year} {2019})}\BibitemShut {NoStop}%
\bibitem [{\citenamefont {Hatano}\ and\ \citenamefont
  {Nelson}(1996)}]{Hatano_nH_PRL96}%
  \BibitemOpen
  \bibfield  {author} {\bibinfo {author} {\bibfnamefont {N.}~\bibnamefont
  {Hatano}}\ and\ \bibinfo {author} {\bibfnamefont {D.~R.}\ \bibnamefont
  {Nelson}},\ }\href {\doibase 10.1103/PhysRevLett.77.570} {\bibfield
  {journal} {\bibinfo  {journal} {Phys. Rev. Lett.}\ }\textbf {\bibinfo
  {volume} {77}},\ \bibinfo {pages} {570} (\bibinfo {year} {1996})}\BibitemShut
  {NoStop}%
\bibitem [{\citenamefont {Hatano}\ and\ \citenamefont
  {Nelson}(1998)}]{Hatano_nH_PRB98}%
  \BibitemOpen
  \bibfield  {author} {\bibinfo {author} {\bibfnamefont {N.}~\bibnamefont
  {Hatano}}\ and\ \bibinfo {author} {\bibfnamefont {D.~R.}\ \bibnamefont
  {Nelson}},\ }\href {\doibase 10.1103/PhysRevB.58.8384} {\bibfield  {journal}
  {\bibinfo  {journal} {Phys. Rev. B}\ }\textbf {\bibinfo {volume} {58}},\
  \bibinfo {pages} {8384} (\bibinfo {year} {1998})}\BibitemShut {NoStop}%
\bibitem [{\citenamefont {Bender}\ and\ \citenamefont
  {Boettcher}(1998)}]{CMBender_nH_PRL98}%
  \BibitemOpen
  \bibfield  {author} {\bibinfo {author} {\bibfnamefont {C.~M.}\ \bibnamefont
  {Bender}}\ and\ \bibinfo {author} {\bibfnamefont {S.}~\bibnamefont
  {Boettcher}},\ }\href {\doibase 10.1103/PhysRevLett.80.5243} {\bibfield
  {journal} {\bibinfo  {journal} {Phys. Rev. Lett.}\ }\textbf {\bibinfo
  {volume} {80}},\ \bibinfo {pages} {5243} (\bibinfo {year}
  {1998})}\BibitemShut {NoStop}%
\bibitem [{\citenamefont {Bender}\ \emph {et~al.}(1999)\citenamefont {Bender},
  \citenamefont {Boettcher},\ and\ \citenamefont
  {Meisinger}}]{Bender_nH_JMP99}%
  \BibitemOpen
  \bibfield  {author} {\bibinfo {author} {\bibfnamefont {C.~M.}\ \bibnamefont
  {Bender}}, \bibinfo {author} {\bibfnamefont {S.}~\bibnamefont {Boettcher}}, \
  and\ \bibinfo {author} {\bibfnamefont {P.~N.}\ \bibnamefont {Meisinger}},\
  }\href {\doibase 10.1063/1.532860} {\bibfield  {journal} {\bibinfo  {journal}
  {Journal of Mathematical Physics}\ }\textbf {\bibinfo {volume} {40}},\
  \bibinfo {pages} {2201} (\bibinfo {year} {1999})}\BibitemShut {NoStop}%
\bibitem [{\citenamefont {Esaki}\ \emph {et~al.}(2011)\citenamefont {Esaki},
  \citenamefont {Sato}, \citenamefont {Hasebe},\ and\ \citenamefont
  {Kohmoto}}]{Esaki_nHTI_PRB11}%
  \BibitemOpen
  \bibfield  {author} {\bibinfo {author} {\bibfnamefont {K.}~\bibnamefont
  {Esaki}}, \bibinfo {author} {\bibfnamefont {M.}~\bibnamefont {Sato}},
  \bibinfo {author} {\bibfnamefont {K.}~\bibnamefont {Hasebe}}, \ and\ \bibinfo
  {author} {\bibfnamefont {M.}~\bibnamefont {Kohmoto}},\ }\href {\doibase
  10.1103/PhysRevB.84.205128} {\bibfield  {journal} {\bibinfo  {journal} {Phys.
  Rev. B}\ }\textbf {\bibinfo {volume} {84}},\ \bibinfo {pages} {205128}
  (\bibinfo {year} {2011})}\BibitemShut {NoStop}%
\bibitem [{\citenamefont {Sato}\ \emph {et~al.}(2012)\citenamefont {Sato},
  \citenamefont {Hasebe}, \citenamefont {Esaki},\ and\ \citenamefont
  {Kohmoto}}]{Sato_nHTI_PTEP12}%
  \BibitemOpen
  \bibfield  {author} {\bibinfo {author} {\bibfnamefont {M.}~\bibnamefont
  {Sato}}, \bibinfo {author} {\bibfnamefont {K.}~\bibnamefont {Hasebe}},
  \bibinfo {author} {\bibfnamefont {K.}~\bibnamefont {Esaki}}, \ and\ \bibinfo
  {author} {\bibfnamefont {M.}~\bibnamefont {Kohmoto}},\ }\href {\doibase
  10.1143/PTP.127.937} {\bibfield  {journal} {\bibinfo  {journal} {Progress of
  Theoretical Physics}\ }\textbf {\bibinfo {volume} {127}},\ \bibinfo {pages}
  {937} (\bibinfo {year} {2012})}\BibitemShut {NoStop}%
\bibitem [{\citenamefont {Fukui}\ and\ \citenamefont
  {Kawakami}(1998)}]{Fukui_nHcorr_PRB98}%
  \BibitemOpen
  \bibfield  {author} {\bibinfo {author} {\bibfnamefont {T.}~\bibnamefont
  {Fukui}}\ and\ \bibinfo {author} {\bibfnamefont {N.}~\bibnamefont
  {Kawakami}},\ }\href {\doibase 10.1103/PhysRevB.58.16051} {\bibfield
  {journal} {\bibinfo  {journal} {Phys. Rev. B}\ }\textbf {\bibinfo {volume}
  {58}},\ \bibinfo {pages} {16051} (\bibinfo {year} {1998})}\BibitemShut
  {NoStop}%
\bibitem [{\citenamefont {Lee}(2016)}]{TELeePRL16_Half_quantized}%
  \BibitemOpen
  \bibfield  {author} {\bibinfo {author} {\bibfnamefont {T.~E.}\ \bibnamefont
  {Lee}},\ }\href {\doibase 10.1103/PhysRevLett.116.133903} {\bibfield
  {journal} {\bibinfo  {journal} {Phys. Rev. Lett.}\ }\textbf {\bibinfo
  {volume} {116}},\ \bibinfo {pages} {133903} (\bibinfo {year}
  {2016})}\BibitemShut {NoStop}%
\bibitem [{\citenamefont {Kat{\=o}}(1966)}]{TKato_EP_book1966}%
  \BibitemOpen
  \bibfield  {author} {\bibinfo {author} {\bibfnamefont {T.}~\bibnamefont
  {Kat{\=o}}},\ }\href@noop {} {\emph {\bibinfo {title} {Perturbation theory
  for linear operators}}},\ Vol.\ \bibinfo {volume} {132}\ (\bibinfo
  {publisher} {Springer},\ \bibinfo {year} {1966})\BibitemShut {NoStop}%
\bibitem [{\citenamefont {Shen}\ \emph {et~al.}(2017)\citenamefont {Shen},
  \citenamefont {Zhen},\ and\ \citenamefont {Fu}}]{HShen2017_non-Hermi}%
  \BibitemOpen
  \bibfield  {author} {\bibinfo {author} {\bibfnamefont {H.}~\bibnamefont
  {Shen}}, \bibinfo {author} {\bibfnamefont {B.}~\bibnamefont {Zhen}}, \ and\
  \bibinfo {author} {\bibfnamefont {L.}~\bibnamefont {Fu}},\ }\href@noop {}
  {\bibfield  {journal} {\bibinfo  {journal} {arXiv preprint arXiv:1706.07435}\
  } (\bibinfo {year} {2017})}\BibitemShut {NoStop}%
\bibitem [{\citenamefont {Xu}\ \emph {et~al.}(2017)\citenamefont {Xu},
  \citenamefont {Wang},\ and\ \citenamefont
  {Duan}}]{YXuPRL17_exceptional_ring}%
  \BibitemOpen
  \bibfield  {author} {\bibinfo {author} {\bibfnamefont {Y.}~\bibnamefont
  {Xu}}, \bibinfo {author} {\bibfnamefont {S.-T.}\ \bibnamefont {Wang}}, \ and\
  \bibinfo {author} {\bibfnamefont {L.-M.}\ \bibnamefont {Duan}},\ }\href
  {\doibase 10.1103/PhysRevLett.118.045701} {\bibfield  {journal} {\bibinfo
  {journal} {Phys. Rev. Lett.}\ }\textbf {\bibinfo {volume} {118}},\ \bibinfo
  {pages} {045701} (\bibinfo {year} {2017})}\BibitemShut {NoStop}%
\bibitem [{\citenamefont {Carlstr\"om}\ \emph {et~al.}(2019)\citenamefont
  {Carlstr\"om}, \citenamefont {St\aa{}lhammar}, \citenamefont {Budich},\ and\
  \citenamefont {Bergholtz}}]{Carlstrom_nHKER_PRB19}%
  \BibitemOpen
  \bibfield  {author} {\bibinfo {author} {\bibfnamefont {J.}~\bibnamefont
  {Carlstr\"om}}, \bibinfo {author} {\bibfnamefont {M.}~\bibnamefont
  {St\aa{}lhammar}}, \bibinfo {author} {\bibfnamefont {J.~C.}\ \bibnamefont
  {Budich}}, \ and\ \bibinfo {author} {\bibfnamefont {E.~J.}\ \bibnamefont
  {Bergholtz}},\ }\href {\doibase 10.1103/PhysRevB.99.161115} {\bibfield
  {journal} {\bibinfo  {journal} {Phys. Rev. B}\ }\textbf {\bibinfo {volume}
  {99}},\ \bibinfo {pages} {161115} (\bibinfo {year} {2019})}\BibitemShut
  {NoStop}%
\bibitem [{\citenamefont {Yao}\ and\ \citenamefont
  {Wang}(2018)}]{SYao_nHSkin-1D_PRL18}%
  \BibitemOpen
  \bibfield  {author} {\bibinfo {author} {\bibfnamefont {S.}~\bibnamefont
  {Yao}}\ and\ \bibinfo {author} {\bibfnamefont {Z.}~\bibnamefont {Wang}},\
  }\href {\doibase 10.1103/PhysRevLett.121.086803} {\bibfield  {journal}
  {\bibinfo  {journal} {Phys. Rev. Lett.}\ }\textbf {\bibinfo {volume} {121}},\
  \bibinfo {pages} {086803} (\bibinfo {year} {2018})}\BibitemShut {NoStop}%
\bibitem [{\citenamefont {Kunst}\ \emph {et~al.}(2018)\citenamefont {Kunst},
  \citenamefont {Edvardsson}, \citenamefont {Budich},\ and\ \citenamefont
  {Bergholtz}}]{KFlore_nHSkin_PRL18}%
  \BibitemOpen
  \bibfield  {author} {\bibinfo {author} {\bibfnamefont {F.~K.}\ \bibnamefont
  {Kunst}}, \bibinfo {author} {\bibfnamefont {E.}~\bibnamefont {Edvardsson}},
  \bibinfo {author} {\bibfnamefont {J.~C.}\ \bibnamefont {Budich}}, \ and\
  \bibinfo {author} {\bibfnamefont {E.~J.}\ \bibnamefont {Bergholtz}},\ }\href
  {\doibase 10.1103/PhysRevLett.121.026808} {\bibfield  {journal} {\bibinfo
  {journal} {Phys. Rev. Lett.}\ }\textbf {\bibinfo {volume} {121}},\ \bibinfo
  {pages} {026808} (\bibinfo {year} {2018})}\BibitemShut {NoStop}%
\bibitem [{\citenamefont {Edvardsson}\ \emph {et~al.}(2019)\citenamefont
  {Edvardsson}, \citenamefont {Kunst},\ and\ \citenamefont
  {Bergholtz}}]{EElizabet_PRBnHSkinHOTI_PRB19}%
  \BibitemOpen
  \bibfield  {author} {\bibinfo {author} {\bibfnamefont {E.}~\bibnamefont
  {Edvardsson}}, \bibinfo {author} {\bibfnamefont {F.~K.}\ \bibnamefont
  {Kunst}}, \ and\ \bibinfo {author} {\bibfnamefont {E.~J.}\ \bibnamefont
  {Bergholtz}},\ }\href {\doibase 10.1103/PhysRevB.99.081302} {\bibfield
  {journal} {\bibinfo  {journal} {Phys. Rev. B}\ }\textbf {\bibinfo {volume}
  {99}},\ \bibinfo {pages} {081302} (\bibinfo {year} {2019})}\BibitemShut
  {NoStop}%
\bibitem [{\citenamefont {Yokomizo}\ and\ \citenamefont
  {Murakami}(2019)}]{Yokomizo_BBC_PRL19}%
  \BibitemOpen
  \bibfield  {author} {\bibinfo {author} {\bibfnamefont {K.}~\bibnamefont
  {Yokomizo}}\ and\ \bibinfo {author} {\bibfnamefont {S.}~\bibnamefont
  {Murakami}},\ }\href {\doibase 10.1103/PhysRevLett.123.066404} {\bibfield
  {journal} {\bibinfo  {journal} {Phys. Rev. Lett.}\ }\textbf {\bibinfo
  {volume} {123}},\ \bibinfo {pages} {066404} (\bibinfo {year}
  {2019})}\BibitemShut {NoStop}%
\bibitem [{\citenamefont {Okuma}\ and\ \citenamefont
  {Sato}(2019)}]{Okuma_nHBBCpg_PRL19}%
  \BibitemOpen
  \bibfield  {author} {\bibinfo {author} {\bibfnamefont {N.}~\bibnamefont
  {Okuma}}\ and\ \bibinfo {author} {\bibfnamefont {M.}~\bibnamefont {Sato}},\
  }\href {\doibase 10.1103/PhysRevLett.123.097701} {\bibfield  {journal}
  {\bibinfo  {journal} {Phys. Rev. Lett.}\ }\textbf {\bibinfo {volume} {123}},\
  \bibinfo {pages} {097701} (\bibinfo {year} {2019})}\BibitemShut {NoStop}%
\bibitem [{\citenamefont {Xiao}\ \emph {et~al.}(2019)\citenamefont {Xiao},
  \citenamefont {Deng}, \citenamefont {Wang}, \citenamefont {Zhu},
  \citenamefont {Wang}, \citenamefont {Yi},\ and\ \citenamefont
  {Xue}}]{Xiao_nHSkin_Exp_arXiv19}%
  \BibitemOpen
  \bibfield  {author} {\bibinfo {author} {\bibfnamefont {L.}~\bibnamefont
  {Xiao}}, \bibinfo {author} {\bibfnamefont {T.}~\bibnamefont {Deng}}, \bibinfo
  {author} {\bibfnamefont {K.}~\bibnamefont {Wang}}, \bibinfo {author}
  {\bibfnamefont {G.}~\bibnamefont {Zhu}}, \bibinfo {author} {\bibfnamefont
  {Z.}~\bibnamefont {Wang}}, \bibinfo {author} {\bibfnamefont {W.}~\bibnamefont
  {Yi}}, \ and\ \bibinfo {author} {\bibfnamefont {P.}~\bibnamefont {Xue}},\
  }\href@noop {} {\bibfield  {journal} {\bibinfo  {journal} {arXiv preprint
  arXiv:1907.12566}\ } (\bibinfo {year} {2019})}\BibitemShut {NoStop}%
\bibitem [{\citenamefont {Martinez~Alvarez}\ \emph {et~al.}(2018)\citenamefont
  {Martinez~Alvarez}, \citenamefont {Barrios~Vargas},\ and\ \citenamefont
  {Foa~Torres}}]{Alvarez_nHSkin_PRB18}%
  \BibitemOpen
  \bibfield  {author} {\bibinfo {author} {\bibfnamefont {V.~M.}\ \bibnamefont
  {Martinez~Alvarez}}, \bibinfo {author} {\bibfnamefont {J.~E.}\ \bibnamefont
  {Barrios~Vargas}}, \ and\ \bibinfo {author} {\bibfnamefont {L.~E.~F.}\
  \bibnamefont {Foa~Torres}},\ }\href {\doibase 10.1103/PhysRevB.97.121401}
  {\bibfield  {journal} {\bibinfo  {journal} {Phys. Rev. B}\ }\textbf {\bibinfo
  {volume} {97}},\ \bibinfo {pages} {121401} (\bibinfo {year}
  {2018})}\BibitemShut {NoStop}%
\bibitem [{\citenamefont {Lee}\ and\ \citenamefont
  {Thomale}(2019)}]{Lee_Skin19}%
  \BibitemOpen
  \bibfield  {author} {\bibinfo {author} {\bibfnamefont {C.~H.}\ \bibnamefont
  {Lee}}\ and\ \bibinfo {author} {\bibfnamefont {R.}~\bibnamefont {Thomale}},\
  }\href {\doibase 10.1103/PhysRevB.99.201103} {\bibfield  {journal} {\bibinfo
  {journal} {Phys. Rev. B}\ }\textbf {\bibinfo {volume} {99}},\ \bibinfo
  {pages} {201103} (\bibinfo {year} {2019})}\BibitemShut {NoStop}%
\bibitem [{\citenamefont {Zhang}\ \emph {et~al.}(2019)\citenamefont {Zhang},
  \citenamefont {Yang},\ and\ \citenamefont {Fang}}]{Zhang_BECskin19}%
  \BibitemOpen
  \bibfield  {author} {\bibinfo {author} {\bibfnamefont {K.}~\bibnamefont
  {Zhang}}, \bibinfo {author} {\bibfnamefont {Z.}~\bibnamefont {Yang}}, \ and\
  \bibinfo {author} {\bibfnamefont {C.}~\bibnamefont {Fang}},\ }\href@noop {}
  {\bibfield  {journal} {\bibinfo  {journal} {arXiv preprint arXiv:1910.01131}\
  } (\bibinfo {year} {2019})}\BibitemShut {NoStop}%
\bibitem [{\citenamefont {Okuma}\ \emph {et~al.}(2019)\citenamefont {Okuma},
  \citenamefont {Kawabata}, \citenamefont {Shiozaki},\ and\ \citenamefont
  {Sato}}]{Okuma_BECskin19}%
  \BibitemOpen
  \bibfield  {author} {\bibinfo {author} {\bibfnamefont {N.}~\bibnamefont
  {Okuma}}, \bibinfo {author} {\bibfnamefont {K.}~\bibnamefont {Kawabata}},
  \bibinfo {author} {\bibfnamefont {K.}~\bibnamefont {Shiozaki}}, \ and\
  \bibinfo {author} {\bibfnamefont {M.}~\bibnamefont {Sato}},\ }\href@noop {}
  {\bibfield  {journal} {\bibinfo  {journal} {arXiv preprint arXiv:1910.02878}\
  } (\bibinfo {year} {2019})}\BibitemShut {NoStop}%
\bibitem [{\citenamefont {Gong}\ \emph {et~al.}(2018)\citenamefont {Gong},
  \citenamefont {Ashida}, \citenamefont {Kawabata}, \citenamefont {Takasan},
  \citenamefont {Higashikawa},\ and\ \citenamefont {Ueda}}]{Gong_class_PRX18}%
  \BibitemOpen
  \bibfield  {author} {\bibinfo {author} {\bibfnamefont {Z.}~\bibnamefont
  {Gong}}, \bibinfo {author} {\bibfnamefont {Y.}~\bibnamefont {Ashida}},
  \bibinfo {author} {\bibfnamefont {K.}~\bibnamefont {Kawabata}}, \bibinfo
  {author} {\bibfnamefont {K.}~\bibnamefont {Takasan}}, \bibinfo {author}
  {\bibfnamefont {S.}~\bibnamefont {Higashikawa}}, \ and\ \bibinfo {author}
  {\bibfnamefont {M.}~\bibnamefont {Ueda}},\ }\href {\doibase
  10.1103/PhysRevX.8.031079} {\bibfield  {journal} {\bibinfo  {journal} {Phys.
  Rev. X}\ }\textbf {\bibinfo {volume} {8}},\ \bibinfo {pages} {031079}
  (\bibinfo {year} {2018})}\BibitemShut {NoStop}%
\bibitem [{\citenamefont {Kawabata}\ \emph
  {et~al.}(2019{\natexlab{a}})\citenamefont {Kawabata}, \citenamefont
  {Higashikawa}, \citenamefont {Gong}, \citenamefont {Ashida},\ and\
  \citenamefont {Ueda}}]{Kawabata_Class_NatComm2019}%
  \BibitemOpen
  \bibfield  {author} {\bibinfo {author} {\bibfnamefont {K.}~\bibnamefont
  {Kawabata}}, \bibinfo {author} {\bibfnamefont {S.}~\bibnamefont
  {Higashikawa}}, \bibinfo {author} {\bibfnamefont {Z.}~\bibnamefont {Gong}},
  \bibinfo {author} {\bibfnamefont {Y.}~\bibnamefont {Ashida}}, \ and\ \bibinfo
  {author} {\bibfnamefont {M.}~\bibnamefont {Ueda}},\ }\href {\doibase
  10.1038/s41467-018-08254-y} {\bibfield  {journal} {\bibinfo  {journal}
  {Nature Communications}\ }\textbf {\bibinfo {volume} {10}},\ \bibinfo {pages}
  {297} (\bibinfo {year} {2019}{\natexlab{a}})}\BibitemShut {NoStop}%
\bibitem [{\citenamefont {Kawabata}\ \emph
  {et~al.}(2019{\natexlab{b}})\citenamefont {Kawabata}, \citenamefont
  {Shiozaki}, \citenamefont {Ueda},\ and\ \citenamefont
  {Sato}}]{Kawabata_gapped_PRX19}%
  \BibitemOpen
  \bibfield  {author} {\bibinfo {author} {\bibfnamefont {K.}~\bibnamefont
  {Kawabata}}, \bibinfo {author} {\bibfnamefont {K.}~\bibnamefont {Shiozaki}},
  \bibinfo {author} {\bibfnamefont {M.}~\bibnamefont {Ueda}}, \ and\ \bibinfo
  {author} {\bibfnamefont {M.}~\bibnamefont {Sato}},\ }\href {\doibase
  10.1103/PhysRevX.9.041015} {\bibfield  {journal} {\bibinfo  {journal} {Phys.
  Rev. X}\ }\textbf {\bibinfo {volume} {9}},\ \bibinfo {pages} {041015}
  (\bibinfo {year} {2019}{\natexlab{b}})}\BibitemShut {NoStop}%
\bibitem [{\citenamefont {Zhou}\ and\ \citenamefont
  {Lee}(2019)}]{Zhou_gapped_class_PRB19}%
  \BibitemOpen
  \bibfield  {author} {\bibinfo {author} {\bibfnamefont {H.}~\bibnamefont
  {Zhou}}\ and\ \bibinfo {author} {\bibfnamefont {J.~Y.}\ \bibnamefont {Lee}},\
  }\href {\doibase 10.1103/PhysRevB.99.235112} {\bibfield  {journal} {\bibinfo
  {journal} {Phys. Rev. B}\ }\textbf {\bibinfo {volume} {99}},\ \bibinfo
  {pages} {235112} (\bibinfo {year} {2019})}\BibitemShut {NoStop}%
\bibitem [{\citenamefont {Budich}\ \emph {et~al.}(2019)\citenamefont {Budich},
  \citenamefont {Carlstr\"om}, \citenamefont {Kunst},\ and\ \citenamefont
  {Bergholtz}}]{Budich_SPERs_PRB19}%
  \BibitemOpen
  \bibfield  {author} {\bibinfo {author} {\bibfnamefont {J.~C.}\ \bibnamefont
  {Budich}}, \bibinfo {author} {\bibfnamefont {J.}~\bibnamefont {Carlstr\"om}},
  \bibinfo {author} {\bibfnamefont {F.~K.}\ \bibnamefont {Kunst}}, \ and\
  \bibinfo {author} {\bibfnamefont {E.~J.}\ \bibnamefont {Bergholtz}},\ }\href
  {\doibase 10.1103/PhysRevB.99.041406} {\bibfield  {journal} {\bibinfo
  {journal} {Phys. Rev. B}\ }\textbf {\bibinfo {volume} {99}},\ \bibinfo
  {pages} {041406} (\bibinfo {year} {2019})}\BibitemShut {NoStop}%
\bibitem [{\citenamefont {Okugawa}\ and\ \citenamefont
  {Yokoyama}(2019)}]{Okugawa_SPERs_PRB19}%
  \BibitemOpen
  \bibfield  {author} {\bibinfo {author} {\bibfnamefont {R.}~\bibnamefont
  {Okugawa}}\ and\ \bibinfo {author} {\bibfnamefont {T.}~\bibnamefont
  {Yokoyama}},\ }\href {\doibase 10.1103/PhysRevB.99.041202} {\bibfield
  {journal} {\bibinfo  {journal} {Phys. Rev. B}\ }\textbf {\bibinfo {volume}
  {99}},\ \bibinfo {pages} {041202} (\bibinfo {year} {2019})}\BibitemShut
  {NoStop}%
\bibitem [{\citenamefont {Zhou}\ \emph {et~al.}(2019)\citenamefont {Zhou},
  \citenamefont {Lee}, \citenamefont {Liu},\ and\ \citenamefont
  {Zhen}}]{Zhou_SPERs_Optica19}%
  \BibitemOpen
  \bibfield  {author} {\bibinfo {author} {\bibfnamefont {H.}~\bibnamefont
  {Zhou}}, \bibinfo {author} {\bibfnamefont {J.~Y.}\ \bibnamefont {Lee}},
  \bibinfo {author} {\bibfnamefont {S.}~\bibnamefont {Liu}}, \ and\ \bibinfo
  {author} {\bibfnamefont {B.}~\bibnamefont {Zhen}},\ }\href {\doibase
  10.1364/OPTICA.6.000190} {\bibfield  {journal} {\bibinfo  {journal} {Optica}\
  }\textbf {\bibinfo {volume} {6}},\ \bibinfo {pages} {190} (\bibinfo {year}
  {2019})}\BibitemShut {NoStop}%
\bibitem [{\citenamefont {Kawabata}\ \emph
  {et~al.}(2019{\natexlab{c}})\citenamefont {Kawabata}, \citenamefont
  {Bessho},\ and\ \citenamefont {Sato}}]{Kawabata_gapless_PRL19}%
  \BibitemOpen
  \bibfield  {author} {\bibinfo {author} {\bibfnamefont {K.}~\bibnamefont
  {Kawabata}}, \bibinfo {author} {\bibfnamefont {T.}~\bibnamefont {Bessho}}, \
  and\ \bibinfo {author} {\bibfnamefont {M.}~\bibnamefont {Sato}},\ }\href
  {\doibase 10.1103/PhysRevLett.123.066405} {\bibfield  {journal} {\bibinfo
  {journal} {Phys. Rev. Lett.}\ }\textbf {\bibinfo {volume} {123}},\ \bibinfo
  {pages} {066405} (\bibinfo {year} {2019}{\natexlab{c}})}\BibitemShut
  {NoStop}%
\bibitem [{\citenamefont {Ghatak}\ \emph {et~al.}(2019)\citenamefont {Ghatak},
  \citenamefont {Brandenbourger}, \citenamefont {van Wezel},\ and\
  \citenamefont {Coulais}}]{Ghatak_mechSkin_arXiv19}%
  \BibitemOpen
  \bibfield  {author} {\bibinfo {author} {\bibfnamefont {A.}~\bibnamefont
  {Ghatak}}, \bibinfo {author} {\bibfnamefont {M.}~\bibnamefont
  {Brandenbourger}}, \bibinfo {author} {\bibfnamefont {J.}~\bibnamefont {van
  Wezel}}, \ and\ \bibinfo {author} {\bibfnamefont {C.}~\bibnamefont
  {Coulais}},\ }\href@noop {} {\bibfield  {journal} {\bibinfo  {journal} {arXiv
  preprint arXiv:1907.11619}\ } (\bibinfo {year} {2019})}\BibitemShut {NoStop}%
\bibitem [{\citenamefont {Colin~Scheibner}(2020)}]{Scheibner_mechEP_arXiv20}%
  \BibitemOpen
  \bibfield  {author} {\bibinfo {author} {\bibfnamefont {V.~V.}\ \bibnamefont
  {Colin~Scheibner}, \bibfnamefont {William T. M.~Irvine}},\ }\href@noop {}
  {\bibfield  {journal} {\bibinfo  {journal} {arXiv preprint arXiv:2001.04969}\
  } (\bibinfo {year} {2020})}\BibitemShut {NoStop}%
\bibitem [{\citenamefont {McClarty}\ and\ \citenamefont
  {Rau}(2019)}]{McClarty_nHbEP_PRB19}%
  \BibitemOpen
  \bibfield  {author} {\bibinfo {author} {\bibfnamefont {P.~A.}\ \bibnamefont
  {McClarty}}\ and\ \bibinfo {author} {\bibfnamefont {J.~G.}\ \bibnamefont
  {Rau}},\ }\href {\doibase 10.1103/PhysRevB.100.100405} {\bibfield  {journal}
  {\bibinfo  {journal} {Phys. Rev. B}\ }\textbf {\bibinfo {volume} {100}},\
  \bibinfo {pages} {100405} (\bibinfo {year} {2019})}\BibitemShut {NoStop}%
\bibitem [{\citenamefont {Bergholtz}\ and\ \citenamefont
  {Budich}(2019)}]{Bergholtz_nHJunc_PRR19}%
  \BibitemOpen
  \bibfield  {author} {\bibinfo {author} {\bibfnamefont {E.~J.}\ \bibnamefont
  {Bergholtz}}\ and\ \bibinfo {author} {\bibfnamefont {J.~C.}\ \bibnamefont
  {Budich}},\ }\href {\doibase 10.1103/PhysRevResearch.1.012003} {\bibfield
  {journal} {\bibinfo  {journal} {Phys. Rev. Research}\ }\textbf {\bibinfo
  {volume} {1}},\ \bibinfo {pages} {012003} (\bibinfo {year}
  {2019})}\BibitemShut {NoStop}%
\bibitem [{\citenamefont {Bergholtz}\ \emph {et~al.}(2019)\citenamefont
  {Bergholtz}, \citenamefont {Budich},\ and\ \citenamefont
  {Kunst}}]{Bergholtz_Review19}%
  \BibitemOpen
  \bibfield  {author} {\bibinfo {author} {\bibfnamefont {E.~J.}\ \bibnamefont
  {Bergholtz}}, \bibinfo {author} {\bibfnamefont {J.~C.}\ \bibnamefont
  {Budich}}, \ and\ \bibinfo {author} {\bibfnamefont {F.~K.}\ \bibnamefont
  {Kunst}},\ }\href@noop {} {\bibfield  {journal} {\bibinfo  {journal} {arXiv
  preprint arXiv:1912.10048}\ } (\bibinfo {year} {2019})}\BibitemShut {NoStop}%
\bibitem [{\citenamefont {Guo}\ \emph {et~al.}(2009)\citenamefont {Guo},
  \citenamefont {Salamo}, \citenamefont {Duchesne}, \citenamefont {Morandotti},
  \citenamefont {Volatier-Ravat}, \citenamefont {Aimez}, \citenamefont
  {Siviloglou},\ and\ \citenamefont {Christodoulides}}]{Guo_nHExp_PRL09}%
  \BibitemOpen
  \bibfield  {author} {\bibinfo {author} {\bibfnamefont {A.}~\bibnamefont
  {Guo}}, \bibinfo {author} {\bibfnamefont {G.~J.}\ \bibnamefont {Salamo}},
  \bibinfo {author} {\bibfnamefont {D.}~\bibnamefont {Duchesne}}, \bibinfo
  {author} {\bibfnamefont {R.}~\bibnamefont {Morandotti}}, \bibinfo {author}
  {\bibfnamefont {M.}~\bibnamefont {Volatier-Ravat}}, \bibinfo {author}
  {\bibfnamefont {V.}~\bibnamefont {Aimez}}, \bibinfo {author} {\bibfnamefont
  {G.~A.}\ \bibnamefont {Siviloglou}}, \ and\ \bibinfo {author} {\bibfnamefont
  {D.~N.}\ \bibnamefont {Christodoulides}},\ }\href {\doibase
  10.1103/PhysRevLett.103.093902} {\bibfield  {journal} {\bibinfo  {journal}
  {Phys. Rev. Lett.}\ }\textbf {\bibinfo {volume} {103}},\ \bibinfo {pages}
  {093902} (\bibinfo {year} {2009})}\BibitemShut {NoStop}%
\bibitem [{\citenamefont {R{\"u}ter}\ \emph {et~al.}(2010)\citenamefont
  {R{\"u}ter}, \citenamefont {Makris}, \citenamefont {El-Ganainy},
  \citenamefont {Christodoulides}, \citenamefont {Segev},\ and\ \citenamefont
  {Kip}}]{Ruter_nHExp_NatPhys10}%
  \BibitemOpen
  \bibfield  {author} {\bibinfo {author} {\bibfnamefont {C.~E.}\ \bibnamefont
  {R{\"u}ter}}, \bibinfo {author} {\bibfnamefont {K.~G.}\ \bibnamefont
  {Makris}}, \bibinfo {author} {\bibfnamefont {R.}~\bibnamefont {El-Ganainy}},
  \bibinfo {author} {\bibfnamefont {D.~N.}\ \bibnamefont {Christodoulides}},
  \bibinfo {author} {\bibfnamefont {M.}~\bibnamefont {Segev}}, \ and\ \bibinfo
  {author} {\bibfnamefont {D.}~\bibnamefont {Kip}},\ }\href@noop {} {\bibfield
  {journal} {\bibinfo  {journal} {Nature physics}\ }\textbf {\bibinfo {volume}
  {6}},\ \bibinfo {pages} {192} (\bibinfo {year} {2010})}\BibitemShut {NoStop}%
\bibitem [{\citenamefont {Szameit}\ \emph {et~al.}(2011)\citenamefont
  {Szameit}, \citenamefont {Rechtsman}, \citenamefont {Bahat-Treidel},\ and\
  \citenamefont {Segev}}]{Szameit_PRA11}%
  \BibitemOpen
  \bibfield  {author} {\bibinfo {author} {\bibfnamefont {A.}~\bibnamefont
  {Szameit}}, \bibinfo {author} {\bibfnamefont {M.~C.}\ \bibnamefont
  {Rechtsman}}, \bibinfo {author} {\bibfnamefont {O.}~\bibnamefont
  {Bahat-Treidel}}, \ and\ \bibinfo {author} {\bibfnamefont {M.}~\bibnamefont
  {Segev}},\ }\href {\doibase 10.1103/PhysRevA.84.021806} {\bibfield  {journal}
  {\bibinfo  {journal} {Phys. Rev. A}\ }\textbf {\bibinfo {volume} {84}},\
  \bibinfo {pages} {021806} (\bibinfo {year} {2011})}\BibitemShut {NoStop}%
\bibitem [{\citenamefont {Regensburger}\ \emph {et~al.}(2012)\citenamefont
  {Regensburger}, \citenamefont {Bersch}, \citenamefont {Miri}, \citenamefont
  {Onishchukov}, \citenamefont {Christodoulides},\ and\ \citenamefont
  {Peschel}}]{Regensburger_nHExp_Nat12}%
  \BibitemOpen
  \bibfield  {author} {\bibinfo {author} {\bibfnamefont {A.}~\bibnamefont
  {Regensburger}}, \bibinfo {author} {\bibfnamefont {C.}~\bibnamefont
  {Bersch}}, \bibinfo {author} {\bibfnamefont {M.-A.}\ \bibnamefont {Miri}},
  \bibinfo {author} {\bibfnamefont {G.}~\bibnamefont {Onishchukov}}, \bibinfo
  {author} {\bibfnamefont {D.~N.}\ \bibnamefont {Christodoulides}}, \ and\
  \bibinfo {author} {\bibfnamefont {U.}~\bibnamefont {Peschel}},\ }\href@noop
  {} {\bibfield  {journal} {\bibinfo  {journal} {Nature}\ }\textbf {\bibinfo
  {volume} {488}},\ \bibinfo {pages} {167} (\bibinfo {year}
  {2012})}\BibitemShut {NoStop}%
\bibitem [{\citenamefont {Zhen}\ \emph {et~al.}(2015)\citenamefont {Zhen},
  \citenamefont {Hsu}, \citenamefont {Igarashi}, \citenamefont {Lu},
  \citenamefont {Kaminer}, \citenamefont {Pick}, \citenamefont {Chua},
  \citenamefont {Joannopoulos},\ and\ \citenamefont
  {Soljacic}}]{BZhen_nH-PHC_Nat15}%
  \BibitemOpen
  \bibfield  {author} {\bibinfo {author} {\bibfnamefont {B.}~\bibnamefont
  {Zhen}}, \bibinfo {author} {\bibfnamefont {C.~W.}\ \bibnamefont {Hsu}},
  \bibinfo {author} {\bibfnamefont {Y.}~\bibnamefont {Igarashi}}, \bibinfo
  {author} {\bibfnamefont {L.}~\bibnamefont {Lu}}, \bibinfo {author}
  {\bibfnamefont {I.}~\bibnamefont {Kaminer}}, \bibinfo {author} {\bibfnamefont
  {A.}~\bibnamefont {Pick}}, \bibinfo {author} {\bibfnamefont {S.-L.}\
  \bibnamefont {Chua}}, \bibinfo {author} {\bibfnamefont {J.~D.}\ \bibnamefont
  {Joannopoulos}}, \ and\ \bibinfo {author} {\bibfnamefont {M.}~\bibnamefont
  {Soljacic}},\ }\href {http://dx.doi.org/10.1038/nature14889} {\bibfield
  {journal} {\bibinfo  {journal} {Nature}\ }\textbf {\bibinfo {volume} {525}},\
  \bibinfo {pages} {354 EP } (\bibinfo {year} {2015})}\BibitemShut {NoStop}%
\bibitem [{\citenamefont {Hassan}\ \emph {et~al.}(2017)\citenamefont {Hassan},
  \citenamefont {Zhen}, \citenamefont {Solja\ifmmode \check{c}\else
  \v{c}\fi{}i\ifmmode~\acute{c}\else \'{c}\fi{}}, \citenamefont {Khajavikhan},\
  and\ \citenamefont {Christodoulides}}]{Hassan_EP_PRL17}%
  \BibitemOpen
  \bibfield  {author} {\bibinfo {author} {\bibfnamefont {A.~U.}\ \bibnamefont
  {Hassan}}, \bibinfo {author} {\bibfnamefont {B.}~\bibnamefont {Zhen}},
  \bibinfo {author} {\bibfnamefont {M.}~\bibnamefont {Solja\ifmmode
  \check{c}\else \v{c}\fi{}i\ifmmode~\acute{c}\else \'{c}\fi{}}}, \bibinfo
  {author} {\bibfnamefont {M.}~\bibnamefont {Khajavikhan}}, \ and\ \bibinfo
  {author} {\bibfnamefont {D.~N.}\ \bibnamefont {Christodoulides}},\ }\href
  {\doibase 10.1103/PhysRevLett.118.093002} {\bibfield  {journal} {\bibinfo
  {journal} {Phys. Rev. Lett.}\ }\textbf {\bibinfo {volume} {118}},\ \bibinfo
  {pages} {093002} (\bibinfo {year} {2017})}\BibitemShut {NoStop}%
\bibitem [{\citenamefont {Feng}\ \emph {et~al.}(2017)\citenamefont {Feng},
  \citenamefont {El-Ganainy},\ and\ \citenamefont
  {Ge}}]{Feng_nH-PHC_NatPhoto17}%
  \BibitemOpen
  \bibfield  {author} {\bibinfo {author} {\bibfnamefont {L.}~\bibnamefont
  {Feng}}, \bibinfo {author} {\bibfnamefont {R.}~\bibnamefont {El-Ganainy}}, \
  and\ \bibinfo {author} {\bibfnamefont {L.}~\bibnamefont {Ge}},\ }\href
  {\doibase 10.1038/s41566-017-0031-1} {\bibfield  {journal} {\bibinfo
  {journal} {Nature Photonics}\ }\textbf {\bibinfo {volume} {11}},\ \bibinfo
  {pages} {752} (\bibinfo {year} {2017})}\BibitemShut {NoStop}%
\bibitem [{\citenamefont {Takata}\ and\ \citenamefont
  {Notomi}(2018)}]{Takata_nH-PHC_PRL18}%
  \BibitemOpen
  \bibfield  {author} {\bibinfo {author} {\bibfnamefont {K.}~\bibnamefont
  {Takata}}\ and\ \bibinfo {author} {\bibfnamefont {M.}~\bibnamefont
  {Notomi}},\ }\href {\doibase 10.1103/PhysRevLett.121.213902} {\bibfield
  {journal} {\bibinfo  {journal} {Phys. Rev. Lett.}\ }\textbf {\bibinfo
  {volume} {121}},\ \bibinfo {pages} {213902} (\bibinfo {year}
  {2018})}\BibitemShut {NoStop}%
\bibitem [{\citenamefont {Zhou}\ \emph {et~al.}(2018)\citenamefont {Zhou},
  \citenamefont {Peng}, \citenamefont {Yoon}, \citenamefont {Hsu},
  \citenamefont {Nelson}, \citenamefont {Fu}, \citenamefont {Joannopoulos},
  \citenamefont {Solja{\v c}i{\'c}},\ and\ \citenamefont
  {Zhen}}]{Zhou_BFarc_PHC_Science18}%
  \BibitemOpen
  \bibfield  {author} {\bibinfo {author} {\bibfnamefont {H.}~\bibnamefont
  {Zhou}}, \bibinfo {author} {\bibfnamefont {C.}~\bibnamefont {Peng}}, \bibinfo
  {author} {\bibfnamefont {Y.}~\bibnamefont {Yoon}}, \bibinfo {author}
  {\bibfnamefont {C.~W.}\ \bibnamefont {Hsu}}, \bibinfo {author} {\bibfnamefont
  {K.~A.}\ \bibnamefont {Nelson}}, \bibinfo {author} {\bibfnamefont
  {L.}~\bibnamefont {Fu}}, \bibinfo {author} {\bibfnamefont {J.~D.}\
  \bibnamefont {Joannopoulos}}, \bibinfo {author} {\bibfnamefont
  {M.}~\bibnamefont {Solja{\v c}i{\'c}}}, \ and\ \bibinfo {author}
  {\bibfnamefont {B.}~\bibnamefont {Zhen}},\ }\href {\doibase
  10.1126/science.aap9859} {\ \textbf {\bibinfo {volume} {359}},\ \bibinfo
  {pages} {1009} (\bibinfo {year} {2018})}\BibitemShut {NoStop}%
\bibitem [{\citenamefont {Takata}\ \emph {et~al.}(2019)\citenamefont {Takata},
  \citenamefont {Nozaki}, \citenamefont {Kuramochi}, \citenamefont {Matsuo},
  \citenamefont {Takeda}, \citenamefont {Fujii}, \citenamefont {Kita},
  \citenamefont {Shinya},\ and\ \citenamefont {Notomi}}]{Takata_nH-PHC_OSA19}%
  \BibitemOpen
  \bibfield  {author} {\bibinfo {author} {\bibfnamefont {K.}~\bibnamefont
  {Takata}}, \bibinfo {author} {\bibfnamefont {K.}~\bibnamefont {Nozaki}},
  \bibinfo {author} {\bibfnamefont {E.}~\bibnamefont {Kuramochi}}, \bibinfo
  {author} {\bibfnamefont {S.}~\bibnamefont {Matsuo}}, \bibinfo {author}
  {\bibfnamefont {K.}~\bibnamefont {Takeda}}, \bibinfo {author} {\bibfnamefont
  {T.}~\bibnamefont {Fujii}}, \bibinfo {author} {\bibfnamefont
  {S.}~\bibnamefont {Kita}}, \bibinfo {author} {\bibfnamefont {A.}~\bibnamefont
  {Shinya}}, \ and\ \bibinfo {author} {\bibfnamefont {M.}~\bibnamefont
  {Notomi}},\ }in\ \href {\doibase 10.1364/FIO.2019.FM4E.3} {\emph {\bibinfo
  {booktitle} {Frontiers in Optics $+$ Laser Science APS/DLS}}}\ (\bibinfo
  {publisher} {Optical Society of America},\ \bibinfo {year} {2019})\ p.\
  \bibinfo {pages} {FM4E.3}\BibitemShut {NoStop}%
\bibitem [{\citenamefont {Ozawa}\ \emph {et~al.}(2019)\citenamefont {Ozawa},
  \citenamefont {Price}, \citenamefont {Amo}, \citenamefont {Goldman},
  \citenamefont {Hafezi}, \citenamefont {Lu}, \citenamefont {Rechtsman},
  \citenamefont {Schuster}, \citenamefont {Simon}, \citenamefont {Zilberberg},\
  and\ \citenamefont {Carusotto}}]{Ozawa_nHPHC_PMP19}%
  \BibitemOpen
  \bibfield  {author} {\bibinfo {author} {\bibfnamefont {T.}~\bibnamefont
  {Ozawa}}, \bibinfo {author} {\bibfnamefont {H.~M.}\ \bibnamefont {Price}},
  \bibinfo {author} {\bibfnamefont {A.}~\bibnamefont {Amo}}, \bibinfo {author}
  {\bibfnamefont {N.}~\bibnamefont {Goldman}}, \bibinfo {author} {\bibfnamefont
  {M.}~\bibnamefont {Hafezi}}, \bibinfo {author} {\bibfnamefont
  {L.}~\bibnamefont {Lu}}, \bibinfo {author} {\bibfnamefont {M.~C.}\
  \bibnamefont {Rechtsman}}, \bibinfo {author} {\bibfnamefont {D.}~\bibnamefont
  {Schuster}}, \bibinfo {author} {\bibfnamefont {J.}~\bibnamefont {Simon}},
  \bibinfo {author} {\bibfnamefont {O.}~\bibnamefont {Zilberberg}}, \ and\
  \bibinfo {author} {\bibfnamefont {I.}~\bibnamefont {Carusotto}},\ }\href
  {\doibase 10.1103/RevModPhys.91.015006} {\bibfield  {journal} {\bibinfo
  {journal} {Rev. Mod. Phys.}\ }\textbf {\bibinfo {volume} {91}},\ \bibinfo
  {pages} {015006} (\bibinfo {year} {2019})}\BibitemShut {NoStop}%
\bibitem [{\citenamefont {Gong}\ \emph {et~al.}(2017)\citenamefont {Gong},
  \citenamefont {Higashikawa},\ and\ \citenamefont
  {Ueda}}]{Gong_ZenoHall_PRL17}%
  \BibitemOpen
  \bibfield  {author} {\bibinfo {author} {\bibfnamefont {Z.}~\bibnamefont
  {Gong}}, \bibinfo {author} {\bibfnamefont {S.}~\bibnamefont {Higashikawa}}, \
  and\ \bibinfo {author} {\bibfnamefont {M.}~\bibnamefont {Ueda}},\ }\href
  {\doibase 10.1103/PhysRevLett.118.200401} {\bibfield  {journal} {\bibinfo
  {journal} {Phys. Rev. Lett.}\ }\textbf {\bibinfo {volume} {118}},\ \bibinfo
  {pages} {200401} (\bibinfo {year} {2017})}\BibitemShut {NoStop}%
\bibitem [{\citenamefont {Liu}\ \emph {et~al.}(2019{\natexlab{a}})\citenamefont
  {Liu}, \citenamefont {Zhang}, \citenamefont {Ai}, \citenamefont {Gong},
  \citenamefont {Kawabata}, \citenamefont {Ueda},\ and\ \citenamefont
  {Nori}}]{Liu_nHSndTI_PRL19}%
  \BibitemOpen
  \bibfield  {author} {\bibinfo {author} {\bibfnamefont {T.}~\bibnamefont
  {Liu}}, \bibinfo {author} {\bibfnamefont {Y.-R.}\ \bibnamefont {Zhang}},
  \bibinfo {author} {\bibfnamefont {Q.}~\bibnamefont {Ai}}, \bibinfo {author}
  {\bibfnamefont {Z.}~\bibnamefont {Gong}}, \bibinfo {author} {\bibfnamefont
  {K.}~\bibnamefont {Kawabata}}, \bibinfo {author} {\bibfnamefont
  {M.}~\bibnamefont {Ueda}}, \ and\ \bibinfo {author} {\bibfnamefont
  {F.}~\bibnamefont {Nori}},\ }\href {\doibase 10.1103/PhysRevLett.122.076801}
  {\bibfield  {journal} {\bibinfo  {journal} {Phys. Rev. Lett.}\ }\textbf
  {\bibinfo {volume} {122}},\ \bibinfo {pages} {076801} (\bibinfo {year}
  {2019}{\natexlab{a}})}\BibitemShut {NoStop}%
\bibitem [{\citenamefont {Hatano}(2019)}]{Hatano_nHopen_MolPhys19}%
  \BibitemOpen
  \bibfield  {author} {\bibinfo {author} {\bibfnamefont {N.}~\bibnamefont
  {Hatano}},\ }\href {\doibase 10.1080/00268976.2019.1593535} {\bibfield
  {journal} {\bibinfo  {journal} {Molecular Physics}\ }\textbf {\bibinfo
  {volume} {117}},\ \bibinfo {pages} {2121} (\bibinfo {year}
  {2019})}\BibitemShut {NoStop}%
\bibitem [{\citenamefont {Yoshida}\ \emph
  {et~al.}(2019{\natexlab{b}})\citenamefont {Yoshida}, \citenamefont {Kudo},\
  and\ \citenamefont {Hatsugai}}]{Yoshida_nHFQH19}%
  \BibitemOpen
  \bibfield  {author} {\bibinfo {author} {\bibfnamefont {T.}~\bibnamefont
  {Yoshida}}, \bibinfo {author} {\bibfnamefont {K.}~\bibnamefont {Kudo}}, \
  and\ \bibinfo {author} {\bibfnamefont {Y.}~\bibnamefont {Hatsugai}},\ }\href
  {\doibase 10.1038/s41598-019-53253-8} {\bibfield  {journal} {\bibinfo
  {journal} {Scientific Reports}\ }\textbf {\bibinfo {volume} {9}},\ \bibinfo
  {pages} {16895} (\bibinfo {year} {2019}{\natexlab{b}})}\BibitemShut {NoStop}%
\bibitem [{\citenamefont {Ashida}\ \emph {et~al.}(2016)\citenamefont {Ashida},
  \citenamefont {Furukawa},\ and\ \citenamefont {Ueda}}]{Ashida_nHbHubb_PRA16}%
  \BibitemOpen
  \bibfield  {author} {\bibinfo {author} {\bibfnamefont {Y.}~\bibnamefont
  {Ashida}}, \bibinfo {author} {\bibfnamefont {S.}~\bibnamefont {Furukawa}}, \
  and\ \bibinfo {author} {\bibfnamefont {M.}~\bibnamefont {Ueda}},\ }\href
  {\doibase 10.1103/PhysRevA.94.053615} {\bibfield  {journal} {\bibinfo
  {journal} {Phys. Rev. A}\ }\textbf {\bibinfo {volume} {94}},\ \bibinfo
  {pages} {053615} (\bibinfo {year} {2016})}\BibitemShut {NoStop}%
\bibitem [{\citenamefont {Ashida}\ \emph {et~al.}(2017)\citenamefont {Ashida},
  \citenamefont {Furukawa},\ and\ \citenamefont
  {Ueda}}]{Ashida_PTcritical_NatComm17}%
  \BibitemOpen
  \bibfield  {author} {\bibinfo {author} {\bibfnamefont {Y.}~\bibnamefont
  {Ashida}}, \bibinfo {author} {\bibfnamefont {S.}~\bibnamefont {Furukawa}}, \
  and\ \bibinfo {author} {\bibfnamefont {M.}~\bibnamefont {Ueda}},\ }\href@noop
  {} {\bibfield  {journal} {\bibinfo  {journal} {Nature communications}\
  }\textbf {\bibinfo {volume} {8}},\ \bibinfo {pages} {15791} (\bibinfo {year}
  {2017})}\BibitemShut {NoStop}%
\bibitem [{\citenamefont {Nakagawa}\ \emph {et~al.}(2018)\citenamefont
  {Nakagawa}, \citenamefont {Kawakami},\ and\ \citenamefont
  {Ueda}}]{Nakagawa_nHKondo_PRL18}%
  \BibitemOpen
  \bibfield  {author} {\bibinfo {author} {\bibfnamefont {M.}~\bibnamefont
  {Nakagawa}}, \bibinfo {author} {\bibfnamefont {N.}~\bibnamefont {Kawakami}},
  \ and\ \bibinfo {author} {\bibfnamefont {M.}~\bibnamefont {Ueda}},\ }\href
  {\doibase 10.1103/PhysRevLett.121.203001} {\bibfield  {journal} {\bibinfo
  {journal} {Phys. Rev. Lett.}\ }\textbf {\bibinfo {volume} {121}},\ \bibinfo
  {pages} {203001} (\bibinfo {year} {2018})}\BibitemShut {NoStop}%
\bibitem [{\citenamefont {Yamamoto}\ \emph {et~al.}(2019)\citenamefont
  {Yamamoto}, \citenamefont {Nakagawa}, \citenamefont {Adachi}, \citenamefont
  {Takasan}, \citenamefont {Ueda},\ and\ \citenamefont
  {Kawakami}}]{Yamamoto_nHBCS_arXiv19}%
  \BibitemOpen
  \bibfield  {author} {\bibinfo {author} {\bibfnamefont {K.}~\bibnamefont
  {Yamamoto}}, \bibinfo {author} {\bibfnamefont {M.}~\bibnamefont {Nakagawa}},
  \bibinfo {author} {\bibfnamefont {K.}~\bibnamefont {Adachi}}, \bibinfo
  {author} {\bibfnamefont {K.}~\bibnamefont {Takasan}}, \bibinfo {author}
  {\bibfnamefont {M.}~\bibnamefont {Ueda}}, \ and\ \bibinfo {author}
  {\bibfnamefont {N.}~\bibnamefont {Kawakami}},\ }\href@noop {} {\bibfield
  {journal} {\bibinfo  {journal} {arXiv preprint arXiv:1903.04720}\ } (\bibinfo
  {year} {2019})}\BibitemShut {NoStop}%
\bibitem [{\citenamefont {Shibata}\ and\ \citenamefont
  {Katsura}(2019)}]{Shitaba_nHopen_PRB19}%
  \BibitemOpen
  \bibfield  {author} {\bibinfo {author} {\bibfnamefont {N.}~\bibnamefont
  {Shibata}}\ and\ \bibinfo {author} {\bibfnamefont {H.}~\bibnamefont
  {Katsura}},\ }\href {\doibase 10.1103/PhysRevB.99.174303} {\bibfield
  {journal} {\bibinfo  {journal} {Phys. Rev. B}\ }\textbf {\bibinfo {volume}
  {99}},\ \bibinfo {pages} {174303} (\bibinfo {year} {2019})}\BibitemShut
  {NoStop}%
\bibitem [{\citenamefont {Scazza}\ \emph {et~al.}(2014)\citenamefont {Scazza},
  \citenamefont {Hofrichter}, \citenamefont {HDan~fer}, \citenamefont
  {De~Groot}, \citenamefont {Bloch},\ and\ \citenamefont
  {FDan~lling}}]{Scazza_2bdlossExp_NatPhys14}%
  \BibitemOpen
  \bibfield  {author} {\bibinfo {author} {\bibfnamefont {F.}~\bibnamefont
  {Scazza}}, \bibinfo {author} {\bibfnamefont {C.}~\bibnamefont {Hofrichter}},
  \bibinfo {author} {\bibfnamefont {M.}~\bibnamefont {HDan~fer}}, \bibinfo
  {author} {\bibfnamefont {P.~C.}\ \bibnamefont {De~Groot}}, \bibinfo {author}
  {\bibfnamefont {I.}~\bibnamefont {Bloch}}, \ and\ \bibinfo {author}
  {\bibfnamefont {S.}~\bibnamefont {FDan~lling}},\ }\href
  {https://doi.org/10.1038/nphys3061} {\bibfield  {journal} {\bibinfo
  {journal} {Nature Physics}\ }\textbf {\bibinfo {volume} {10}},\ \bibinfo
  {pages} {779 EP } (\bibinfo {year} {2014})},\ \bibinfo {note}
  {article}\BibitemShut {NoStop}%
\bibitem [{\citenamefont {Pagano}\ \emph {et~al.}(2015)\citenamefont {Pagano},
  \citenamefont {Mancini}, \citenamefont {Cappellini}, \citenamefont {Livi},
  \citenamefont {Sias}, \citenamefont {Catani}, \citenamefont {Inguscio},\ and\
  \citenamefont {Fallani}}]{Pagano_2bdloss_PRL15}%
  \BibitemOpen
  \bibfield  {author} {\bibinfo {author} {\bibfnamefont {G.}~\bibnamefont
  {Pagano}}, \bibinfo {author} {\bibfnamefont {M.}~\bibnamefont {Mancini}},
  \bibinfo {author} {\bibfnamefont {G.}~\bibnamefont {Cappellini}}, \bibinfo
  {author} {\bibfnamefont {L.}~\bibnamefont {Livi}}, \bibinfo {author}
  {\bibfnamefont {C.}~\bibnamefont {Sias}}, \bibinfo {author} {\bibfnamefont
  {J.}~\bibnamefont {Catani}}, \bibinfo {author} {\bibfnamefont
  {M.}~\bibnamefont {Inguscio}}, \ and\ \bibinfo {author} {\bibfnamefont
  {L.}~\bibnamefont {Fallani}},\ }\href {\doibase
  10.1103/PhysRevLett.115.265301} {\bibfield  {journal} {\bibinfo  {journal}
  {Phys. Rev. Lett.}\ }\textbf {\bibinfo {volume} {115}},\ \bibinfo {pages}
  {265301} (\bibinfo {year} {2015})}\BibitemShut {NoStop}%
\bibitem [{\citenamefont {H\"ofer}\ \emph {et~al.}(2015)\citenamefont
  {H\"ofer}, \citenamefont {Riegger}, \citenamefont {Scazza}, \citenamefont
  {Hofrichter}, \citenamefont {Fernandes}, \citenamefont {Parish},
  \citenamefont {Levinsen}, \citenamefont {Bloch},\ and\ \citenamefont
  {F\"olling}}]{Hoefer_2bdlossExp_PRL15}%
  \BibitemOpen
  \bibfield  {author} {\bibinfo {author} {\bibfnamefont {M.}~\bibnamefont
  {H\"ofer}}, \bibinfo {author} {\bibfnamefont {L.}~\bibnamefont {Riegger}},
  \bibinfo {author} {\bibfnamefont {F.}~\bibnamefont {Scazza}}, \bibinfo
  {author} {\bibfnamefont {C.}~\bibnamefont {Hofrichter}}, \bibinfo {author}
  {\bibfnamefont {D.~R.}\ \bibnamefont {Fernandes}}, \bibinfo {author}
  {\bibfnamefont {M.~M.}\ \bibnamefont {Parish}}, \bibinfo {author}
  {\bibfnamefont {J.}~\bibnamefont {Levinsen}}, \bibinfo {author}
  {\bibfnamefont {I.}~\bibnamefont {Bloch}}, \ and\ \bibinfo {author}
  {\bibfnamefont {S.}~\bibnamefont {F\"olling}},\ }\href {\doibase
  10.1103/PhysRevLett.115.265302} {\bibfield  {journal} {\bibinfo  {journal}
  {Phys. Rev. Lett.}\ }\textbf {\bibinfo {volume} {115}},\ \bibinfo {pages}
  {265302} (\bibinfo {year} {2015})}\BibitemShut {NoStop}%
\bibitem [{\citenamefont {Riegger}\ \emph {et~al.}(2018)\citenamefont
  {Riegger}, \citenamefont {Darkwah~Oppong}, \citenamefont {H\"ofer},
  \citenamefont {Fernandes}, \citenamefont {Bloch},\ and\ \citenamefont
  {F\"olling}}]{Riegger_2bdlossExp_PRL18}%
  \BibitemOpen
  \bibfield  {author} {\bibinfo {author} {\bibfnamefont {L.}~\bibnamefont
  {Riegger}}, \bibinfo {author} {\bibfnamefont {N.}~\bibnamefont
  {Darkwah~Oppong}}, \bibinfo {author} {\bibfnamefont {M.}~\bibnamefont
  {H\"ofer}}, \bibinfo {author} {\bibfnamefont {D.~R.}\ \bibnamefont
  {Fernandes}}, \bibinfo {author} {\bibfnamefont {I.}~\bibnamefont {Bloch}}, \
  and\ \bibinfo {author} {\bibfnamefont {S.}~\bibnamefont {F\"olling}},\ }\href
  {\doibase 10.1103/PhysRevLett.120.143601} {\bibfield  {journal} {\bibinfo
  {journal} {Phys. Rev. Lett.}\ }\textbf {\bibinfo {volume} {120}},\ \bibinfo
  {pages} {143601} (\bibinfo {year} {2018})}\BibitemShut {NoStop}%
\bibitem [{\citenamefont {Tomita}\ \emph {et~al.}(2017)\citenamefont {Tomita},
  \citenamefont {Nakajima}, \citenamefont {Danshita}, \citenamefont {Takasu},\
  and\ \citenamefont {Takahashi}}]{Tomita_oneLoss_Science17}%
  \BibitemOpen
  \bibfield  {author} {\bibinfo {author} {\bibfnamefont {T.}~\bibnamefont
  {Tomita}}, \bibinfo {author} {\bibfnamefont {S.}~\bibnamefont {Nakajima}},
  \bibinfo {author} {\bibfnamefont {I.}~\bibnamefont {Danshita}}, \bibinfo
  {author} {\bibfnamefont {Y.}~\bibnamefont {Takasu}}, \ and\ \bibinfo {author}
  {\bibfnamefont {Y.}~\bibnamefont {Takahashi}},\ }\href {\doibase
  10.1126/sciadv.1701513} {\ \textbf {\bibinfo {volume} {3}} (\bibinfo {year}
  {2017}),\ 10.1126/sciadv.1701513}\BibitemShut {NoStop}%
\bibitem [{\citenamefont {Horio}\ \emph {et~al.}(2016)\citenamefont {Horio},
  \citenamefont {Adachi}, \citenamefont {Mori}, \citenamefont {Takahashi},
  \citenamefont {Yoshida}, \citenamefont {Suzuki}, \citenamefont {Ambolode},
  \citenamefont {Okazaki}, \citenamefont {Ono}, \citenamefont {Kumigashira},
  \citenamefont {Anzai}, \citenamefont {Arita}, \citenamefont {Namatame},
  \citenamefont {Taniguchi}, \citenamefont {Ootsuki}, \citenamefont {Sawada},
  \citenamefont {Takahashi}, \citenamefont {Mizokawa}, \citenamefont {Koike},\
  and\ \citenamefont {Fujimori}}]{Horio_PGapCu_NatCom16}%
  \BibitemOpen
  \bibfield  {author} {\bibinfo {author} {\bibfnamefont {M.}~\bibnamefont
  {Horio}}, \bibinfo {author} {\bibfnamefont {T.}~\bibnamefont {Adachi}},
  \bibinfo {author} {\bibfnamefont {Y.}~\bibnamefont {Mori}}, \bibinfo {author}
  {\bibfnamefont {A.}~\bibnamefont {Takahashi}}, \bibinfo {author}
  {\bibfnamefont {T.}~\bibnamefont {Yoshida}}, \bibinfo {author} {\bibfnamefont
  {H.}~\bibnamefont {Suzuki}}, \bibinfo {author} {\bibfnamefont {L.~C.~C.}\
  \bibnamefont {Ambolode}}, \bibinfo {author} {\bibfnamefont {K.}~\bibnamefont
  {Okazaki}}, \bibinfo {author} {\bibfnamefont {K.}~\bibnamefont {Ono}},
  \bibinfo {author} {\bibfnamefont {H.}~\bibnamefont {Kumigashira}}, \bibinfo
  {author} {\bibfnamefont {H.}~\bibnamefont {Anzai}}, \bibinfo {author}
  {\bibfnamefont {M.}~\bibnamefont {Arita}}, \bibinfo {author} {\bibfnamefont
  {H.}~\bibnamefont {Namatame}}, \bibinfo {author} {\bibfnamefont
  {M.}~\bibnamefont {Taniguchi}}, \bibinfo {author} {\bibfnamefont
  {D.}~\bibnamefont {Ootsuki}}, \bibinfo {author} {\bibfnamefont
  {K.}~\bibnamefont {Sawada}}, \bibinfo {author} {\bibfnamefont
  {M.}~\bibnamefont {Takahashi}}, \bibinfo {author} {\bibfnamefont
  {T.}~\bibnamefont {Mizokawa}}, \bibinfo {author} {\bibfnamefont
  {Y.}~\bibnamefont {Koike}}, \ and\ \bibinfo {author} {\bibfnamefont
  {A.}~\bibnamefont {Fujimori}},\ }\href {\doibase 10.1038/ncomms10567}
  {\bibfield  {journal} {\bibinfo  {journal} {Nature Communications}\ }\textbf
  {\bibinfo {volume} {7}},\ \bibinfo {pages} {10567} (\bibinfo {year}
  {2016})}\BibitemShut {NoStop}%
\bibitem [{\citenamefont {Tan}\ \emph {et~al.}(2015)\citenamefont {Tan},
  \citenamefont {Hsu}, \citenamefont {Zeng}, \citenamefont {Hatnean},
  \citenamefont {Harrison}, \citenamefont {Zhu}, \citenamefont {Hartstein},
  \citenamefont {Kiourlappou}, \citenamefont {Srivastava}, \citenamefont
  {Johannes}, \citenamefont {Murphy}, \citenamefont {Park}, \citenamefont
  {Balicas}, \citenamefont {Lonzarich}, \citenamefont {Balakrishnan},\ and\
  \citenamefont {Sebastian}}]{BTan_Science15_OscillationSmB6}%
  \BibitemOpen
  \bibfield  {author} {\bibinfo {author} {\bibfnamefont {B.~S.}\ \bibnamefont
  {Tan}}, \bibinfo {author} {\bibfnamefont {Y.-T.}\ \bibnamefont {Hsu}},
  \bibinfo {author} {\bibfnamefont {B.}~\bibnamefont {Zeng}}, \bibinfo {author}
  {\bibfnamefont {M.~C.}\ \bibnamefont {Hatnean}}, \bibinfo {author}
  {\bibfnamefont {N.}~\bibnamefont {Harrison}}, \bibinfo {author}
  {\bibfnamefont {Z.}~\bibnamefont {Zhu}}, \bibinfo {author} {\bibfnamefont
  {M.}~\bibnamefont {Hartstein}}, \bibinfo {author} {\bibfnamefont
  {M.}~\bibnamefont {Kiourlappou}}, \bibinfo {author} {\bibfnamefont
  {A.}~\bibnamefont {Srivastava}}, \bibinfo {author} {\bibfnamefont {M.~D.}\
  \bibnamefont {Johannes}}, \bibinfo {author} {\bibfnamefont {T.~P.}\
  \bibnamefont {Murphy}}, \bibinfo {author} {\bibfnamefont {J.-H.}\
  \bibnamefont {Park}}, \bibinfo {author} {\bibfnamefont {L.}~\bibnamefont
  {Balicas}}, \bibinfo {author} {\bibfnamefont {G.~G.}\ \bibnamefont
  {Lonzarich}}, \bibinfo {author} {\bibfnamefont {G.}~\bibnamefont
  {Balakrishnan}}, \ and\ \bibinfo {author} {\bibfnamefont {S.~E.}\
  \bibnamefont {Sebastian}},\ }\href {\doibase 10.1126/science.aaa7974} {\
  \textbf {\bibinfo {volume} {349}},\ \bibinfo {pages} {287} (\bibinfo {year}
  {2015})}\BibitemShut {NoStop}%
\bibitem [{\citenamefont {Xiang}\ \emph {et~al.}(2018)\citenamefont {Xiang},
  \citenamefont {Kasahara}, \citenamefont {Asaba}, \citenamefont {Lawson},
  \citenamefont {Tinsman}, \citenamefont {Chen}, \citenamefont {Sugimoto},
  \citenamefont {Kawaguchi}, \citenamefont {Sato}, \citenamefont {Li},
  \citenamefont {Yao}, \citenamefont {Chen}, \citenamefont {Iga}, \citenamefont
  {Singleton}, \citenamefont {Matsuda},\ and\ \citenamefont
  {Li}}]{ZXiang_OsciYbB12_Science18}%
  \BibitemOpen
  \bibfield  {author} {\bibinfo {author} {\bibfnamefont {Z.}~\bibnamefont
  {Xiang}}, \bibinfo {author} {\bibfnamefont {Y.}~\bibnamefont {Kasahara}},
  \bibinfo {author} {\bibfnamefont {T.}~\bibnamefont {Asaba}}, \bibinfo
  {author} {\bibfnamefont {B.}~\bibnamefont {Lawson}}, \bibinfo {author}
  {\bibfnamefont {C.}~\bibnamefont {Tinsman}}, \bibinfo {author} {\bibfnamefont
  {L.}~\bibnamefont {Chen}}, \bibinfo {author} {\bibfnamefont {K.}~\bibnamefont
  {Sugimoto}}, \bibinfo {author} {\bibfnamefont {S.}~\bibnamefont {Kawaguchi}},
  \bibinfo {author} {\bibfnamefont {Y.}~\bibnamefont {Sato}}, \bibinfo {author}
  {\bibfnamefont {G.}~\bibnamefont {Li}}, \bibinfo {author} {\bibfnamefont
  {S.}~\bibnamefont {Yao}}, \bibinfo {author} {\bibfnamefont {Y.~L.}\
  \bibnamefont {Chen}}, \bibinfo {author} {\bibfnamefont {F.}~\bibnamefont
  {Iga}}, \bibinfo {author} {\bibfnamefont {J.}~\bibnamefont {Singleton}},
  \bibinfo {author} {\bibfnamefont {Y.}~\bibnamefont {Matsuda}}, \ and\
  \bibinfo {author} {\bibfnamefont {L.}~\bibnamefont {Li}},\ }\href {\doibase
  10.1126/science.aap9607} {\ \textbf {\bibinfo {volume} {362}},\ \bibinfo
  {pages} {65} (\bibinfo {year} {2018})}\BibitemShut {NoStop}%
\bibitem [{\citenamefont {Liu}\ \emph {et~al.}(2018)\citenamefont {Liu},
  \citenamefont {Hartstein}, \citenamefont {Wallace}, \citenamefont {Davies},
  \citenamefont {Hatnean}, \citenamefont {Johannes}, \citenamefont
  {Shitsevalova}, \citenamefont {Balakrishnan},\ and\ \citenamefont
  {Sebastian}}]{HLiu_IOP18_OscillationYbB12}%
  \BibitemOpen
  \bibfield  {author} {\bibinfo {author} {\bibfnamefont {H.}~\bibnamefont
  {Liu}}, \bibinfo {author} {\bibfnamefont {M.}~\bibnamefont {Hartstein}},
  \bibinfo {author} {\bibfnamefont {G.~J.}\ \bibnamefont {Wallace}}, \bibinfo
  {author} {\bibfnamefont {A.~J.}\ \bibnamefont {Davies}}, \bibinfo {author}
  {\bibfnamefont {M.~C.}\ \bibnamefont {Hatnean}}, \bibinfo {author}
  {\bibfnamefont {M.~D.}\ \bibnamefont {Johannes}}, \bibinfo {author}
  {\bibfnamefont {N.}~\bibnamefont {Shitsevalova}}, \bibinfo {author}
  {\bibfnamefont {G.}~\bibnamefont {Balakrishnan}}, \ and\ \bibinfo {author}
  {\bibfnamefont {S.~E.}\ \bibnamefont {Sebastian}},\ }\href
  {http://stacks.iop.org/0953-8984/30/i=16/a=16LT01} {\bibfield  {journal}
  {\bibinfo  {journal} {Journal of Physics: Condensed Matter}\ }\textbf
  {\bibinfo {volume} {30}},\ \bibinfo {pages} {16LT01} (\bibinfo {year}
  {2018})}\BibitemShut {NoStop}%
\bibitem [{she()}]{shen_nu_ftnt}%
  \BibitemOpen
  \href@noop {} {}\bibinfo {note} {{ This can be seen as follows.
Firstly, by diagonalizing the Hamiltonian, we rewrite the vorticity as
\begin{eqnarray}
\nu &=& \frac{1}{4\pi i} \oint \! d\bm{k} \cdot \bm{\nabla}_{\bm{k}} \sum_n \log [E_n(\bm{k})-E_0], \nonumber
\end{eqnarray}
where $E_n$ ($n=1,\cdots,\mathrm{dim} H$) denotes the energy eigenvalues of $H$.
Substituting Eq.~(\ref{eq: 2x2 nonHermi}) to the above equation, we obtain
\begin{eqnarray}
\nu &=& \frac{1}{4\pi i} \oint \! d\bm{k} \cdot \bm{\nabla}_{\bm{k}}\sum_{n=\pm}\log [E_n(\bm{k})-E_0], \nonumber
\end{eqnarray}
where $E_\pm$ is the energy eigenvalues [see Eq.~(\ref{eq: 2x2 Epm})].
This can be rewritten as
\begin{eqnarray}
\nu &=& \frac{1}{4\pi i} \oint \! d\bm{k} \cdot \bm{\nabla}_{\bm{k}}[\log \Delta(\bm{k})+\log \{-\Delta(\bm{k})\} ] \nonumber \\
    &=& \frac{1}{4\pi i} \oint \! d\bm{k} \cdot \bm{\nabla}_{\bm{k}}[2\log \Delta(\bm{k})] \nonumber \\
    &=& \frac{1}{2\pi} \oint \! d\bm{k} \cdot \bm{\nabla}_{\bm{k}}\mathrm{arg} (E_+-E_-), \nonumber
\end{eqnarray}
with $\Delta:=(E_+-E_-)/2=\sqrt{b^2-d^2+2i\bm{b}\cdot\bm{d}}$. Here, we have omitted the term proportional to $b_0+id_{0}$ by assuming that it is canceled with $E_0$.
The last line of the above equation corresponds to the right hand side of Eq.~(\ref{eq: vorticity Shen app})

  }}\BibitemShut {NoStop}%
\bibitem [{nu_()}]{nu_w_to_nu_ftnt}%
  \BibitemOpen
  \href@noop {} {}\bibinfo {note} {{ Eq.~(\ref{eq: vorticity}) is obtained as follows. 
With $\Sigma=\1\otimes \rho_3$, Eq.~(\ref{eq: winding H'}) is rewritten as
\begin{eqnarray}
\nu_W 
&=& \frac{1}{4\pi i} \oint  \! d^2\bm{k} \cdot \mathrm{tr}[
%
\left(
\begin{array}{cc}
\1 &  \\
 & -\1
\end{array}
\right)_\rho
%
\left(
\begin{array}{cc}
0 & H(\bm{k}) \\
H^\dagger(\bm{k}) & 0
\end{array}
\right)^{-1}_\rho
%
\bm{\nabla}_{\bm{k}}
%
\left(
\begin{array}{cc}
0 & H(\bm{k}) \\
H^\dagger(\bm{k}) & 0
\end{array}
\right)_\rho
] \nonumber \\
&=& \frac{1}{4\pi i} \oint  \! d^2\bm{k} \cdot \mathrm{tr}[ \bm{\nabla}_{\bm{k}} \log H^\dagger(\bm{k}) - \bm{\nabla}_{\bm{k}} \log H(\bm{k}) ] \nonumber \\
&=& \frac{-1}{2\pi} \oint  \! d^2\bm{k} \cdot \bm{\nabla}_{\bm{k}} \log \mathrm{det} H(\bm{k}). \nonumber
\end{eqnarray}
The last line corresponds to Eq.~(\ref{eq: vorticity}) up to the prefactor; $\nu=-\nu_W/2$

  }}\BibitemShut {NoStop}%
\bibitem [{\citenamefont {Abrikosov}\ \emph {et~al.}(2012)\citenamefont
  {Abrikosov}, \citenamefont {Gorkov},\ and\ \citenamefont
  {Dzyaloshinski}}]{AGD}%
  \BibitemOpen
  \bibfield  {author} {\bibinfo {author} {\bibfnamefont {A.~A.}\ \bibnamefont
  {Abrikosov}}, \bibinfo {author} {\bibfnamefont {L.~P.}\ \bibnamefont
  {Gorkov}}, \ and\ \bibinfo {author} {\bibfnamefont {I.~E.}\ \bibnamefont
  {Dzyaloshinski}},\ }\href@noop {} {\emph {\bibinfo {title} {Methods of
  quantum field theory in statistical physics}}}\ (\bibinfo  {publisher}
  {Courier Corporation},\ \bibinfo {year} {2012})\BibitemShut {NoStop}%
\bibitem [{\citenamefont {Metzner}\ and\ \citenamefont
  {Vollhardt}(1989)}]{WMetznerPRL89_DMFT}%
  \BibitemOpen
  \bibfield  {author} {\bibinfo {author} {\bibfnamefont {W.}~\bibnamefont
  {Metzner}}\ and\ \bibinfo {author} {\bibfnamefont {D.}~\bibnamefont
  {Vollhardt}},\ }\href {\doibase 10.1103/PhysRevLett.62.324} {\bibfield
  {journal} {\bibinfo  {journal} {Phys. Rev. Lett.}\ }\textbf {\bibinfo
  {volume} {62}},\ \bibinfo {pages} {324} (\bibinfo {year} {1989})}\BibitemShut
  {NoStop}%
\bibitem [{\citenamefont {M{\"u}ller-Hartmann}(1989)}]{MHartmannZP89_DMFT}%
  \BibitemOpen
  \bibfield  {author} {\bibinfo {author} {\bibfnamefont {E.}~\bibnamefont
  {M{\"u}ller-Hartmann}},\ }\href {\doibase 10.1007/BF01311397} {\bibfield
  {journal} {\bibinfo  {journal} {Zeitschrift f{\"u}r Physik B Condensed
  Matter}\ }\textbf {\bibinfo {volume} {74}},\ \bibinfo {pages} {507} (\bibinfo
  {year} {1989})}\BibitemShut {NoStop}%
\bibitem [{\citenamefont {Georges}\ and\ \citenamefont
  {Kotliar}(1992)}]{Kotliar_DMFT_92}%
  \BibitemOpen
  \bibfield  {author} {\bibinfo {author} {\bibfnamefont {A.}~\bibnamefont
  {Georges}}\ and\ \bibinfo {author} {\bibfnamefont {G.}~\bibnamefont
  {Kotliar}},\ }\href {\doibase 10.1103/PhysRevB.45.6479} {\bibfield  {journal}
  {\bibinfo  {journal} {Phys. Rev. B}\ }\textbf {\bibinfo {volume} {45}},\
  \bibinfo {pages} {6479} (\bibinfo {year} {1992})}\BibitemShut {NoStop}%
\bibitem [{\citenamefont {Georges}\ \emph {et~al.}(1996)\citenamefont
  {Georges}, \citenamefont {Kotliar}, \citenamefont {Krauth},\ and\
  \citenamefont {Rozenberg}}]{AGeorgesRMP96_DMFT}%
  \BibitemOpen
  \bibfield  {author} {\bibinfo {author} {\bibfnamefont {A.}~\bibnamefont
  {Georges}}, \bibinfo {author} {\bibfnamefont {G.}~\bibnamefont {Kotliar}},
  \bibinfo {author} {\bibfnamefont {W.}~\bibnamefont {Krauth}}, \ and\ \bibinfo
  {author} {\bibfnamefont {M.~J.}\ \bibnamefont {Rozenberg}},\ }\href {\doibase
  10.1103/RevModPhys.68.13} {\bibfield  {journal} {\bibinfo  {journal} {Rev.
  Mod. Phys.}\ }\textbf {\bibinfo {volume} {68}},\ \bibinfo {pages} {13}
  (\bibinfo {year} {1996})}\BibitemShut {NoStop}%
\bibitem [{DMF()}]{DMFT_ftnt}%
  \BibitemOpen
  \href@noop {} {}\bibinfo {note} {{ The action of the effective impurity model is written as
\begin{subequations}
\begin{eqnarray}
\mathcal{Z}_{\mathrm{imp}} &=& \int \! \prod_{s} \mathcal{D}\bar{c}_{bs}\mathcal{D}c_{bs} \mathrm{tr}_S \exp(-\mathcal{S}_{\mathrm{imp}}), \nonumber \\
-\mathcal{S}_{\mathrm{imp}} &=& \int \! d\tau d\tau' \sum_{s} \bar{c}_{bs} \mathcal{G}^{-1}_{s} (\tau-\tau') c_{bs} -H_{\mathrm{int}}\delta(\tau-\tau'), \nonumber
\end{eqnarray}
\end{subequations}
where $H_{\mathrm{int}}$ denotes the local interaction term $H_{\mathrm{int}}:=J\bm{s}_{0b}\cdot\bm{S}$. $\delta(\tau-\tau')$ is the delta function. $\mathrm{tr}_S$ denotes taking trace for the localized spin.
$\mathcal{G}_{s} (\tau-\tau')$ denotes the Green's function of the effective bath. $\bar{c}_{bs}$ ($c_{bs}$) is a Grassmannian variable which corresponds to the creation operator $c^\dagger_{0bs}$ (annihilation operator $c_{0bs}$) at site $i=0$.
%
Solving the above model with an impurity solver, we obtain the self-energy $\Sigma_{bs}$, which allows us to compute the Green's function as
\begin{eqnarray}
\mathcal{G}^{-1}_{s} (i\omega_n)&=& \left[\frac{1}{N}\sum_{\bm{k}}\frac{1}{i\omega_n -h(\bm{k})-\Sigma_{s}(i\omega_n)}\right]+\Sigma_{s}(i\omega_n), \nonumber
\end{eqnarray}
with the Matsubara frequency $\omega_n=(2n+1)\pi T$. Here, $N$ denotes the number of unit cells. Computing the effective Green's function $\mathcal{G}^{-1}_{s}$ yields the self-energy $\Sigma_{s}(i\omega_n)$

  }}\BibitemShut {NoStop}%
\bibitem [{\citenamefont {Wilson}(1975)}]{KWilsonRMP75_NRG}%
  \BibitemOpen
  \bibfield  {author} {\bibinfo {author} {\bibfnamefont {K.~G.}\ \bibnamefont
  {Wilson}},\ }\href {\doibase 10.1103/RevModPhys.47.773} {\bibfield  {journal}
  {\bibinfo  {journal} {Rev. Mod. Phys.}\ }\textbf {\bibinfo {volume} {47}},\
  \bibinfo {pages} {773} (\bibinfo {year} {1975})}\BibitemShut {NoStop}%
\bibitem [{\citenamefont {Peters}\ \emph {et~al.}(2006)\citenamefont {Peters},
  \citenamefont {Pruschke},\ and\ \citenamefont {Anders}}]{RPetersPRB06_NRG}%
  \BibitemOpen
  \bibfield  {author} {\bibinfo {author} {\bibfnamefont {R.}~\bibnamefont
  {Peters}}, \bibinfo {author} {\bibfnamefont {T.}~\bibnamefont {Pruschke}}, \
  and\ \bibinfo {author} {\bibfnamefont {F.~B.}\ \bibnamefont {Anders}},\
  }\href {\doibase 10.1103/PhysRevB.74.245114} {\bibfield  {journal} {\bibinfo
  {journal} {Phys. Rev. B}\ }\textbf {\bibinfo {volume} {74}},\ \bibinfo
  {pages} {245114} (\bibinfo {year} {2006})}\BibitemShut {NoStop}%
\bibitem [{\citenamefont {Bulla}\ \emph {et~al.}(2008)\citenamefont {Bulla},
  \citenamefont {Costi},\ and\ \citenamefont {Pruschke}}]{RBullaRMP08_NRG}%
  \BibitemOpen
  \bibfield  {author} {\bibinfo {author} {\bibfnamefont {R.}~\bibnamefont
  {Bulla}}, \bibinfo {author} {\bibfnamefont {T.~A.}\ \bibnamefont {Costi}}, \
  and\ \bibinfo {author} {\bibfnamefont {T.}~\bibnamefont {Pruschke}},\ }\href
  {\doibase 10.1103/RevModPhys.80.395} {\bibfield  {journal} {\bibinfo
  {journal} {Rev. Mod. Phys.}\ }\textbf {\bibinfo {volume} {80}},\ \bibinfo
  {pages} {395} (\bibinfo {year} {2008})}\BibitemShut {NoStop}%
\bibitem [{\citenamefont {Hirsch}\ and\ \citenamefont
  {Fye}(1986)}]{Hirsh_QMC_PRL86}%
  \BibitemOpen
  \bibfield  {author} {\bibinfo {author} {\bibfnamefont {J.~E.}\ \bibnamefont
  {Hirsch}}\ and\ \bibinfo {author} {\bibfnamefont {R.~M.}\ \bibnamefont
  {Fye}},\ }\href {\doibase 10.1103/PhysRevLett.56.2521} {\bibfield  {journal}
  {\bibinfo  {journal} {Phys. Rev. Lett.}\ }\textbf {\bibinfo {volume} {56}},\
  \bibinfo {pages} {2521} (\bibinfo {year} {1986})}\BibitemShut {NoStop}%
\bibitem [{\citenamefont {Werner}\ \emph {et~al.}(2006)\citenamefont {Werner},
  \citenamefont {Comanac}, \citenamefont {de' Medici}, \citenamefont {Troyer},\
  and\ \citenamefont {Millis}}]{Werner_CTQMC_PRL2006}%
  \BibitemOpen
  \bibfield  {author} {\bibinfo {author} {\bibfnamefont {P.}~\bibnamefont
  {Werner}}, \bibinfo {author} {\bibfnamefont {A.}~\bibnamefont {Comanac}},
  \bibinfo {author} {\bibfnamefont {L.}~\bibnamefont {de' Medici}}, \bibinfo
  {author} {\bibfnamefont {M.}~\bibnamefont {Troyer}}, \ and\ \bibinfo {author}
  {\bibfnamefont {A.~J.}\ \bibnamefont {Millis}},\ }\href {\doibase
  10.1103/PhysRevLett.97.076405} {\bibfield  {journal} {\bibinfo  {journal}
  {Phys. Rev. Lett.}\ }\textbf {\bibinfo {volume} {97}},\ \bibinfo {pages}
  {076405} (\bibinfo {year} {2006})}\BibitemShut {NoStop}%
\bibitem [{\citenamefont {Werner}\ and\ \citenamefont
  {Millis}(2006)}]{Werner_CTQMC_CTQMC2006}%
  \BibitemOpen
  \bibfield  {author} {\bibinfo {author} {\bibfnamefont {P.}~\bibnamefont
  {Werner}}\ and\ \bibinfo {author} {\bibfnamefont {A.~J.}\ \bibnamefont
  {Millis}},\ }\href {\doibase 10.1103/PhysRevB.74.155107} {\bibfield
  {journal} {\bibinfo  {journal} {Phys. Rev. B}\ }\textbf {\bibinfo {volume}
  {74}},\ \bibinfo {pages} {155107} (\bibinfo {year} {2006})}\BibitemShut
  {NoStop}%
\bibitem [{\citenamefont {Ruderman}\ and\ \citenamefont
  {Kittel}(1954)}]{Ruderman_RKKY_PR54}%
  \BibitemOpen
  \bibfield  {author} {\bibinfo {author} {\bibfnamefont {M.~A.}\ \bibnamefont
  {Ruderman}}\ and\ \bibinfo {author} {\bibfnamefont {C.}~\bibnamefont
  {Kittel}},\ }\href {\doibase 10.1103/PhysRev.96.99} {\bibfield  {journal}
  {\bibinfo  {journal} {Phys. Rev.}\ }\textbf {\bibinfo {volume} {96}},\
  \bibinfo {pages} {99} (\bibinfo {year} {1954})}\BibitemShut {NoStop}%
\bibitem [{\citenamefont {Kasuya}(1956)}]{Kasuya_RKKY_PTP56}%
  \BibitemOpen
  \bibfield  {author} {\bibinfo {author} {\bibfnamefont {T.}~\bibnamefont
  {Kasuya}},\ }\href {\doibase 10.1143/PTP.16.45} {\bibfield  {journal}
  {\bibinfo  {journal} {Progress of Theoretical Physics}\ }\textbf {\bibinfo
  {volume} {16}},\ \bibinfo {pages} {45} (\bibinfo {year} {1956})}\BibitemShut
  {NoStop}%
\bibitem [{\citenamefont {Yosida}(1957)}]{Yosida_PR57}%
  \BibitemOpen
  \bibfield  {author} {\bibinfo {author} {\bibfnamefont {K.}~\bibnamefont
  {Yosida}},\ }\href {\doibase 10.1103/PhysRev.106.893} {\bibfield  {journal}
  {\bibinfo  {journal} {Phys. Rev.}\ }\textbf {\bibinfo {volume} {106}},\
  \bibinfo {pages} {893} (\bibinfo {year} {1957})}\BibitemShut {NoStop}%
\bibitem [{\citenamefont {Schnyder}\ \emph {et~al.}(2008)\citenamefont
  {Schnyder}, \citenamefont {Ryu}, \citenamefont {Furusaki},\ and\
  \citenamefont {Ludwig}}]{Schnyder_classification_free_2008}%
  \BibitemOpen
  \bibfield  {author} {\bibinfo {author} {\bibfnamefont {A.~P.}\ \bibnamefont
  {Schnyder}}, \bibinfo {author} {\bibfnamefont {S.}~\bibnamefont {Ryu}},
  \bibinfo {author} {\bibfnamefont {A.}~\bibnamefont {Furusaki}}, \ and\
  \bibinfo {author} {\bibfnamefont {A.~W.~W.}\ \bibnamefont {Ludwig}},\ }\href
  {\doibase 10.1103/PhysRevB.78.195125} {\bibfield  {journal} {\bibinfo
  {journal} {Phys. Rev. B}\ }\textbf {\bibinfo {volume} {78}},\ \bibinfo
  {pages} {195125} (\bibinfo {year} {2008})}\BibitemShut {NoStop}%
\bibitem [{\citenamefont {Kitaev}(2009)}]{Kitaev_classification_free_2009}%
  \BibitemOpen
  \bibfield  {author} {\bibinfo {author} {\bibfnamefont {A.}~\bibnamefont
  {Kitaev}},\ }\href {\doibase 10.1063/1.3149495} {\bibfield  {journal}
  {\bibinfo  {journal} {AIP Conf. Proc.}\ }\textbf {\bibinfo {volume} {1134}},\
  \bibinfo {pages} {22} (\bibinfo {year} {2009})}\BibitemShut {NoStop}%
\bibitem [{\citenamefont {Ryu}\ \emph {et~al.}(2010)\citenamefont {Ryu},
  \citenamefont {Schnyder}, \citenamefont {Furusaki},\ and\ \citenamefont
  {Ludwig}}]{Ryu_classification_free_2010}%
  \BibitemOpen
  \bibfield  {author} {\bibinfo {author} {\bibfnamefont {S.}~\bibnamefont
  {Ryu}}, \bibinfo {author} {\bibfnamefont {A.~P.}\ \bibnamefont {Schnyder}},
  \bibinfo {author} {\bibfnamefont {A.}~\bibnamefont {Furusaki}}, \ and\
  \bibinfo {author} {\bibfnamefont {A.~W.~W.}\ \bibnamefont {Ludwig}},\ }\href
  {http://stacks.iop.org/1367-2630/12/i=6/a=065010} {\bibfield  {journal}
  {\bibinfo  {journal} {New J. Phys.}\ }\textbf {\bibinfo {volume} {12}},\
  \bibinfo {pages} {065010} (\bibinfo {year} {2010})}\BibitemShut {NoStop}%
\bibitem [{\citenamefont {Chiu}\ \emph {et~al.}(2016)\citenamefont {Chiu},
  \citenamefont {Teo}, \citenamefont {Schnyder},\ and\ \citenamefont
  {Ryu}}]{Chiu_class_RMP16}%
  \BibitemOpen
  \bibfield  {author} {\bibinfo {author} {\bibfnamefont {C.-K.}\ \bibnamefont
  {Chiu}}, \bibinfo {author} {\bibfnamefont {J.~C.~Y.}\ \bibnamefont {Teo}},
  \bibinfo {author} {\bibfnamefont {A.~P.}\ \bibnamefont {Schnyder}}, \ and\
  \bibinfo {author} {\bibfnamefont {S.}~\bibnamefont {Ryu}},\ }\href {\doibase
  10.1103/RevModPhys.88.035005} {\bibfield  {journal} {\bibinfo  {journal}
  {Rev. Mod. Phys.}\ }\textbf {\bibinfo {volume} {88}},\ \bibinfo {pages}
  {035005} (\bibinfo {year} {2016})}\BibitemShut {NoStop}%
\bibitem [{\citenamefont {Yamazaki}\ \emph {et~al.}(2010)\citenamefont
  {Yamazaki}, \citenamefont {Taie}, \citenamefont {Sugawa},\ and\ \citenamefont
  {Takahashi}}]{RYamazaki_PRL10}%
  \BibitemOpen
  \bibfield  {author} {\bibinfo {author} {\bibfnamefont {R.}~\bibnamefont
  {Yamazaki}}, \bibinfo {author} {\bibfnamefont {S.}~\bibnamefont {Taie}},
  \bibinfo {author} {\bibfnamefont {S.}~\bibnamefont {Sugawa}}, \ and\ \bibinfo
  {author} {\bibfnamefont {Y.}~\bibnamefont {Takahashi}},\ }\href {\doibase
  10.1103/PhysRevLett.105.050405} {\bibfield  {journal} {\bibinfo  {journal}
  {Phys. Rev. Lett.}\ }\textbf {\bibinfo {volume} {105}},\ \bibinfo {pages}
  {050405} (\bibinfo {year} {2010})}\BibitemShut {NoStop}%
\bibitem [{\citenamefont {Clark}\ \emph {et~al.}(2015)\citenamefont {Clark},
  \citenamefont {Ha}, \citenamefont {Xu},\ and\ \citenamefont
  {Chin}}]{LWClark_PRL15}%
  \BibitemOpen
  \bibfield  {author} {\bibinfo {author} {\bibfnamefont {L.~W.}\ \bibnamefont
  {Clark}}, \bibinfo {author} {\bibfnamefont {L.-C.}\ \bibnamefont {Ha}},
  \bibinfo {author} {\bibfnamefont {C.-Y.}\ \bibnamefont {Xu}}, \ and\ \bibinfo
  {author} {\bibfnamefont {C.}~\bibnamefont {Chin}},\ }\href {\doibase
  10.1103/PhysRevLett.115.155301} {\bibfield  {journal} {\bibinfo  {journal}
  {Phys. Rev. Lett.}\ }\textbf {\bibinfo {volume} {115}},\ \bibinfo {pages}
  {155301} (\bibinfo {year} {2015})}\BibitemShut {NoStop}%
\bibitem [{gam()}]{gamma_En_ftnt}%
  \BibitemOpen
  \href@noop {} {}\bibinfo {note} {{ This fact can be understood as follows.
Suppose that the Hamiltonian is chiral symmetric [see Eq.~(\ref{eq: SPERs chiral symm generic})]. 
Then, with the eigenvalue $E_n$ and the right eigenvector $|\varphi^R_n \rangle$ ($H|\varphi^R _n \rangle  = |\varphi^R _n \rangle E_n $, $n\in \mathbb{Z}$), we obtain the relation
\begin{eqnarray}
H^\dagger U^\dagger_\Gamma|\varphi^R _n \rangle  &=& -U^\dagger_\Gamma |\varphi^R _n \rangle E_n. \nonumber
\end{eqnarray}
Here, we have used Eq.~(\ref{eq: SPERs chiral symm generic}).
Noticing that the eigenvalues problem of the left eigenvector $|\varphi^L _n \rangle $ ($n\in \mathbb{Z}$) is written as
$
H^\dagger |\varphi^L _n \rangle  
= 
|\varphi^L _n \rangle E^*_n,
$
we can see that the vector $U^\dagger_\Gamma |\varphi^R _n \rangle$ is the left eigenvector with the eigenvalue $E_n$

  }}\BibitemShut {NoStop}%
\bibitem [{\citenamefont {Zhang}\ \emph {et~al.}(1993)\citenamefont {Zhang},
  \citenamefont {Rozenberg},\ and\ \citenamefont {Kotliar}}]{Zhang_IPT_93}%
  \BibitemOpen
  \bibfield  {author} {\bibinfo {author} {\bibfnamefont {X.~Y.}\ \bibnamefont
  {Zhang}}, \bibinfo {author} {\bibfnamefont {M.~J.}\ \bibnamefont
  {Rozenberg}}, \ and\ \bibinfo {author} {\bibfnamefont {G.}~\bibnamefont
  {Kotliar}},\ }\href {\doibase 10.1103/PhysRevLett.70.1666} {\bibfield
  {journal} {\bibinfo  {journal} {Phys. Rev. Lett.}\ }\textbf {\bibinfo
  {volume} {70}},\ \bibinfo {pages} {1666} (\bibinfo {year}
  {1993})}\BibitemShut {NoStop}%
\bibitem [{\citenamefont {Kajueter}\ and\ \citenamefont
  {Kotliar}(1996)}]{Kajueter_IPT_96}%
  \BibitemOpen
  \bibfield  {author} {\bibinfo {author} {\bibfnamefont {H.}~\bibnamefont
  {Kajueter}}\ and\ \bibinfo {author} {\bibfnamefont {G.}~\bibnamefont
  {Kotliar}},\ }\href {\doibase 10.1103/PhysRevLett.77.131} {\bibfield
  {journal} {\bibinfo  {journal} {Phys. Rev. Lett.}\ }\textbf {\bibinfo
  {volume} {77}},\ \bibinfo {pages} {131} (\bibinfo {year} {1996})}\BibitemShut
  {NoStop}%
\bibitem [{sus()}]{suscep_RPA_ftnt}%
  \BibitemOpen
  \href@noop {} {}\bibinfo {note} {{ The local magnetic susceptibility $\chi^s_\alpha$ is computed as follows.
With the RPA, the matrix of the susceptibility is written as
\begin{eqnarray}
\chi^{\mathrm{RPA}}(i\epsilon_n,\bm{q}) &=& (\1-\chi^0 U )\chi^0, \nonumber 
\end{eqnarray}
with $U=\mathrm{diag}(U_A,U_B)$ and $\xi^0$ defined as 
\begin{eqnarray}
\chi^0_{\alpha\beta}(i\epsilon_n,\bm{q}) &=& -\frac{T}{N} \sum_{\bm{k},m} G_{\alpha\beta}(i\omega_m+i\epsilon_n,\bm{q}+\bm{k}) G_{\beta\alpha}(i\omega_n,\bm{k}).\nonumber 
\end{eqnarray}
Here, $\omega_n$ and $\epsilon_n$ denote the Matsubara frequency [$\omega_n=(2n+1)\pi T$ and $\epsilon_n=2n\pi T$ with $n \in \mathbb{Z}$].
$N$ denotes the number of unit cells.
The local magnetic susceptibility $\chi^s_\alpha$ is obtained as
\begin{eqnarray}
\chi^s_A&=& (\chi^{\mathrm{RPA}}_{AA}+\chi^{\mathrm{RPA}}_{AB})/2, \nonumber  \\
\chi^s_B&=& (\chi^{\mathrm{RPA}}_{BB}+\chi^{\mathrm{RPA}}_{BA})/2, \nonumber 
\end{eqnarray}
with $\bm{q}=0$. We set $\epsilon_n \to 0$ instead of doing analytic continuation

  }}\BibitemShut {NoStop}%
\bibitem [{TP_()}]{TP_GF_ftnt}%
  \BibitemOpen
  \href@noop {} {}\bibinfo {note} {{ Eq.~(\ref{eq: Class PT symm GF}) can be obtained as follows.  
Firstly, we note that the following relations hold
\begin{subequations}
\begin{eqnarray}
it\hat{H} &=& it \hat{U}_{PT} \hat{H}^* \hat{U}^\dagger_{PT}, \nonumber \\
%
 \hat{U}^\dagger_{PT} \hat{c}^\dagger_{i\alpha} \hat{U}_{PT} &=& \sum_{\alpha'} \hat{c}^\dagger_{-i\alpha'} U^\dagger_{\alpha'\alpha}, \nonumber\\
%
%
\langle n^* | \hat{A}  | m^* \rangle
&=&
\sum_{\{i\alpha\},\{j\beta\}}
\langle n^* | \{i\alpha\} \rangle \langle \{i\alpha\} |\hat{A}| \{j\beta\} \rangle \langle \{j\beta\}| m^*\rangle
\nonumber \\
&=&
\sum_{\{i\alpha\},\{j\beta\}}
\langle m | \{j\beta\} \rangle
\langle \{j\beta\} |\hat{A}^T| \{i\alpha\} \rangle
\langle  \{i\alpha\} | n\rangle 
\nonumber\\
&=&
\langle m | \hat{A}^T  | n \rangle,
\nonumber
\end{eqnarray}
%
\end{subequations}
where $|\{i\alpha\}\rangle$ and $|\{j\beta\}\rangle$ denote the states generated by applying the creation operators $\hat{c}^\dagger_{i\alpha}$ on the vacuum $|0\rangle$. $|n\rangle$ and $|m \rangle$ are arbitrary states.
$|n^* \rangle := \sum_{\{i\alpha \}} | \{ i\alpha \}\rangle \langle n | \{ i\alpha \} \rangle $.
%
By using the above relations, the correlation function $\langle \hat{c}_{i\alpha}(t)\hat{c}^\dagger_{j\gamma} \rangle$ is rewritten as
\begin{eqnarray}
\langle \hat{c}_{i\alpha}(t)\hat{c}^\dagger_{j\gamma} \rangle
&=& 
Z^{-1}
\mathrm{tr}[ e^{-\beta \hat{H}} e^{it\hat{H}} \hat{c}_{i\alpha} e^{-it\hat{H}} \hat{c}^\dagger_{j\gamma} ] \nonumber \\
&=& 
Z^{-1}
\mathrm{tr}[ e^{-\beta \hat{H}^*} e^{it\hat{H}^*} \hat{U}^\dagger_{PT} \hat{c}_{i\alpha} \hat{U}_{PT} e^{-it\hat{H}^*} \hat{U}^\dagger_{PT} \hat{c}^\dagger_{j\gamma} \hat{U}_{PT} ] \nonumber \\
&=& 
Z^{-1}
\sum_{\alpha'\gamma'}
\mathrm{tr}[ e^{-\beta \hat{H}^*} e^{it\hat{H}^*} U_{PT,\alpha \alpha'}\hat{c}_{-i\alpha'}  e^{-it\hat{H}^*} \hat{c}^\dagger_{-j\gamma'} U^\dagger_{PT,\gamma' \gamma} ] \nonumber \\
&=& 
Z^{-1}
\sum_{\alpha'\gamma'}
U_{PT,\alpha \alpha'}
U^\dagger_{PT,\gamma' \gamma}
\mathrm{tr}[ 
\hat{c}_{-j\gamma'}
e^{-it\hat{H}}
\hat{c}^\dagger_{-i\alpha'}
e^{it\hat{H}}
e^{-\beta \hat{H}}
] \nonumber \\
&=&
\sum_{\alpha'\gamma'}
U_{PT,\alpha \alpha'}
U^\dagger_{PT,\gamma' \gamma}
\langle 
\hat{c}_{-j\gamma'}(t)
\hat{c}^\dagger_{-i\alpha'}
\rangle,
\nonumber
\end{eqnarray}
which is equivalent to
\begin{eqnarray}
\langle \hat{c}_{\bm{k}\alpha}(t)\hat{c}^\dagger_{\bm{k}\gamma} \rangle
 &=&
\sum_{\alpha'\gamma'}
U_{PT,\alpha \alpha'}
U^\dagger_{PT,\gamma' \gamma}
\langle \hat{c}_{\bm{k}\gamma'}(t)\hat{c}^\dagger_{\bm{k}\alpha'} \rangle.
\nonumber
\end{eqnarray}
%
Here, $Z$ denotes the partition function.
In a similar way, we obtain
\begin{eqnarray}
\langle \hat{c}^\dagger_{\bm{k}\gamma}\hat{c}_{\bm{k}\alpha}(t) \rangle
 &=&
\sum_{\alpha'\gamma'}
U_{PT,\alpha \alpha'}
U^\dagger_{PT,\gamma' \gamma}
\langle 
\hat{c}^\dagger_{\bm{k}\alpha'}
\hat{c}_{\bm{k}\gamma'}(t)
\rangle.
\nonumber
\end{eqnarray}
Remembering that the definition of the retarded Green's function [see Eq.~(\ref{eq: defs of GR(t,k)})], we end up with Eq.~(\ref{eq: Class PT symm GF})

  }}\BibitemShut {NoStop}%
\bibitem [{CP_()}]{CP_GF_ftnt}%
  \BibitemOpen
  \href@noop {} {}\bibinfo {note} {{ Eq.~(\ref{eq: CP GR to GA}) can be obtained by the following calculations.
Firstly, we note that the following relation holds.
\begin{eqnarray}
 \hat{U}^\dagger_{CP}\hat{c}^\dagger_{i\alpha} \hat{U}_{CP} &=&  \sum_{\alpha'} \hat{c}_{-i\alpha'} U^\dagger_{\alpha'\alpha}. \nonumber
\end{eqnarray}
%
By using the above relations, $\langle \hat{c}_{i\alpha}(t)\hat{c}^\dagger_{j\gamma} \rangle$ is rewritten as
\begin{eqnarray}
\langle \hat{c}_{i\alpha}(t)\hat{c}^\dagger_{j\gamma} \rangle 
&=&
Z^{-1}
\mathrm{tr}[ e^{-\beta \hat{H}} e^{it\hat{H}} \hat{c}_{i\alpha} e^{-it\hat{H}} \hat{c}^\dagger_{j\gamma} ] \nonumber \\
&=&
Z^{-1}
\mathrm{tr}[ e^{-\beta \hat{U}_{CP} \hat{H} \hat{U}^\dagger_{CP} } e^{it \hat{U}_{CP} \hat{H} \hat{U}^\dagger_{CP} } \hat{c}_{i\alpha} e^{-it \hat{U}_{CP} \hat{H} \hat{U}^\dagger_{CP} } \hat{c}^\dagger_{j\gamma} ] \nonumber \\
&=&
Z^{-1}
\mathrm{tr}[ e^{-\beta \hat{H} } e^{it \hat{H}  } \hat{U}^\dagger_{CP} \hat{c}_{i\alpha} \hat{U}_{CP} e^{-it \hat{H} } \hat{U}^\dagger_{CP} \hat{c}^\dagger_{j\gamma} \hat{U}_{CP} ] \nonumber \\
&=&
Z^{-1}
\sum_{\alpha'\gamma'}
\mathrm{tr}[ e^{-\beta \hat{H} } e^{it \hat{H}  } U_{CP,\alpha \alpha'} \hat{c}^\dagger_{-i\alpha'}  e^{-it \hat{H} } \hat{c}_{-j\gamma'} U^\dagger_{CP,\gamma'\gamma}  ] \nonumber \\
&=&
\sum_{\alpha'\gamma'}
U_{CP,\alpha \alpha'}
U^\dagger_{CP,\gamma'\gamma}
\langle \hat{c}^\dagger_{-i\alpha'}(t) \hat{c}_{-j\gamma'} \rangle, \nonumber
\end{eqnarray}
which is equivalent to 
\begin{eqnarray}
\langle \hat{c}_{\bm{k}\alpha}(t)\hat{c}^\dagger_{\bm{k}\gamma} \rangle 
&=& 
\sum_{\alpha'\gamma'}
U_{CP,\alpha \alpha'}
U^\dagger_{CP,\gamma'\gamma}
\langle \hat{c}^\dagger_{\bm{k}\alpha'} \hat{c}_{\bm{k}\gamma'}(-t) \rangle. \nonumber
\end{eqnarray}
Here, $Z$ denotes the partition function.
In a similar way, we obtain
\begin{eqnarray}
\langle \hat{c}^\dagger_{\bm{k}\gamma} \hat{c}_{\bm{k}\alpha}(t) \rangle 
&=& 
\sum_{\alpha'\gamma'}
U_{CP,\alpha \alpha'}
U^\dagger_{CP,\gamma'\gamma}
\langle  \hat{c}_{\bm{k}\gamma'}(-t) \hat{c}^\dagger_{\bm{k}\alpha'} \rangle. \nonumber
\end{eqnarray}
Namely, the above calculations yield the following relation between the relarded and the advanced Green's function:
\begin{eqnarray}
\langle \hat{c}_{\bm{k}\alpha}(t)\hat{c}^\dagger_{\bm{k}\gamma} + \hat{c}^\dagger_{\bm{k}\gamma} \hat{c}_{\bm{k}\alpha}(t) \rangle\theta(t)
&=& 
\sum_{\alpha'\gamma'}
U_{CP,\alpha \alpha'}
U^\dagger_{CP,\gamma'\gamma}
\langle \hat{c}_{\bm{k}\gamma'}(t')\hat{c}^\dagger_{\bm{k}\alpha'} + \hat{c}^\dagger_{\bm{k}\alpha'} \hat{c}_{\bm{k}\gamma'}(t') \rangle\theta(-t'). \nonumber 
\end{eqnarray}
with $t'=-t$.
We note that the right (left) hand side of the above equation corresponds to $iG^R$ ($-iG^A$), respectively.
Applying the Fourier transformation, we obtain Eq.~(\ref{eq: CP GR to GA})
%

  }}\BibitemShut {NoStop}%
\bibitem [{GAG()}]{GAGR_ftnt}%
  \BibitemOpen
  \href@noop {} {}\bibinfo {note} {{ Eq.~(\ref{eq: GAdag=GR}) can be obtained by making use of Hermiticity of the Hamiltonian.
With the Lehmann representation, the Green's function can be written as 
\begin{eqnarray}
G_{\alpha\beta}(\omega+i\delta,\bm{k})&=& Z^{-1} \sum_{nm} e^{-\beta E_n } \frac{e^{\beta(E_n-E_m)}+1}{(\omega+i\delta+E_n-E_m)} \langle n|\hat{c}_{\bm{k}\alpha}|m \rangle \langle m|\hat{c}^\dagger_{\bm{k}\beta}|n \rangle, \nonumber 
\end{eqnarray}
where $|n\rangle$'s are eigenstates of the Hamiltonian $\hat{H}$. $Z$ denotes the partition function $Z:=\sum_n e^{-\beta E_n}$.
With this representation, we can see that the following relation holds:
\begin{eqnarray}
G^*_{\beta \alpha}(\omega+i\delta,\bm{k})
&=& Z^{-1} \sum_{nm} e^{-\beta E_n } \frac{e^{\beta(E_n-E_m)}+1}{(\omega+i\delta+E_n-E_m)^*} \langle n|\hat{c}_{\bm{k}\beta}|m \rangle^* \langle m|\hat{c}^\dagger_{\bm{k}\alpha}|n \rangle^* \nonumber \\
&=& Z^{-1} \sum_{nm} e^{-\beta E_n } \frac{e^{\beta(E_n-E_m)}+1}{(\omega-i\delta+E_n-E_m)} \langle n|\hat{c}_{\bm{k}\alpha}|m \rangle \langle m|\hat{c}^\dagger_{\bm{k}\beta}|n \rangle  \nonumber \\
&=& G_{\alpha\beta}(\omega-i\delta,\bm{k}), \nonumber
\end{eqnarray}
which is nothing but the relation shown in Eq.~(\ref{eq: GAdag=GR})

  }}\BibitemShut {NoStop}%
\bibitem [{Gam()}]{Gamma_GF_ftnt}%
  \BibitemOpen
  \href@noop {} {}\bibinfo {note} {{ Eq.~(\ref{eq: Class Gamma symm GF}) can be obtained as follows.
  Firstly, we note that the following relations hold:
\begin{eqnarray}
it\hat{H} &=& it \hat{U}_\Gamma \hat{H}^{*} \hat{U}^\dagger_\Gamma, \nonumber \\ 
\hat{U}_\Gamma \hat{c}_{i\alpha} \hat{U}^\dagger_\Gamma &=& \sum_{\beta} U_{\Gamma,\alpha\beta} \hat{c}^\dagger_{i\beta}, \nonumber \\
 \langle n ^* |\hat{A}| m^*\rangle &=& \langle m |\hat{A}^T| n \rangle, \nonumber
\end{eqnarray}
where $|n\rangle$ and $|m\rangle$ are arbitrary states. 
$| n^* \rangle $ is defined as $| n^* \rangle := \sum_{\{ i\alpha \} }| \{ i \alpha\}\rangle  \langle n | \{ i \alpha\}\rangle $ with the states $| \{ i \alpha\}\rangle$ obtained by applying the operators $\hat{c}^\dagger_{i\alpha}$ to the vacuum.
By using the above relations, $\langle \hat{c}_{i\alpha}(t) \hat{c}^\dagger_{j\gamma}\rangle$ is rewritten as
\begin{eqnarray}
\langle \hat{c}_{i\alpha}(t) \hat{c}^\dagger_{j\gamma}\rangle &=& Z^{-1} \mathrm{tr}[e^{-\beta \hat{H}} e^{ it \hat{H}} \hat{c}_{i\alpha} e^{ -it \hat{H}} \hat{c}^\dagger_{j\gamma}  ] \nonumber \\
                                                  &=& Z^{-1} \mathrm{tr}[e^{-\beta \hat{H}^*} e^{ it \hat{H}^*} \hat{U}^\dagger_\Gamma \hat{c}_{i\alpha} \hat{U}_\Gamma e^{ -it \hat{H}^*} \hat{U}^\dagger_\Gamma \hat{c}^\dagger_{j\gamma}\hat{U}_\Gamma  ] \nonumber \\
                                                  &=& Z^{-1} \sum_{\alpha'\gamma'} \mathrm{tr}[e^{-\beta \hat{H}^*} e^{ it \hat{H}^*} U_{\Gamma,\alpha\alpha'} \hat{c}^\dagger_{i\alpha'} e^{ -it \hat{H}^*}  \hat{c}_{i\gamma'} U^\dagger_{\Gamma,\gamma'\gamma} ] \nonumber \\
                                                  &=& Z^{-1} \sum_{\alpha'\gamma'} U_{\Gamma,\alpha\alpha'} U^\dagger_{\Gamma,\gamma'\gamma} \mathrm{tr}[ e^{-\beta \hat{H}} e^{ it \hat{H}} \hat{c}^\dagger_{i\gamma'} e^{ -it \hat{H}} \hat{c}_{i\alpha'} ] \nonumber \\
                                                  &=& \sum_{\alpha'\gamma'} U_{\Gamma,\alpha\alpha'} U^\dagger_{\Gamma,\gamma'\gamma} \langle \hat{c}^\dagger_{i\gamma'} \hat{c}_{i\alpha'}(-t) \rangle, \nonumber
\end{eqnarray}
%
which is equivalent to 
\begin{eqnarray}
\langle \hat{c}_{\bm{k}\alpha}(t) \hat{c}^\dagger_{\bm{k}\gamma}\rangle &=& \sum_{\alpha'\gamma'} U_{\Gamma,\alpha\alpha'} U^\dagger_{\Gamma,\gamma'\gamma} \langle \hat{c}^\dagger_{\bm{k}\gamma'} \hat{c}_{\bm{k}\alpha'}(-t) \rangle. \nonumber 
\end{eqnarray}
%
Here, $Z$ denotes the partition function.
In a similar way, we obtain 
\begin{eqnarray}
\langle \hat{c}^\dagger_{\bm{k}\gamma} \hat{c}_{\bm{k}\alpha}(t) \rangle &=& \sum_{\alpha'\gamma'} U_{\Gamma,\alpha\alpha'} U^\dagger_{\Gamma,\gamma'\gamma} \langle \hat{c}_{\bm{k}\alpha'}(-t) \hat{c}^\dagger_{\bm{k}\gamma'}  \rangle. \nonumber
\end{eqnarray}
%
Namely, the above calculation yields the following relation between the retarded and the advanced Green's function:
\begin{eqnarray}
\langle \hat{c}_{\bm{k}\alpha}(t) \hat{c}^\dagger_{\bm{k}\gamma} + \hat{c}^\dagger_{\bm{k}\gamma} \hat{c}_{\bm{k}\alpha}(t) \rangle \theta(t)
&=& 
\sum_{\alpha'\gamma'} 
U_{\Gamma,\alpha\alpha'} U^\dagger_{\Gamma,\gamma'\gamma} \langle \hat{c}_{\bm{k}\alpha'}(t') \hat{c}^\dagger_{\bm{k}\gamma'} + \hat{c}^\dagger_{\bm{k}\gamma'} \hat{c}_{\bm{k}\alpha'}(t') \hat{c}^\dagger_{\bm{k}\gamma'}  \rangle \theta(-t'), \nonumber 
\end{eqnarray}
with $t'=-t$.
We note that the right (left) hand side of the above equation corresponds to $iG^R$ ($-iG^A$), respectively.
With the Fourier transformation and Eq.~(\ref{eq: GAdag=GR}) we obtain Eq.~(\ref{eq: Class Gamma symm GF})

  }}\BibitemShut {NoStop}%
\bibitem [{\citenamefont {Yoshida}\ and\ \citenamefont
  {Hatsugai}(2019)}]{Yoshida_SPERs_mech19}%
  \BibitemOpen
  \bibfield  {author} {\bibinfo {author} {\bibfnamefont {T.}~\bibnamefont
  {Yoshida}}\ and\ \bibinfo {author} {\bibfnamefont {Y.}~\bibnamefont
  {Hatsugai}},\ }\href {\doibase 10.1103/PhysRevB.100.054109} {\bibfield
  {journal} {\bibinfo  {journal} {Phys. Rev. B}\ }\textbf {\bibinfo {volume}
  {100}},\ \bibinfo {pages} {054109} (\bibinfo {year} {2019})}\BibitemShut
  {NoStop}%
\bibitem [{\citenamefont {Liu}\ \emph {et~al.}(2019{\natexlab{b}})\citenamefont
  {Liu}, \citenamefont {Jiang},\ and\ \citenamefont
  {Chen}}]{Liu_MirroPointClass_PRB19}%
  \BibitemOpen
  \bibfield  {author} {\bibinfo {author} {\bibfnamefont {C.-H.}\ \bibnamefont
  {Liu}}, \bibinfo {author} {\bibfnamefont {H.}~\bibnamefont {Jiang}}, \ and\
  \bibinfo {author} {\bibfnamefont {S.}~\bibnamefont {Chen}},\ }\href {\doibase
  10.1103/PhysRevB.99.125103} {\bibfield  {journal} {\bibinfo  {journal} {Phys.
  Rev. B}\ }\textbf {\bibinfo {volume} {99}},\ \bibinfo {pages} {125103}
  (\bibinfo {year} {2019}{\natexlab{b}})}\BibitemShut {NoStop}%
\bibitem [{\citenamefont {Teo}\ and\ \citenamefont
  {Kane}(2010)}]{Teo_disloc_class_PRB10}%
  \BibitemOpen
  \bibfield  {author} {\bibinfo {author} {\bibfnamefont {J.~C.~Y.}\
  \bibnamefont {Teo}}\ and\ \bibinfo {author} {\bibfnamefont {C.~L.}\
  \bibnamefont {Kane}},\ }\href {\doibase 10.1103/PhysRevB.82.115120}
  {\bibfield  {journal} {\bibinfo  {journal} {Phys. Rev. B}\ }\textbf {\bibinfo
  {volume} {82}},\ \bibinfo {pages} {115120} (\bibinfo {year}
  {2010})}\BibitemShut {NoStop}%
\bibitem [{\citenamefont {Chiu}\ and\ \citenamefont
  {Schnyder}(2014)}]{Chiu_gapless_class_PRB14}%
  \BibitemOpen
  \bibfield  {author} {\bibinfo {author} {\bibfnamefont {C.-K.}\ \bibnamefont
  {Chiu}}\ and\ \bibinfo {author} {\bibfnamefont {A.~P.}\ \bibnamefont
  {Schnyder}},\ }\href {\doibase 10.1103/PhysRevB.90.205136} {\bibfield
  {journal} {\bibinfo  {journal} {Phys. Rev. B}\ }\textbf {\bibinfo {volume}
  {90}},\ \bibinfo {pages} {205136} (\bibinfo {year} {2014})}\BibitemShut
  {NoStop}%
\bibitem [{\citenamefont {Morimoto}\ and\ \citenamefont
  {Furusaki}(2013)}]{TMorimoto_AZclass_PRB13}%
  \BibitemOpen
  \bibfield  {author} {\bibinfo {author} {\bibfnamefont {T.}~\bibnamefont
  {Morimoto}}\ and\ \bibinfo {author} {\bibfnamefont {A.}~\bibnamefont
  {Furusaki}},\ }\href {\doibase 10.1103/PhysRevB.88.125129} {\bibfield
  {journal} {\bibinfo  {journal} {Phys. Rev. B}\ }\textbf {\bibinfo {volume}
  {88}},\ \bibinfo {pages} {125129} (\bibinfo {year} {2013})}\BibitemShut
  {NoStop}%
\bibitem [{\citenamefont {Shiozaki}\ and\ \citenamefont
  {Sato}(2014)}]{KShiozaki_AZclass_PRB14}%
  \BibitemOpen
  \bibfield  {author} {\bibinfo {author} {\bibfnamefont {K.}~\bibnamefont
  {Shiozaki}}\ and\ \bibinfo {author} {\bibfnamefont {M.}~\bibnamefont
  {Sato}},\ }\href {\doibase 10.1103/PhysRevB.90.165114} {\bibfield  {journal}
  {\bibinfo  {journal} {Phys. Rev. B}\ }\textbf {\bibinfo {volume} {90}},\
  \bibinfo {pages} {165114} (\bibinfo {year} {2014})}\BibitemShut {NoStop}%
\bibitem [{\citenamefont {Bzdu\ifmmode~\check{s}\else \v{s}\fi{}ek}\ and\
  \citenamefont {Sigrist}(2017)}]{Bzdusek_AZ+I_PRB17}%
  \BibitemOpen
  \bibfield  {author} {\bibinfo {author} {\bibfnamefont {T.~c.~v.}\
  \bibnamefont {Bzdu\ifmmode~\check{s}\else \v{s}\fi{}ek}}\ and\ \bibinfo
  {author} {\bibfnamefont {M.}~\bibnamefont {Sigrist}},\ }\href {\doibase
  10.1103/PhysRevB.96.155105} {\bibfield  {journal} {\bibinfo  {journal} {Phys.
  Rev. B}\ }\textbf {\bibinfo {volume} {96}},\ \bibinfo {pages} {155105}
  (\bibinfo {year} {2017})}\BibitemShut {NoStop}%
\bibitem [{\citenamefont {Takimoto}(2011)}]{Takimoto_SmB6_JPSJ11}%
  \BibitemOpen
  \bibfield  {author} {\bibinfo {author} {\bibfnamefont {T.}~\bibnamefont
  {Takimoto}},\ }\href {\doibase 10.1143/JPSJ.80.123710} {\bibfield  {journal}
  {\bibinfo  {journal} {Journal of the Physical Society of Japan}\ }\textbf
  {\bibinfo {volume} {80}},\ \bibinfo {pages} {123710} (\bibinfo {year}
  {2011})}\BibitemShut {NoStop}%
\bibitem [{\citenamefont {Neupane}\ \emph {et~al.}(2013)\citenamefont
  {Neupane}, \citenamefont {Alidoust}, \citenamefont {Xu}, \citenamefont
  {Kondo}, \citenamefont {Ishida}, \citenamefont {Kim}, \citenamefont {Liu},
  \citenamefont {Belopolski}, \citenamefont {Jo}, \citenamefont {Chang},
  \citenamefont {Jeng}, \citenamefont {Durakiewicz}, \citenamefont {Balicas},
  \citenamefont {Lin}, \citenamefont {Bansil}, \citenamefont {Shin},
  \citenamefont {Fisk},\ and\ \citenamefont {Hasan}}]{Neupane_SmB6_NatComm13}%
  \BibitemOpen
  \bibfield  {author} {\bibinfo {author} {\bibfnamefont {M.}~\bibnamefont
  {Neupane}}, \bibinfo {author} {\bibfnamefont {N.}~\bibnamefont {Alidoust}},
  \bibinfo {author} {\bibfnamefont {S.-Y.}\ \bibnamefont {Xu}}, \bibinfo
  {author} {\bibfnamefont {T.}~\bibnamefont {Kondo}}, \bibinfo {author}
  {\bibfnamefont {Y.}~\bibnamefont {Ishida}}, \bibinfo {author} {\bibfnamefont
  {D.~J.}\ \bibnamefont {Kim}}, \bibinfo {author} {\bibfnamefont
  {C.}~\bibnamefont {Liu}}, \bibinfo {author} {\bibfnamefont {I.}~\bibnamefont
  {Belopolski}}, \bibinfo {author} {\bibfnamefont {Y.~J.}\ \bibnamefont {Jo}},
  \bibinfo {author} {\bibfnamefont {T.-R.}\ \bibnamefont {Chang}}, \bibinfo
  {author} {\bibfnamefont {H.-T.}\ \bibnamefont {Jeng}}, \bibinfo {author}
  {\bibfnamefont {T.}~\bibnamefont {Durakiewicz}}, \bibinfo {author}
  {\bibfnamefont {L.}~\bibnamefont {Balicas}}, \bibinfo {author} {\bibfnamefont
  {H.}~\bibnamefont {Lin}}, \bibinfo {author} {\bibfnamefont {A.}~\bibnamefont
  {Bansil}}, \bibinfo {author} {\bibfnamefont {S.}~\bibnamefont {Shin}},
  \bibinfo {author} {\bibfnamefont {Z.}~\bibnamefont {Fisk}}, \ and\ \bibinfo
  {author} {\bibfnamefont {M.~Z.}\ \bibnamefont {Hasan}},\ }\href {\doibase
  10.1038/ncomms3991} {\bibfield  {journal} {\bibinfo  {journal} {Nature
  Communications}\ }\textbf {\bibinfo {volume} {4}},\ \bibinfo {pages} {2991}
  (\bibinfo {year} {2013})}\BibitemShut {NoStop}%
\bibitem [{\citenamefont {Jiang}\ \emph {et~al.}(2013)\citenamefont {Jiang},
  \citenamefont {Li}, \citenamefont {Zhang}, \citenamefont {Sun}, \citenamefont
  {Chen}, \citenamefont {Ye}, \citenamefont {Xu}, \citenamefont {Ge},
  \citenamefont {Tan}, \citenamefont {Niu}, \citenamefont {Xia}, \citenamefont
  {Xie}, \citenamefont {Li}, \citenamefont {Chen}, \citenamefont {Wen},\ and\
  \citenamefont {Feng}}]{Jiang_SmB6_NatComm13}%
  \BibitemOpen
  \bibfield  {author} {\bibinfo {author} {\bibfnamefont {J.}~\bibnamefont
  {Jiang}}, \bibinfo {author} {\bibfnamefont {S.}~\bibnamefont {Li}}, \bibinfo
  {author} {\bibfnamefont {T.}~\bibnamefont {Zhang}}, \bibinfo {author}
  {\bibfnamefont {Z.}~\bibnamefont {Sun}}, \bibinfo {author} {\bibfnamefont
  {F.}~\bibnamefont {Chen}}, \bibinfo {author} {\bibfnamefont {Z.~R.}\
  \bibnamefont {Ye}}, \bibinfo {author} {\bibfnamefont {M.}~\bibnamefont {Xu}},
  \bibinfo {author} {\bibfnamefont {Q.~Q.}\ \bibnamefont {Ge}}, \bibinfo
  {author} {\bibfnamefont {S.~Y.}\ \bibnamefont {Tan}}, \bibinfo {author}
  {\bibfnamefont {X.~H.}\ \bibnamefont {Niu}}, \bibinfo {author} {\bibfnamefont
  {M.}~\bibnamefont {Xia}}, \bibinfo {author} {\bibfnamefont {B.~P.}\
  \bibnamefont {Xie}}, \bibinfo {author} {\bibfnamefont {Y.~F.}\ \bibnamefont
  {Li}}, \bibinfo {author} {\bibfnamefont {X.~H.}\ \bibnamefont {Chen}},
  \bibinfo {author} {\bibfnamefont {H.~H.}\ \bibnamefont {Wen}}, \ and\
  \bibinfo {author} {\bibfnamefont {D.~L.}\ \bibnamefont {Feng}},\ }\href
  {\doibase 10.1038/ncomms4010} {\bibfield  {journal} {\bibinfo  {journal}
  {Nature Communications}\ }\textbf {\bibinfo {volume} {4}},\ \bibinfo {pages}
  {3010} (\bibinfo {year} {2013})}\BibitemShut {NoStop}%
\bibitem [{\citenamefont {Xu}\ \emph {et~al.}(2013)\citenamefont {Xu},
  \citenamefont {Shi}, \citenamefont {Biswas}, \citenamefont {Matt},
  \citenamefont {Dhaka}, \citenamefont {Huang}, \citenamefont {Plumb},
  \citenamefont {Radovi\ifmmode~\acute{c}\else \'{c}\fi{}}, \citenamefont
  {Dil}, \citenamefont {Pomjakushina}, \citenamefont {Conder}, \citenamefont
  {Amato}, \citenamefont {Salman}, \citenamefont {Paul}, \citenamefont {Mesot},
  \citenamefont {Ding},\ and\ \citenamefont {Shi}}]{Xu_SmB6_PRB13}%
  \BibitemOpen
  \bibfield  {author} {\bibinfo {author} {\bibfnamefont {N.}~\bibnamefont
  {Xu}}, \bibinfo {author} {\bibfnamefont {X.}~\bibnamefont {Shi}}, \bibinfo
  {author} {\bibfnamefont {P.~K.}\ \bibnamefont {Biswas}}, \bibinfo {author}
  {\bibfnamefont {C.~E.}\ \bibnamefont {Matt}}, \bibinfo {author}
  {\bibfnamefont {R.~S.}\ \bibnamefont {Dhaka}}, \bibinfo {author}
  {\bibfnamefont {Y.}~\bibnamefont {Huang}}, \bibinfo {author} {\bibfnamefont
  {N.~C.}\ \bibnamefont {Plumb}}, \bibinfo {author} {\bibfnamefont
  {M.}~\bibnamefont {Radovi\ifmmode~\acute{c}\else \'{c}\fi{}}}, \bibinfo
  {author} {\bibfnamefont {J.~H.}\ \bibnamefont {Dil}}, \bibinfo {author}
  {\bibfnamefont {E.}~\bibnamefont {Pomjakushina}}, \bibinfo {author}
  {\bibfnamefont {K.}~\bibnamefont {Conder}}, \bibinfo {author} {\bibfnamefont
  {A.}~\bibnamefont {Amato}}, \bibinfo {author} {\bibfnamefont
  {Z.}~\bibnamefont {Salman}}, \bibinfo {author} {\bibfnamefont {D.~M.}\
  \bibnamefont {Paul}}, \bibinfo {author} {\bibfnamefont {J.}~\bibnamefont
  {Mesot}}, \bibinfo {author} {\bibfnamefont {H.}~\bibnamefont {Ding}}, \ and\
  \bibinfo {author} {\bibfnamefont {M.}~\bibnamefont {Shi}},\ }\href {\doibase
  10.1103/PhysRevB.88.121102} {\bibfield  {journal} {\bibinfo  {journal} {Phys.
  Rev. B}\ }\textbf {\bibinfo {volume} {88}},\ \bibinfo {pages} {121102}
  (\bibinfo {year} {2013})}\BibitemShut {NoStop}%
\bibitem [{\citenamefont {Peters}\ \emph {et~al.}(2016)\citenamefont {Peters},
  \citenamefont {Yoshida}, \citenamefont {Sakakibara},\ and\ \citenamefont
  {Kawakami}}]{Peters_SmB6_PRB16}%
  \BibitemOpen
  \bibfield  {author} {\bibinfo {author} {\bibfnamefont {R.}~\bibnamefont
  {Peters}}, \bibinfo {author} {\bibfnamefont {T.}~\bibnamefont {Yoshida}},
  \bibinfo {author} {\bibfnamefont {H.}~\bibnamefont {Sakakibara}}, \ and\
  \bibinfo {author} {\bibfnamefont {N.}~\bibnamefont {Kawakami}},\ }\href
  {\doibase 10.1103/PhysRevB.93.235159} {\bibfield  {journal} {\bibinfo
  {journal} {Phys. Rev. B}\ }\textbf {\bibinfo {volume} {93}},\ \bibinfo
  {pages} {235159} (\bibinfo {year} {2016})}\BibitemShut {NoStop}%
\bibitem [{\citenamefont {Peters}\ \emph {et~al.}(2018)\citenamefont {Peters},
  \citenamefont {Yoshida},\ and\ \citenamefont
  {Kawakami}}]{Peters_SmB6mag_PRB18}%
  \BibitemOpen
  \bibfield  {author} {\bibinfo {author} {\bibfnamefont {R.}~\bibnamefont
  {Peters}}, \bibinfo {author} {\bibfnamefont {T.}~\bibnamefont {Yoshida}}, \
  and\ \bibinfo {author} {\bibfnamefont {N.}~\bibnamefont {Kawakami}},\ }\href
  {\doibase 10.1103/PhysRevB.98.075104} {\bibfield  {journal} {\bibinfo
  {journal} {Phys. Rev. B}\ }\textbf {\bibinfo {volume} {98}},\ \bibinfo
  {pages} {075104} (\bibinfo {year} {2018})}\BibitemShut {NoStop}%
\bibitem [{\citenamefont {Thunstr{\"o}m}\ and\ \citenamefont
  {Held}(2019)}]{Thunstrom_SmB6_arXiv19}%
  \BibitemOpen
  \bibfield  {author} {\bibinfo {author} {\bibfnamefont {P.}~\bibnamefont
  {Thunstr{\"o}m}}\ and\ \bibinfo {author} {\bibfnamefont {K.}~\bibnamefont
  {Held}},\ }\href@noop {} {\bibfield  {journal} {\bibinfo  {journal} {arXiv
  preprint arXiv:1907.03899}\ } (\bibinfo {year} {2019})}\BibitemShut {NoStop}%
\bibitem [{\citenamefont {Weng}\ \emph {et~al.}(2014)\citenamefont {Weng},
  \citenamefont {Zhao}, \citenamefont {Wang}, \citenamefont {Fang},\ and\
  \citenamefont {Dai}}]{Weng_YbB12_PRL14}%
  \BibitemOpen
  \bibfield  {author} {\bibinfo {author} {\bibfnamefont {H.}~\bibnamefont
  {Weng}}, \bibinfo {author} {\bibfnamefont {J.}~\bibnamefont {Zhao}}, \bibinfo
  {author} {\bibfnamefont {Z.}~\bibnamefont {Wang}}, \bibinfo {author}
  {\bibfnamefont {Z.}~\bibnamefont {Fang}}, \ and\ \bibinfo {author}
  {\bibfnamefont {X.}~\bibnamefont {Dai}},\ }\href {\doibase
  10.1103/PhysRevLett.112.016403} {\bibfield  {journal} {\bibinfo  {journal}
  {Phys. Rev. Lett.}\ }\textbf {\bibinfo {volume} {112}},\ \bibinfo {pages}
  {016403} (\bibinfo {year} {2014})}\BibitemShut {NoStop}%
\bibitem [{\citenamefont {Hagiwara}\ \emph {et~al.}(2016)\citenamefont
  {Hagiwara}, \citenamefont {Ohtsubo}, \citenamefont {Matsunami}, \citenamefont
  {Ideta}, \citenamefont {Tanaka}, \citenamefont {Miyazaki}, \citenamefont
  {Rault}, \citenamefont {Fteivre}, \citenamefont {Bertran}, \citenamefont
  {Taleb-Ibrahimi}, \citenamefont {Yukawa}, \citenamefont {Kobayashi},
  \citenamefont {Horiba}, \citenamefont {Kumigashira}, \citenamefont {Sumida},
  \citenamefont {Okuda}, \citenamefont {Iga},\ and\ \citenamefont
  {Kimura}}]{Hagiwara_YbB12_NatComm16}%
  \BibitemOpen
  \bibfield  {author} {\bibinfo {author} {\bibfnamefont {K.}~\bibnamefont
  {Hagiwara}}, \bibinfo {author} {\bibfnamefont {Y.}~\bibnamefont {Ohtsubo}},
  \bibinfo {author} {\bibfnamefont {M.}~\bibnamefont {Matsunami}}, \bibinfo
  {author} {\bibfnamefont {S.-i.}\ \bibnamefont {Ideta}}, \bibinfo {author}
  {\bibfnamefont {K.}~\bibnamefont {Tanaka}}, \bibinfo {author} {\bibfnamefont
  {H.}~\bibnamefont {Miyazaki}}, \bibinfo {author} {\bibfnamefont {J.~E.}\
  \bibnamefont {Rault}}, \bibinfo {author} {\bibfnamefont {P.~L.}\ \bibnamefont
  {Fteivre}}, \bibinfo {author} {\bibfnamefont {F.}~\bibnamefont {Bertran}},
  \bibinfo {author} {\bibfnamefont {A.}~\bibnamefont {Taleb-Ibrahimi}},
  \bibinfo {author} {\bibfnamefont {R.}~\bibnamefont {Yukawa}}, \bibinfo
  {author} {\bibfnamefont {M.}~\bibnamefont {Kobayashi}}, \bibinfo {author}
  {\bibfnamefont {K.}~\bibnamefont {Horiba}}, \bibinfo {author} {\bibfnamefont
  {H.}~\bibnamefont {Kumigashira}}, \bibinfo {author} {\bibfnamefont
  {K.}~\bibnamefont {Sumida}}, \bibinfo {author} {\bibfnamefont
  {T.}~\bibnamefont {Okuda}}, \bibinfo {author} {\bibfnamefont
  {F.}~\bibnamefont {Iga}}, \ and\ \bibinfo {author} {\bibfnamefont {S.-i.}\
  \bibnamefont {Kimura}},\ }\href {\doibase 10.1038/ncomms12690} {\bibfield
  {journal} {\bibinfo  {journal} {Nature Communications}\ }\textbf {\bibinfo
  {volume} {7}},\ \bibinfo {pages} {12690} (\bibinfo {year}
  {2016})}\BibitemShut {NoStop}%
\bibitem [{\citenamefont {Helbig}\ \emph {et~al.}(2019)\citenamefont {Helbig},
  \citenamefont {Hofmann}, \citenamefont {Imhof}, \citenamefont {Abdelghany},
  \citenamefont {Kiessling}, \citenamefont {Molenkamp}, \citenamefont {Lee},
  \citenamefont {Szameit}, \citenamefont {Greiter},\ and\ \citenamefont
  {Thomale}}]{Helbig_elecirSkin_19}%
  \BibitemOpen
  \bibfield  {author} {\bibinfo {author} {\bibfnamefont {T.}~\bibnamefont
  {Helbig}}, \bibinfo {author} {\bibfnamefont {T.}~\bibnamefont {Hofmann}},
  \bibinfo {author} {\bibfnamefont {S.}~\bibnamefont {Imhof}}, \bibinfo
  {author} {\bibfnamefont {M.}~\bibnamefont {Abdelghany}}, \bibinfo {author}
  {\bibfnamefont {T.}~\bibnamefont {Kiessling}}, \bibinfo {author}
  {\bibfnamefont {L.~W.}\ \bibnamefont {Molenkamp}}, \bibinfo {author}
  {\bibfnamefont {C.~H.}\ \bibnamefont {Lee}}, \bibinfo {author} {\bibfnamefont
  {A.}~\bibnamefont {Szameit}}, \bibinfo {author} {\bibfnamefont
  {M.}~\bibnamefont {Greiter}}, \ and\ \bibinfo {author} {\bibfnamefont
  {R.}~\bibnamefont {Thomale}},\ }\href@noop {} {\bibfield  {journal} {\bibinfo
   {journal} {arXiv preprint arXiv:1907.11562}\ } (\bibinfo {year}
  {2019})}\BibitemShut {NoStop}%
\bibitem [{\citenamefont {Hofmann}\ \emph {et~al.}(2019)\citenamefont
  {Hofmann}, \citenamefont {Helbig}, \citenamefont {Schindler}, \citenamefont
  {Salgo}, \citenamefont {Brzezi{\'n}ska}, \citenamefont {Greiter},
  \citenamefont {Kiessling}, \citenamefont {Wolf}, \citenamefont {Vollhardt},
  \citenamefont {Kaba{\v{s}}i} \emph {et~al.}}]{Hofmann_ExpRecipSkin_19}%
  \BibitemOpen
  \bibfield  {author} {\bibinfo {author} {\bibfnamefont {T.}~\bibnamefont
  {Hofmann}}, \bibinfo {author} {\bibfnamefont {T.}~\bibnamefont {Helbig}},
  \bibinfo {author} {\bibfnamefont {F.}~\bibnamefont {Schindler}}, \bibinfo
  {author} {\bibfnamefont {N.}~\bibnamefont {Salgo}}, \bibinfo {author}
  {\bibfnamefont {M.}~\bibnamefont {Brzezi{\'n}ska}}, \bibinfo {author}
  {\bibfnamefont {M.}~\bibnamefont {Greiter}}, \bibinfo {author} {\bibfnamefont
  {T.}~\bibnamefont {Kiessling}}, \bibinfo {author} {\bibfnamefont
  {D.}~\bibnamefont {Wolf}}, \bibinfo {author} {\bibfnamefont {A.}~\bibnamefont
  {Vollhardt}}, \bibinfo {author} {\bibfnamefont {A.}~\bibnamefont
  {Kaba{\v{s}}i}},  \emph {et~al.},\ }\href@noop {} {\bibfield  {journal}
  {\bibinfo  {journal} {arXiv preprint arXiv:1908.02759}\ } (\bibinfo {year}
  {2019})}\BibitemShut {NoStop}%
\bibitem [{\citenamefont {Yoshida}\ \emph
  {et~al.}(2019{\natexlab{c}})\citenamefont {Yoshida}, \citenamefont
  {Mizoguchi},\ and\ \citenamefont {Hatsugai}}]{Yoshida_MSkin_arXiv19}%
  \BibitemOpen
  \bibfield  {author} {\bibinfo {author} {\bibfnamefont {T.}~\bibnamefont
  {Yoshida}}, \bibinfo {author} {\bibfnamefont {T.}~\bibnamefont {Mizoguchi}},
  \ and\ \bibinfo {author} {\bibfnamefont {Y.}~\bibnamefont {Hatsugai}},\
  }\href@noop {} {\bibfield  {journal} {\bibinfo  {journal} {arXiv preprint
  arXiv:1912.12022}\ } (\bibinfo {year} {2019}{\natexlab{c}})}\BibitemShut
  {NoStop}%
\bibitem [{\citenamefont {Jiang}\ \emph {et~al.}(2019)\citenamefont {Jiang},
  \citenamefont {Lang}, \citenamefont {Yang}, \citenamefont {Zhu},\ and\
  \citenamefont {Chen}}]{Jiang_SkinDisord_PRB19}%
  \BibitemOpen
  \bibfield  {author} {\bibinfo {author} {\bibfnamefont {H.}~\bibnamefont
  {Jiang}}, \bibinfo {author} {\bibfnamefont {L.-J.}\ \bibnamefont {Lang}},
  \bibinfo {author} {\bibfnamefont {C.}~\bibnamefont {Yang}}, \bibinfo {author}
  {\bibfnamefont {S.-L.}\ \bibnamefont {Zhu}}, \ and\ \bibinfo {author}
  {\bibfnamefont {S.}~\bibnamefont {Chen}},\ }\href {\doibase
  10.1103/PhysRevB.100.054301} {\bibfield  {journal} {\bibinfo  {journal}
  {Phys. Rev. B}\ }\textbf {\bibinfo {volume} {100}},\ \bibinfo {pages}
  {054301} (\bibinfo {year} {2019})}\BibitemShut {NoStop}%
\bibitem [{\citenamefont {Matsumoto}\ \emph {et~al.}(2019)\citenamefont
  {Matsumoto}, \citenamefont {Kawabata}, \citenamefont {Ashida}, \citenamefont
  {Furukawa},\ and\ \citenamefont {Ueda}}]{Matsumoto_nHtoric_arXiv19}%
  \BibitemOpen
  \bibfield  {author} {\bibinfo {author} {\bibfnamefont {N.}~\bibnamefont
  {Matsumoto}}, \bibinfo {author} {\bibfnamefont {K.}~\bibnamefont {Kawabata}},
  \bibinfo {author} {\bibfnamefont {Y.}~\bibnamefont {Ashida}}, \bibinfo
  {author} {\bibfnamefont {S.}~\bibnamefont {Furukawa}}, \ and\ \bibinfo
  {author} {\bibfnamefont {M.}~\bibnamefont {Ueda}},\ }\href@noop {} {\bibfield
   {journal} {\bibinfo  {journal} {arXiv preprint arXiv:1912.09045}\ }
  (\bibinfo {year} {2019})}\BibitemShut {NoStop}%
\bibitem [{\citenamefont {Guo}\ \emph {et~al.}(2020)\citenamefont {Guo},
  \citenamefont {Wang},\ and\ \citenamefont {Kou}}]{Guo_nHToric_arXiv20}%
  \BibitemOpen
  \bibfield  {author} {\bibinfo {author} {\bibfnamefont {C.-X.}\ \bibnamefont
  {Guo}}, \bibinfo {author} {\bibfnamefont {X.-R.}\ \bibnamefont {Wang}}, \
  and\ \bibinfo {author} {\bibfnamefont {S.-P.}\ \bibnamefont {Kou}},\
  }\href@noop {} {\bibfield  {journal} {\bibinfo  {journal} {arXiv preprint
  arXiv:2001.04209}\ } (\bibinfo {year} {2020})}\BibitemShut {NoStop}%
\end{thebibliography}
%


\end{document}